\documentclass{jfm}
\usepackage{graphicx}
\usepackage{color,bm,amsmath,hyperref,ulem}
\usepackage[square,numbers,sort&compress]{natbib}
\usepackage[most]{tcolorbox}
\usepackage[figuresright]{rotating}
\usepackage[]{subfig}
\usepackage{supertabular}
\usepackage{booktabs}
\usepackage{commath}
\usepackage{amsfonts}
\usepackage{ulem}
\title{Sedimentation of  spheroids in Newtonian fluids with spatially varying viscosity }

\author{Vishal Anand and Vivek Narsimhan \footnote{To whom the correspondence should be addressed. \href{mailto:vnarsim@purdue.edu}{\texttt{vnarsim@purdue.edu}}}\\
\textit{Davidson School of Chemical Engineering, Purdue University,}\\ \textit{West Lafayette, Indiana 47907, USA}}

\begin{document}

\maketitle

\begin{abstract}
This paper examines the rigid body motion of a spheroid sedimenting in a Newtonian fluid with a spatially varying viscosity field. The fluid is at zero Reynolds number, and the viscosity varies linearly in space in an arbitrary direction with respect to the external force.  First, we obtain the correction to the spheroid’s rigid body motion in the limit of small viscosity gradients, using a perturbation expansion combined with the reciprocal theorem.  Next, we determine the general form of the particle’s mobility tensor relating its rigid body motion to an external force and torque.  The viscosity gradient does not alter the force/translation and torque/rotation relationships, but introduces new force/rotation and torque/translation couplings that are determined for a wide range of particle aspect ratios.  Finally, we discuss results for the spheroid’s rotation and center-of-mass trajectory during sedimentation. {A steady orientation arises at long time whose value depends  on the viscosity gradient direction and particle shape}.  These results are significantly different than when no viscosity gradient is present, where the particle stays at its initial orientation for all times.    We summarize the observations for prolate and oblate spheroids for different viscosity gradient directions and provide plots for the orientation and center of mass trajectory versus time.  We also provide guidelines to extend the analysis when the viscosity gradient exhibits a more complicated spatial behavior.
\end{abstract}

\section{Introduction}
\label{sec:intro_transient}
Fluids with inhomogeneous viscosity fields are ubiquitous around us. For example, certain biological fluids like mucus and extracellular microbial polymers are mixtures of fluids with different viscosities \citep{EColi_Berg}, and therefore exhibit variable viscosity, either with \citep{HelicalSwimmer_ViscosityGradients} or without sharp viscosity gradients \citep{Du_PRE_2012}. Similarly, gradients in temperature, salinity, or concentration may induce spatial variation in  viscosity, most commonly observed in marine ecosystems \citep{Arrigo_Robinson_Worthen_Dunbar_DiTullo_VanWoert_Science_99}. Finally, suspensions of particles in Newtonian fluids  (both active and passive) may be treated at the continuum level as fluids with viscosity varying with local volume fraction \citep{Rafai_Effective_Microswimmer,Hatwalne_PRL_2004}.  

\begin{figure}
    \centering
    \includegraphics[width=0.88\linewidth]{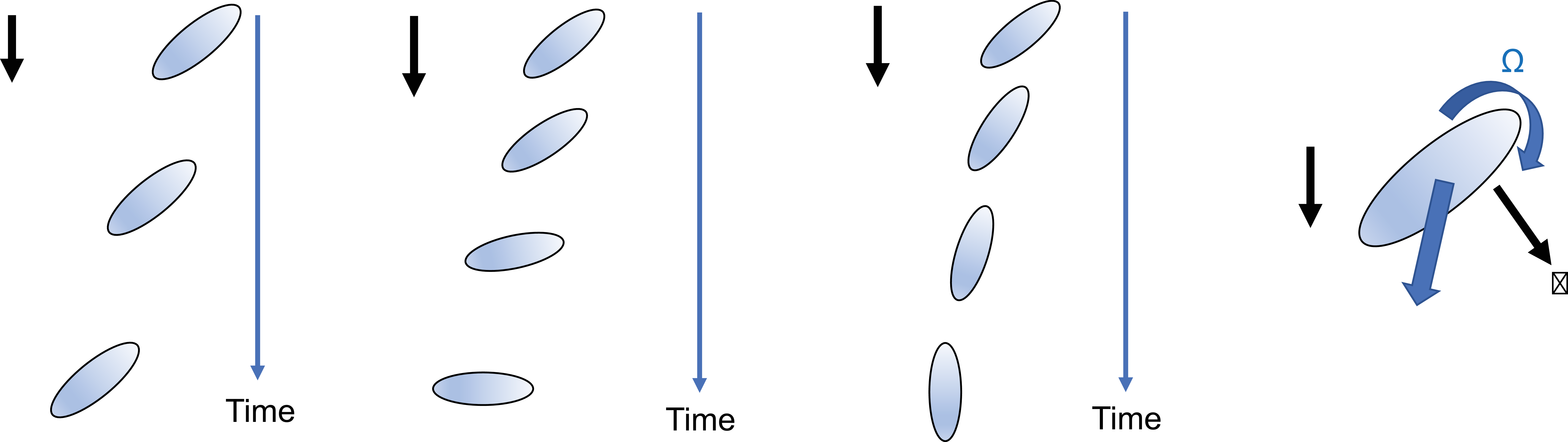}
    \caption{Illustration of spheroid orientation and trajectory during sedimentation in (a) Stokes flow (zero Reynolds number), (b) fluid with finite inertia, and (c) polymeric fluid with normal stresses (large Elasticity number).  This paper investigates the behavior when viscosity stratification is present – i.e., case (d)}
    \label{fig:example_sedimentation}
\end{figure}

In this manuscript, we will examine an idealized problem of a single spheroid sedimenting in a spatially varying viscosity field.  We will discuss the dynamics that are observed, and how they differ from other situations studied in the literature.  By now, it is well-known that in Stokes flow, a spheroid in gravity does not change its orientation due to the particle symmetry and the reversibility of the Stokes equations.  If the orientation starts out neither parallel or perpendicular to the gravity direction, the particle will move in a straight diagonal line, the direction of which is determined by the resistances parallel and perpendicular to the particle’s orientation vector (Fig. \ref{fig:example_sedimentation}a).  These dynamics will change only when symmetry breaking is present in the system.  One way in which symmetry breaking occurs is if fluid inertia is present \citep{Cox_1965,Khayat_Cox_1989,Auguste_JFM_2013}, or if the suspending fluid has normal stresses due to the presence of polymers \citep{Kim86,Galdi_JNNFM_2000,Galdi_MMAS_2011}.  For example, small fluid inertia generates a torque that orients the spheroid’s longest axis perpendicular to the external force -- the so-called ``broad side on'' configuration \citep{Dabade_Subramanian_2015}. Conversely, fluid viscoelasticity orients the spheroid such that its longest axis is along the force direction – i.e., an ``edge wise'' configuration \citep{Dabade_Subramanian_2015, Kim86}.  These effects markedly change the particle trajectory as well as the sedimentation speed (Fig. \ref{fig:example_sedimentation}), since the particle’s drag coefficient is a function of orientation and is minimized when the longest axis is along the force direction.

Another way in which symmetry breaking could occur is if there is a stratified fluid – i.e., variations in density, viscosity, or other fluid properties that alter the force and torque on the particle \citep{ More_Ardekani_Review}.  This area of research is relatively modern, and most of the efforts have examined the effect of density stratification on particle dynamics {including both solid particles \citep{DDA_JFM_2014,ADD_PRF_2017} and liquid droplets \citep{Droplets_Stratification_2020}}. When density increases along the gravity direction, it is found that the drag on a sphere is enhanced as confirmed by theory \citep{MCM_Density_JFM_2018}, experiments \citep{Yick_JFM_2009,LP_Density_JFM_1984} and simulations \citep{HKO-POF-2009,MABA_DensityJFM_2021}. The buoyancy force also leads to continuous deceleration and absence of a terminal velocity \citep{DDA_JFM_2014}. For anisotropic particles like spheroids, there has been some research to understand their settling behavior in density stratified fluids. Using a reciprocal theorem based approach, Varanasi and Subramanian \citep{Varanasi_Subramanian_2022} showed that the hydrostatic torque due to buoyancy originating from density stratification tends to rotate the particle in a broad side on configuration (similar to inertia),  which had earlier been also shown by Dandekar et al \citep{Dandekar_Shaik_Ardekani_JFM2020}. In the cited papers, it was assumed that the fluid density is not altered by the presence of the particle, and gives rise to a so-called “hydrostatic torque”.  However, the particle itself can alter the density field, and this additional effect can modify the particle torque \citep{Varanasi_Subramanian_2022, MABA_DensityJFM_2021}.  For example, density is often linked to a scalar field like temperature, which depends on a convection-diffusion equation.  Depending on the Peclet number, the density around the particle may or may not be coupled with the fluid flow.  In the low Peclet number limit, this additional torque is opposite the hydrostatic torque \citep{Varanasi_Subramanian_2022, MABA_DensityJFM_2021}.

Despite the advances in understanding microhydrodynamics of particles in density stratified fluids, there is a relative lack of literature examining viscosity stratified fluids, even though there is recent evidence suggesting that these effects would be more important than those due to variations in density in a variety of applications \citep{Dandekar_Ardekani_Swimming_Sheet_Viscosity,Jacquemin_Chemistry}.  For example, viscosity gradients are present in the swimming of micro-organisms, and it is of much interest to biologists to understand how organisms move in such complex environments \citep{Hatwalne_PRL_2004,Liebchen_PRL_2018,Rafai_Effective_Microswimmer,Sokolov_PRL_2009,SCS_PhytoPlankton_2017}, as well as roboticists who design microrobots in such fluids \citep{Zhuang_Sitti_2017,Nelson_Kalkiaoos_Abbott_AnnualReview,Kim_Lee_LeeNelson_Zhang_Choi_2016,Cilia_Swimming_2,Cilia_1,Microrobots_Science_Robotics}. Some questions that arise are: how does a spatially varying fluid viscosity affect the common swimming speed, propulsion, and efficiency \citep{Swimming_Cost_ShearThinning}? Do microswimmers orient themselves in preferable positions in response to the viscosity gradients \citep{Swimming_Orientation}? The common approach is to leverage a prototypical swimmer model (squirmers \citep{Vaseem_Elfring_Viscosity,Datt_Elfring_Viscosity_Gradient}, swimming sheet \citep{Dandekar_Ardekani_Swimming_Sheet_Viscosity, Eastham_Schoele_2020PRF}, Purcell's  swimmer \citep{Pak_PurcellSwimmer}, cilia \citep{Cilia_1,Cilia_Swimming_2}) and then couple it to the Stokes flow field with a variable viscosity.  

Currently, work has been performed on the motion of a single sphere in a viscosity varying fluid \citep{Datt_Elfring_Viscosity_Gradient}, but the effect of particle shape has yet to be considered.  We note that the authors in the cited paper found that viscosity gradients give rise to force/rotation and torque/translation coupling for the sphere’s motion, which would otherwise not exist if the viscosity gradient were absent.  This type of coupling is likely to give rise to unique rotational dynamics for orientable particles, which we will investigate in this paper. {We note that similar to the case of density stratified flows, the motion of the particle may induce changes in viscosity in the fluid around the particle that could lead to novel hydrodynamic phenomena. For the case of spheres, as shown by \citep{Vaseem_Elfring_Viscosity}, the changes in viscosity induced by particle motion do not lead to qualitative changes in the particle dynamics. Thus, in our paper, we do not account for changes in viscosity induced by the motion of the particle and only consider a spatially varying, time invariant viscosity field. However, later in Sec.\ref{sec:applicability}, we expound the probable strategy to incorporate the changes in viscosity due to particle motion, in the low Peclet number ($Pe$) limit.}

% Currently, work has been performed on the motion of a single sphere in a viscosity varying fluid \citep{Datt_Elfring_Viscosity_Gradient}, but the effect of particle shape has yet to be considered.  We note that the authors in the cited paper found that viscosity gradients give rise to force/rotation and torque/translation coupling for the sphere’s motion, which would otherwise not exist if the viscosity gradient were absent.  This type of coupling is likely to give rise to unique rotational dynamics for orientable particles, which we will investigate in this paper. \textcolor{red}{Similar to the case of density stratified flows, the motion of the particle may induce changes in viscosity in the fluid around the particle. Due to the additional changes in viscosity, there is a possibility that some novel hydrodynamic phenomena may be unearthed. For the case of spheres, as shown by \citep{Vaseem_Elfring_Viscosity}, the changes in viscosity induced by particle motion do not lead to any novel effects. In our paper, we do not account for changes in viscosity induced by the motion of the particle and only consider a spatially varying, time invariant viscosity field. However, later in Sec.\ref{sec:applicability}, we expound the probable strategy to incorporate the changes in viscosity due to particle motion, in the low Peclet number ($Pe$) limit.}

% \textcolor{red}{Motile phytoplanktons are known to travel through the depths of oceans both ways in search of nutrients during nights and in search of sunlight during the daytime \cite{SCS_PhytoPlankton_2017} }

With this motivation in mind, this manuscript will examine a problem of a single spheroid sedimenting in a Newtonian fluid with a spatially varying viscosity field.  The viscosity field varies linearly in space, and its gradient points in an arbitrary direction with respect to the direction of sedimentation (external force). Sec. \ref{sec:prob_statement} outlines the particle geometry and equations of motion.  Sec. \ref{sec:simulation} numerically solves for the particle’s rigid body motion in the limit of weak viscosity gradient using the reciprocal theorem.  Sec. \ref{sec:theory} uses the principles of symmetry to obtain a general expression for the particle mobility tensor relating the particle’s rigid body motion with the force and torque on the spheroid.  The force/translation and torque/rotation relationships are unaltered due to the presence of a viscosity gradient, but the viscosity gradient gives rise to new force/rotation and torque/translation coupling terms that depend on three undetermined coefficients. We determine the values of these coefficients numerically, and thus are able to solve the rigid body problem for arbitrary set of forcing, viscosity gradient direction, and particle geometry. Sec. \ref{sec:results} discusses some illustrative examples, wherein the orientations and trajectories of settling spheroids are analysed for different directions of the viscosity gradient.  We find that depending on the viscosity gradient direction, particle shape (prolate vs. oblate spheroid), and particle aspect ratio, the spheroid can take on different steady orientation angles.  The section concludes on how to extend the analysis to more complicated situations, followed by Sec. \ref{sec:conclusion} which summarizes all results.

We note that although this work primarily focuses on passive particles in viscosity stratified fluids, the results here will likely be important in a variety of contexts beyond this work.  For example, scientists are interested in quantifying the swimming of particles in viscosity varying fluids, and the mobility relationships developed here can be used for such applications.  Furthermore, understanding the rotation behavior and velocity field from a single, orientable particle can help understand their far-field hydrodynamic interactions in a dilute suspension, which is important in understanding concentration instabilities that arise in fibrous suspensions  \citep{Koch_Shaqfeh_JFM_1989, Herzhaft_experimental_1999, Bulter_Shaqfeh_Simulation_2002,Kuusela_Simulation_2003, Koch_Shaqfeh_Stable,Nicolai_1998_POF, Shin_Koch_Subramanian_1,Shin_Koch_Subramanian_2, Ramanathan_Saintillan_2012}.  {The force/rotation and torque/translation couplings shown in our paper has also been very recently shown in sedimentation of spheroids with non-uniform density, where the center of mass of the particle does not match its gravitational center \citep{Nissanka_Ma_Burton_JFM_2023} and therefore we expect that the theory developed in this paper may also aid in the understanding of the sedimentation of such mass polar spheroids as well. }We will not comment on this point further, noting that the work acts as a stepping stone for these more complicated problems when viscosity gradients are present.

\section{Problem Statement} \label{sec:prob_statement}

\begin{figure}
\centering
\subfloat[Prolate spheroid]{\includegraphics[width=0.45\linewidth]{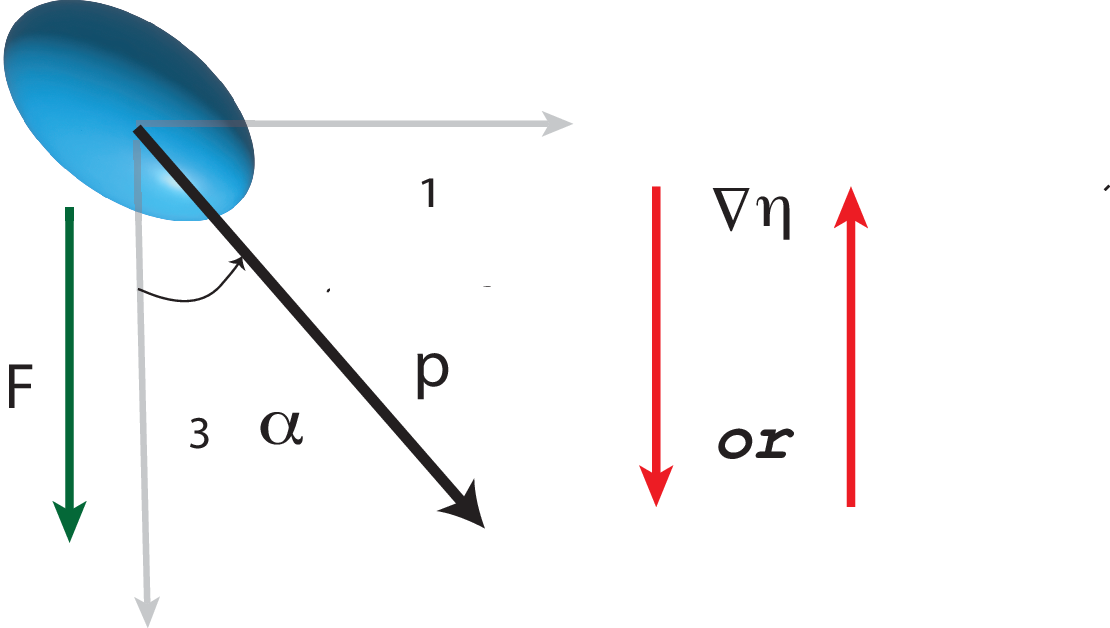}}
\subfloat[Oblate spheroid]{\includegraphics[width=0.45\linewidth]{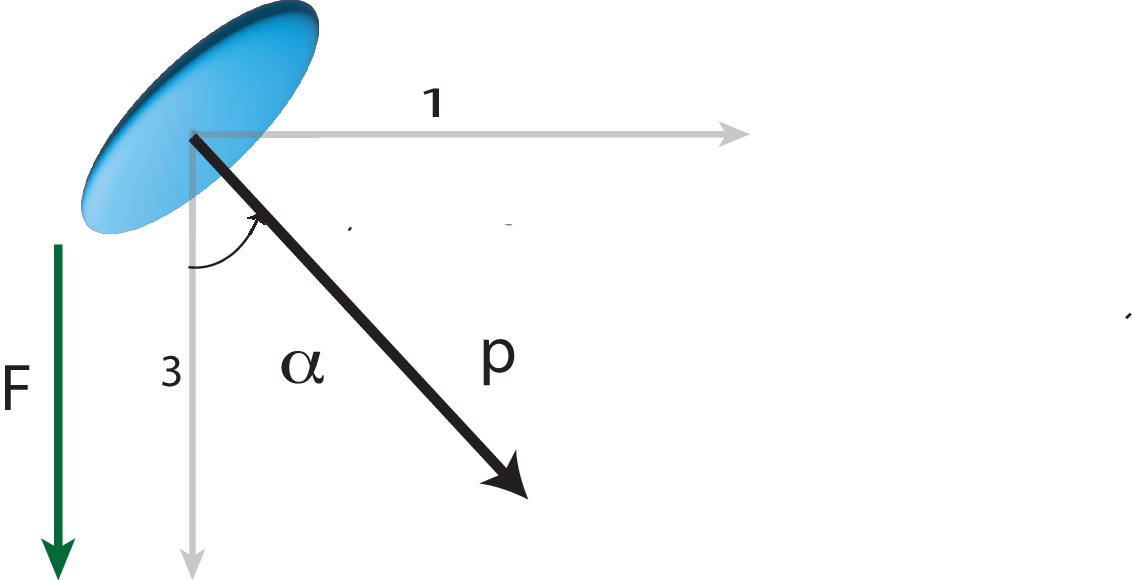}}
\caption{Schematic of a prolate and oblate spheroid falling under an external force acting in the 3-direction.  The viscosity gradient is along the 3-direction (parallel or anti-parallel).  The particle’s orientation vector $\boldsymbol{p}$ makes a polar angle $\alpha \in [0,\pi]$ with respect to the sedimentation direction. }
\label{fig:Schematic_Parallel}
\end{figure}

\begin{figure}
\centering
\subfloat[Prolate spheroid]{\includegraphics[width=0.45\linewidth]{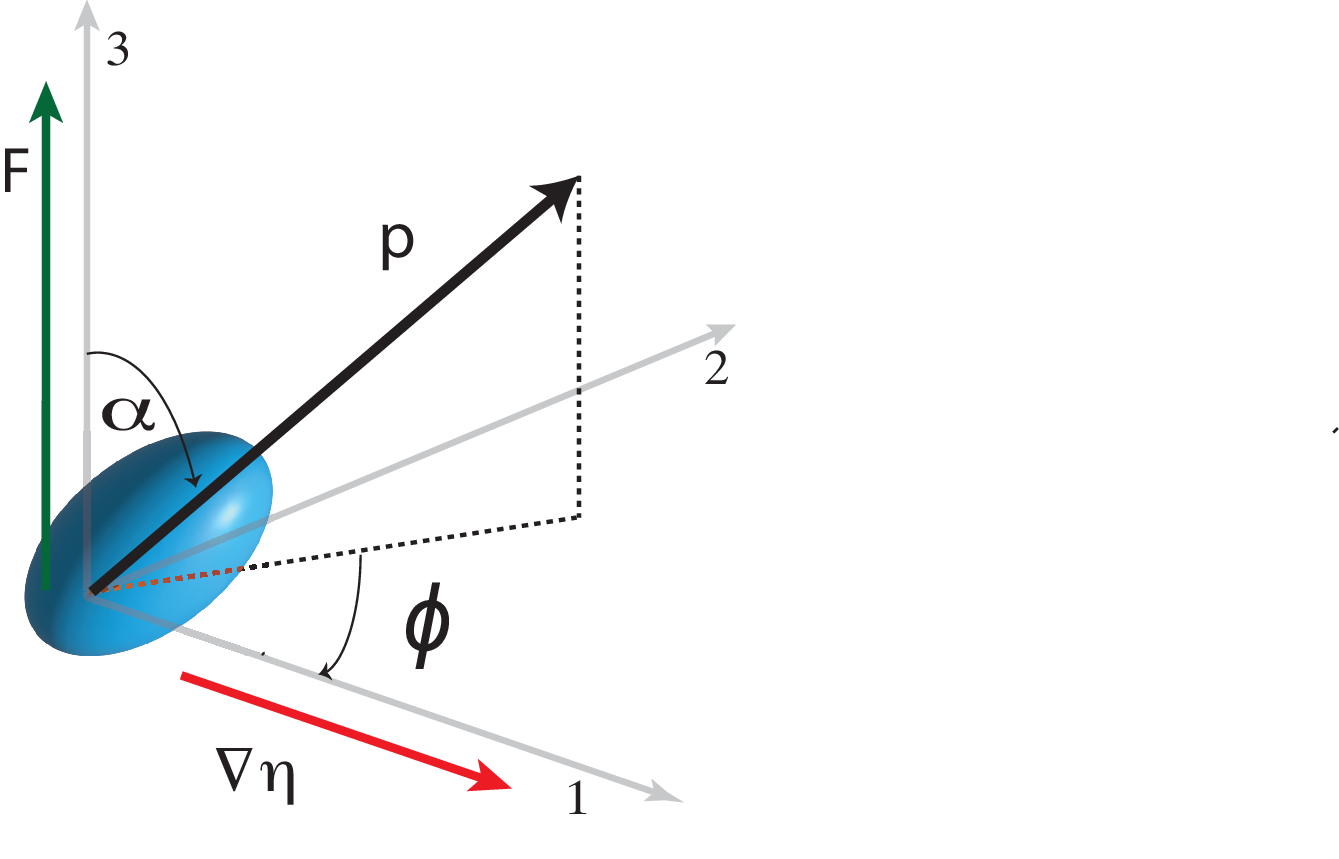}}
\subfloat[Oblate spheroid]{\includegraphics[width=0.45\linewidth]{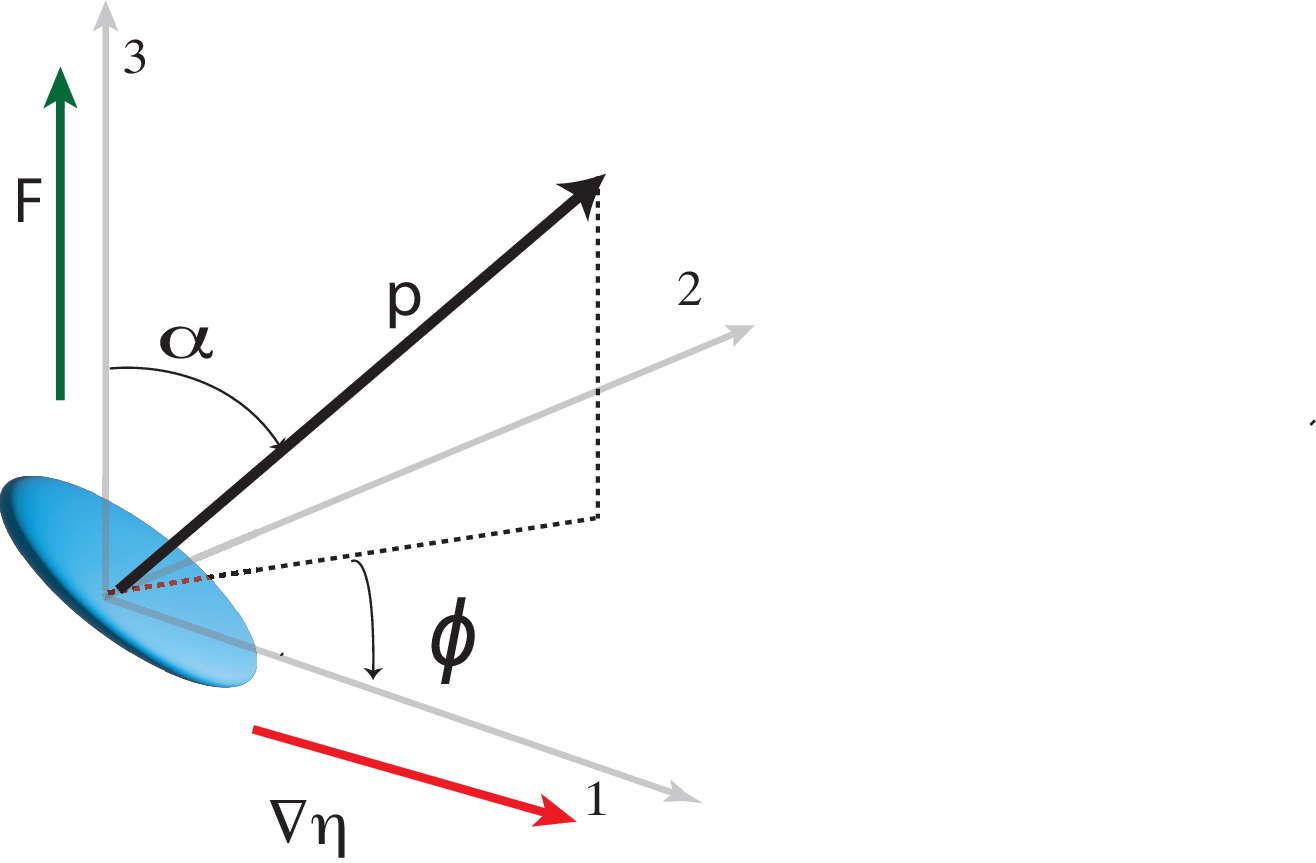}}
\caption{Schematic of a prolate and oblate spheroid falling under an external force $\boldsymbol{F}$ acting in the $3$-direction, while the viscosity varies spatially in the $1$-direction. The particle’s orientation $\boldsymbol{p}$ makes a polar angle $\alpha \in [0,\pi]$ with respect to the 3-direction, and makes an azimuthal angle $\phi \in [0,2\pi)$ in the 1-2 plane.}
\label{fig:Schematic_Perpendicular}
\end{figure}

\subsection{Problem Geometry}

The schematic of our system is shown in Figs.~\ref{fig:Schematic_Parallel} and \ref{fig:Schematic_Perpendicular}. We consider a torque-free spheroid under an external force $\boldsymbol{F}$ in a Newtonian fluid with a constant viscosity gradient $\boldsymbol{\nabla} \eta$. The force is in the positive 3-direction.  The viscosity gradient $\boldsymbol{\nabla} \eta$ can be co-linear with the force (Fig.~\ref{fig:Schematic_Parallel}, where $\boldsymbol{\nabla} \eta$  is in the $\pm$3-direction) or perpendicular to the force (Fig.~\ref{fig:Schematic_Perpendicular}, where $\nabla \eta$ is in the $+$1-direction). The spheroid has three semi-major axes of lengths ($a, b, c$), with $a\neq b = c$.  The initial center of mass of the spheroid is $(x_{01},x_{02},x_{03})=(0,0,0)$.

We will define the spheroid’s orientation vector $\boldsymbol{p}$ as the direction along its unequal axis (i.e., the $a$-axis).  Two different cases arise. A prolate spheroid has $\boldsymbol{p}$ along its longest axis, while an oblate spheroid has $\boldsymbol{p}$ along its shortest axis. Another way to parameterize the particle shape is through an aspect ratio parameter $A_R$ and equivalent radius $R$. Here, $A_R$ is the ratio $a/b$, while $R$ is the radius of an equivalent sphere with the same volume.
\begin{equation}
\label{eq:aspectratio}
    A_R =\frac{a}{b} \qquad
    R =(abc)^{1/3}   
\end{equation}

The two systems of parameterization are connected by the following relationship:
\begin{equation}
\label{eq:part_def_2}
a=R A_{R}^{2 / 3}, \quad b=c= R A_R^{-1/3} 
\end{equation}
 
Evidently, a prolate particle has its aspect ratio parameter $A_R > 1$, while an oblate particle has its aspect ratio parameter $A_R < 1$.
Figs. ~\ref{fig:Schematic_Parallel} and \ref{fig:Schematic_Perpendicular} describe the polar and azimuthal angles $\alpha \in [0,\pi]$ and $ \phi \in [0,2\pi)$  for the particle orientation.  The next section discusses the equations of motion and the rheology of the fluid.

\subsection{Equations of motion and fluid rheology}
\label{sec:rheology}

The fluid surrounding the particle is incompressible and Newtonian.  The fluid also has negligible inertia – in other words, the Reynolds number based on the particle’s largest length scale 
$Re=(\rho_f U L_{\text{max}})/\eta_0 \approx 0 $. Here, $\rho_f$ is the density of the fluid surrounding the particle, $U$ is the translation speed of the particle, $L_{\text{max}} =max(a,b)$ is the largest axis of the particle, and $\eta_0$ is fluid's viscosity at the origin if the particle were absent. 

When these conditions hold, the momentum and mass balance equations in the fluid are given as:
\begin{equation}
\label{eq:stress_balance}
\frac{\partial \sigma_{ij}}{\partial x_j}=0; \qquad \frac{\partial v_i}{\partial x_i}=0
\end{equation}
where $\sigma_{ij}$ is the stress tensor and $v_i$ is the  velocity field. Einstein summation convention is assumed -- i.e., repeated indices are summed. The stress tensor takes the following form:

\begin{equation}
\label{eq:stress_definition}
    \sigma_{ij} =-p\delta_{ij}+\eta(\boldsymbol{x})\dot{\gamma}_{ij}
\end{equation}
where $p$ is the pressure,  $\dot{\gamma}_{ij} = \frac{\partial v_j}{\partial x_i}+\frac{\partial v_i}{\partial x_j}$ is twice the strain rate tensor, and $\eta$ is the viscosity of the medium. In this problem, the viscosity is independent of the strain rate but exhibits a spatial dependence. The viscosity field is:
\begin{equation}
\label{eq:viscosity_variation_general}
    \eta(\boldsymbol{x}) =\eta_{0}\left( 1 +\frac{\beta}{R} \boldsymbol{\hat{d}} \cdot \boldsymbol{x} \right)
\end{equation}

In the above equation, $\eta_0$ is the viscosity at the origin and $\nabla \eta = \frac{\eta_0}{R} \beta \boldsymbol{\hat{d}}$ is a constant viscosity gradient with dimensionless magnitude $\beta$ and unit direction $\boldsymbol{\hat{d}}$. 

The goal of the problem is to solve Eqs.~\eqref{eq:stress_balance}, \eqref{eq:stress_definition}, and \eqref{eq:viscosity_variation_general}  for the stress and velocity around the particle.  The equations have to be solved with the following boundary conditions:

% \begin{subequations} \label{eq:velocity_boundary_1}
% \begin{align} 
%     v_i &\rightarrow 0, & &|x_i| \rightarrow \infty \\
% v_i&=U_i+\epsilon_{ijk}\Omega_j  (x_k-x_k^{cm}), & &x_i \in S_p
% \end{align}
% \end{subequations}

\begin{subequations}
\begin{equation} \label{eq:velocity_boundary_1}
    v_i \rightarrow 0, \qquad |x_i| \rightarrow \infty
\end{equation}
\begin{equation}
v_i=U_i+\epsilon_{ijk}\omega_j  (x_k-x_k^{cm}),    \qquad x_i \in S_p
\end{equation}
\end{subequations}
where $(U_i,\omega_i)$ are the rigid body velocities of the particle, $S_p$ is the particle surface, $x_k^{cm}$ is the center of mass, and $\epsilon_{ijk}$ is the Levi-Civita symbol. An additional constraint is that the particle's external force and torque are specified.  These are:

 \begin{equation} \label{eqn:force_torque_defn}
     F_i = -\int_{S_p}\sigma_{ij}n_jdS, \quad 
     T_i = -\int_{S_p}\epsilon_{ikl}(x_k-x_k^{\text{cm}})\sigma_{lj}n_jdS,
 \end{equation}
where $n_i$ is the outward-pointing vector on the particle surface. For this problem, $T_i =0$.

 In this problem, we specify the viscosity field to have a constant gradient, while for other problems the viscosity field is often found by solving a scalar quantity like temperature or concentration that is a solution to a convection-diffusion equation around the particle.  For such problems in the limit of small Peclet number (one-way coupling), the results will be very similar to the problem formulated here, albeit with minor quantitative differences.  A more detailed discussion will be provided at the end of the manuscript (Sec. \ref{sec:applicability}).

\subsection{Non dimensionalization, dimensionless numbers and perturbation expansion}
 Unless otherwise noted, all quantities from here on out will be written in non-dimensional form.  Lengths will be scaled by the average particle size $R$, forces by its magnitude $F$, and viscosities by its value at the origin $\eta_0$.  Velocities will be scaled by the Newtonian sedimentation velocity $U =\frac{F}{6\pi \eta_0 R}$ , times by $t_c=R/U$, strain rates and rotational velocities by $\dot{\gamma}_c =1/t_c$, stresses and pressures by $\eta_0\dot{\gamma}_c$, and torques (if present) by $FR$.

The dynamics of the spheroid will depend on the following dimensionless quantities – the particle aspect ratio parameter $A_R$, the particle orientation $\boldsymbol{p}$ (characterized by angles $\alpha$ and $\phi$), and the non-dimensional viscosity gradient $\boldsymbol{\nabla} \eta$ (characterized by magnitude $\beta$ and direction $\boldsymbol{\hat{d}}$):

\begin{equation}
    A_R = \frac{b}{a}; \qquad \boldsymbol{p} = [\sin\alpha \cos \phi, 
 \sin \alpha \sin \phi, \cos \alpha]; \qquad \nabla \eta = \beta \boldsymbol{\hat{d}}
\end{equation}
 
In dimensionless form, the viscosity of the fluid in Figs. \ref{fig:Schematic_Parallel} and \ref{fig:Schematic_Perpendicular} is the following:

\begin{equation}
    \eta =1\pm \beta x_3 \qquad
    \eta = 1+\beta x_1
\end{equation}
where the first {equation namely, $\eta =1 \pm \beta x_3$} corresponds to the {situation} where the viscosity gradient is parallel ($+3$ direction) or anti-parallel ($-3$ direction) to the external force, and the second {equation namely, $\eta =1 + \beta x_1$ corresponds to the case} where the viscosity gradient is perpendicular to the external force.  For a general viscosity gradient $\boldsymbol{\nabla} \eta$, the particle motion will be a superposition of the solutions for the two cases listed above.  We will examine particle dynamics in the limit of small viscosity gradient:

\begin{equation}
    Re \ll \beta \ll 1
\end{equation}

The above condition indicates that one can neglect fluid inertia and perform a regular perturbation expansion in $\beta$.  We will solve for the rigid body motion up to $\mathcal{O}(\beta)$, both numerically and semi-analytically using symmetry arguments listed in the next sections.

\section{Numerical solution to particle dynamics}
\label{sec:simulation}
\subsection{Reciprocal theorem}

We will determine the rigid body motion of the spheroid by performing a perturbation expansion in the non-dimensional viscosity gradient $\beta \ll 1$.
We perturb the dependent variables as follows:

\begin{equation} \label{eq:perturbation_expansion}
\begin{split}
{\{v_i, p, \sigma_{ij},\dot{\gamma}_{ij},F_i, U_i, \omega_i \}}= &\left\{v_{i}^{(0)}, p^{(0)}, \sigma_{ij}^{(0)},\dot{\gamma}_{ij}^{(0)},F_i^{(0)},U_i^{(0)}, \omega_i^{(0)}\right\} \\
&+\beta\left\{u_{i}^{(1)}, p^{(1)}, \tau_{ij}^{(1)},\dot{\gamma}_{ij}^{(1)},F_i^{(1)},U_i^{(1)}, \omega_i^{(1)}\right\}+\ldots    
\end{split}
\end{equation}
and solve for the momentum and mass balances Eqs.~\eqref{eq:stress_balance} - \eqref{eqn:force_torque_defn} at each order in $\beta$. At leading order, the spheroid sediments in a zero Reynolds number fluid with a constant, non-dimensional viscosity $\eta =1$ and a non-dimensional external force $F = 1$:
\begin{equation}
\label{eq:stokes_leading_order}
    \frac{\partial ^2 v_i^{(0)}}{\partial x_k \partial x_k}-\frac{\partial p^{(0)}}{\partial x_i} =0 ; \qquad \frac{\partial v_i^{(0)}}{\partial x_i} =0 ;\qquad F_i =\delta_{i3} ;\qquad T_i =0
\end{equation}

The solution to the above problem is given in many classical texts (for example see \citep{KimKarilla2005}).  The velocity field is presented in Appendix A, while the rigid body motion satisfies the classical resistance relationship:

\begin{equation}
\label{eq:Resistance_Order_Leading}
\left(\begin{array}{cc}
\boldsymbol{R}^{FU} & \boldsymbol{R}^{F\omega}   \\
\boldsymbol{R}^{TU} & \boldsymbol{R}^{T\omega} \\
\end{array}\right) \cdot \left(\begin{array}{c}
\boldsymbol{U}^{(0)} \\
\omega^{(0)} \\
\end{array}\right)=\left(\begin{array}{c}
\boldsymbol{F} \\
\boldsymbol{T} \\
\end{array}\right)
\end{equation}
In this equation, $(\boldsymbol{R}^{FU},\boldsymbol{R}^{F\omega},\boldsymbol{R}^{TU},\boldsymbol{R}^{T\omega})$ are the resistance tensors for a spheroid, which are given in Appendix B. The external force and torque are given in Eq.~\eqref{eq:stokes_leading_order} .

At the next order of approximation $\mathcal{O}(\beta)$, the momentum and mass balance equations become Stokes flow with an extra fluid body force $b_i$:
\begin{equation}
    \frac{\partial ^2v_i^{(1)}}{\partial x_k\partial x_k} -\frac{\partial p^{(1)}}{\partial x_i} +b_i =0 ;\qquad \frac{\partial v_i^{(1)}}{\partial x_i} =0
\end{equation}

Here the body force is due to the spatially varying viscosity field:
\begin{equation}
\label{eq:body_force_defined}
    b_i =\frac{\partial\tau_{ij}^{ex}}{\partial x_j}; \qquad \tau_{ij}^{ex} = (\boldsymbol{\hat{d}} \cdot \boldsymbol{x})\dot{\gamma}_{ij}^{(0)}
\end{equation}
In the above Eq.~\eqref{eq:body_force_defined}, $\tau_{ij}^{ex}$ is the extra stress tensor, and $\dot{\gamma}^{(0)}_{ij}$ is twice the rate of strain tensor from the leading order velocity field.

We employ the reciprocal theorem to solve for the translational and rotational velocity for the $\mathcal{O}(\beta)$ problem.  This theorem has a storied history in the Stokes flow community, as is often used to solve for the rigid body motion of particles in Stokes flow with a fluid body force.  The derivation is stated in Appendix C and we present the main results below.  In brief, the translational and rotational velocities follow a resistance relationship similar to Eq.~\eqref{eq:Resistance_Order_Leading}, except the forces and torques are replaced by an effective viscosity-stratified force and torque:

\begin{equation}
\label{eq:Resistance_Order_Beta}
\left(\begin{array}{cc}
\boldsymbol{R}^{FU} & \boldsymbol{R}^{F\omega}   \\
\boldsymbol{R}^{TU} & \boldsymbol{R}^{T\omega} \\
\end{array}\right) \cdot \left(\begin{array}{c}
\boldsymbol{U}^{(1)} \\
\omega^{(1)} \\
\end{array}\right)=\left(\begin{array}{c}
\boldsymbol{F}^{vs} \\
\boldsymbol{T}^{vs} \\
\end{array}\right)
\end{equation}

The viscosity-stratified force and torque are given as follows:
\begin{equation}
\label{eq:PolyForceTorque}
    F_i^{vs} = -\int_{V_{out}}\frac{\partial v_{ki}^{trans}}{\partial x_j}\tau_{kj}^{ex} dV ; \qquad T_i^{vs}= -\int_{V_{out}}\frac{\partial v_{ki}^{rot}}{\partial x_j}\tau_{kj}^{ex} dV
\end{equation}
where the integrals are evaluated over the volume $V_{out}$ outside the particle.  The quantities $v_{ki}^{trans}$ and $v_{ki}^{rot}$ are the Stokes flow velocity fields around a spheroid in the $k$ direction due to unit translation or unit rotation in the $i$ direction.  These quantities are derived from the same velocity fields listed in Appendix A.

\subsection{Numerical implementation:}

The volume integrals in the Eq.~\eqref{eq:PolyForceTorque} are difficult to evaluate analytically.
A custom-made MATLAB code was written to calculate the spheroid’s rigid body motion.  This code is similar to the approach used in our prior papers to investigate the motion of ellipsoids in weakly viscoelastic fluids \citep{Wang_Narsimhan_POF}, except that here the extra stress is modified to account for the viscosity gradient.  First, we transform from the laboratory frame to the particle frame of reference where that the origin is at the particle’s center of mass and the Cartesian coordinate axes align with the particle’s principle axes.  We then evaluate the volume integrals in Eq.~\eqref{eq:PolyForceTorque}  for the viscosity-stratified force and torque, using an elliptical coordinate system and performing Gaussian quadrature via Legendre polynomials. {A mesh convergence study showed that $60$ elements in the radial direction, $80$ elements in the polar direction and $100$  elements in the azimuthal direction yielded a sufficiently accurate mesh for our analysis.} We then solve the matrix equations Eq.~\eqref{eq:Resistance_Order_Leading} and Eq.~\eqref{eq:Resistance_Order_Beta} for the rigid body motions at $\mathcal{O}(1)$ and $\mathcal{O}(\beta)$, and transform back to the laboratory frame.  The particle’s center of mass and orientation are evolved by solving the rigid body dynamics:
\begin{equation}
    \frac{d \boldsymbol{x}^{cm}}{dt}  = \boldsymbol{U}; \qquad \quad              \frac{d \boldsymbol{p}}{dt} = \boldsymbol{\omega} \times \boldsymbol{p}    
\end{equation}
% \begin{align}
% \frac{d x^{cm}}{dt}  &= U \\              \frac{d p}{dt} &= \omega \times p
% \end{align}
We use a forward Euler scheme with $\Delta t = 0.01$. More details are found in our prior publications \citep{Wang_Narsimhan_POF,anand_narsimhan_2023}.
\subsubsection{Verification of code}

\label{sec:code_verify}
For the case of a sphere sedimenting in a linear, imposed viscosity gradient, we refer to the work by \citep{Datt_Elfring_Viscosity_Gradient}. Specifically, Eqs.$(7,8)$ in \citep{Datt_Elfring_Viscosity_Gradient} are the resistance relationships for the external force $\boldsymbol{F}$ and torque $\boldsymbol{T}$ on a sphere of radius $a$ in a fluid with a constant viscosity gradient $\boldsymbol{\nabla} \eta$, with translational velocity $\boldsymbol{U}$ and rotational velocity $\boldsymbol{\omega}$. For convenience, these equations are reproduced in dimensional form here:

\begin{equation*}
\begin{aligned}
&\boldsymbol{F}= 6 \pi a \eta_{0} \boldsymbol{U} - 2 \pi a^{3} \boldsymbol{\nabla} \eta \times \boldsymbol{\omega} \\
&\boldsymbol{T}= 2 \pi a^{3} \boldsymbol{\nabla} \eta \times \boldsymbol{U} + 8 \pi \eta_{0} a^{3} \boldsymbol{\omega}
\end{aligned}
\end{equation*}

\begin{figure}
\centering
\subfloat[$\boldsymbol{F} \propto \boldsymbol{\hat{x}}$, $\boldsymbol{\nabla} \eta \propto \beta \boldsymbol{\hat{y}}$]{\includegraphics[width=0.5\linewidth]{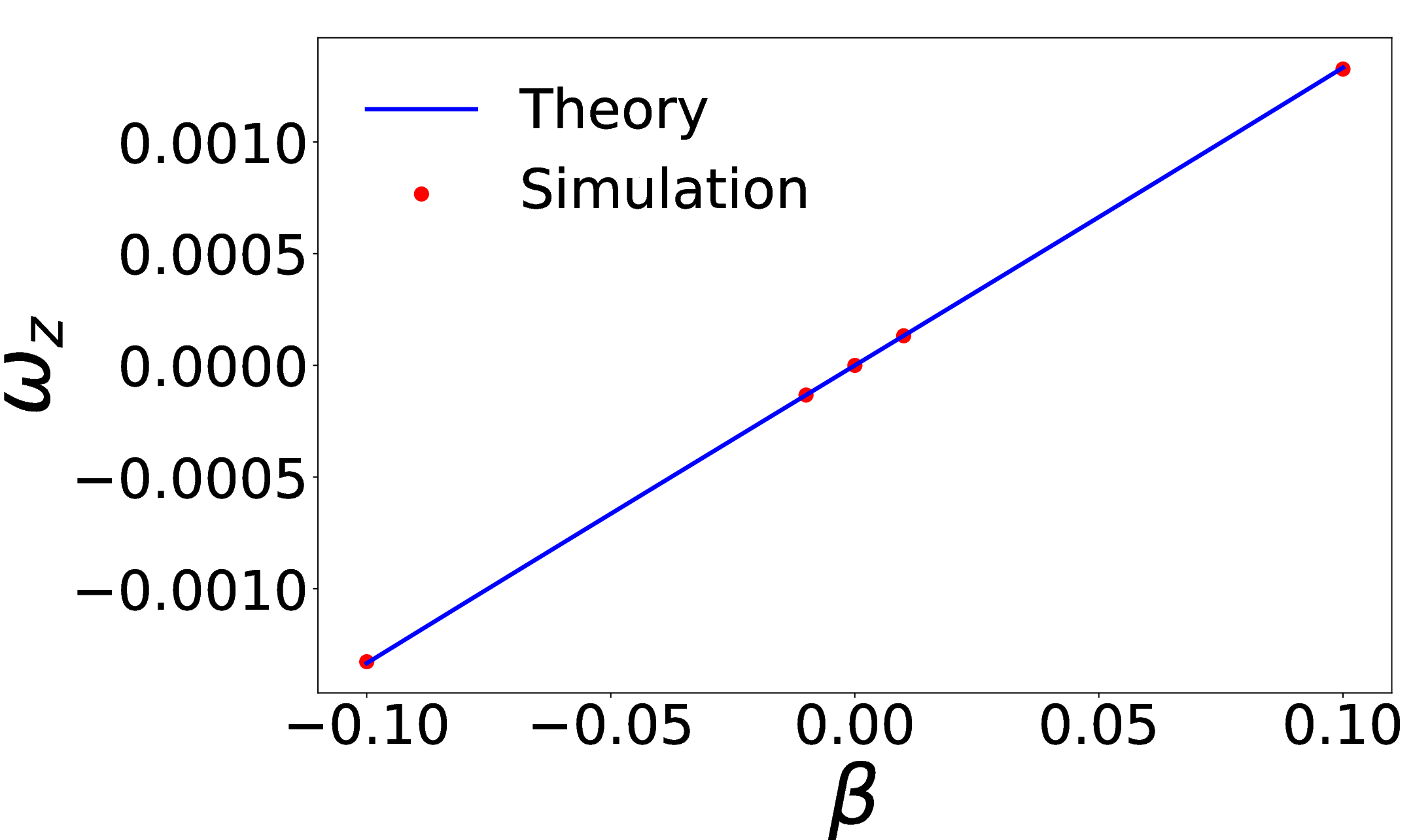}}
\subfloat[$\boldsymbol{F} \propto \boldsymbol{\hat{x}}$, $\boldsymbol{\nabla} \eta \propto \beta \boldsymbol{\hat{x}}$]{\includegraphics[width=0.5\linewidth]{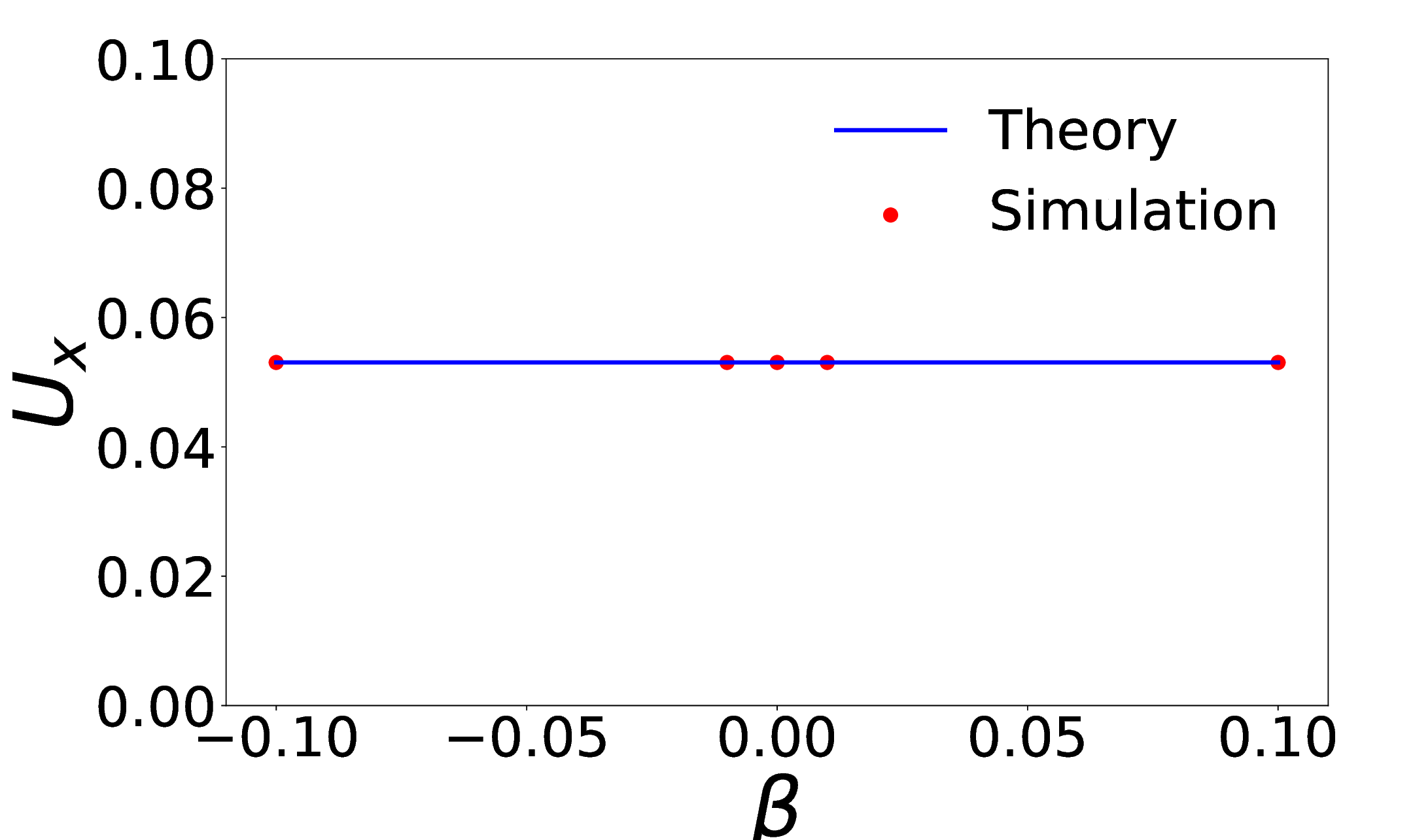}}

\caption{Code validation for a sphere sedimenting in a fluid with a prescribed viscosity gradient in the (a) $y$-direction and (b) $x$-direction. For all the cases, the external force is a unit vector acting in the $x$-direction, while the external torque is $\boldsymbol{T}=0$. The radius and fluid viscosity are $a = 1$ and $\eta_0 = 1$, respectively. The results of the theory are from \citep{Datt_Elfring_Viscosity_Gradient}, expanded in Sec.~\ref{sec:code_verify}.}
\label{fig:Sphere_Verification}
\end{figure}

\begin{enumerate}
    \item \underline{Spatial variation in $y$ direction:}  For a torque-free $(\boldsymbol{T} = 0)$ sphere sedimenting in the $x$-direction where the dimensional viscosity gradient is along the $y$-direction $\boldsymbol{\nabla}\eta = \beta \frac{\eta_0}{a} \boldsymbol{\hat{y}}$, the above equations give us:
\begin{align}
\label{eq:Velocity_Verification_y}
      U_x = \frac{F_x}{\pi a \eta_0 (6-0.5\beta^2)}; \qquad  U_y = U_z = 0  \\
\label{eq:Orientation_Verification_y}
    \omega_x = \omega_y = 0 \qquad \omega_z = 0.25\beta \frac{F_x}{\pi \eta_0 a^2 (6-0.5\beta^2)}
\end{align}
 
The analytical result for $\omega_z$ in Eq.~\eqref{eq:Orientation_Verification_y} is compared against the results of the numerical simulation and the comparison is shown in Fig. \ref{fig:Sphere_Verification}(a) showing an accurate match.

\item \underline{Spatial variation in $x$ direction:} Similarly, for a torque-free sphere sedimenting in the $x$-direction where the dimensional viscosity gradient is along the $x$-direction $\boldsymbol{\nabla} \eta = \beta \frac{\eta_0}{a} \boldsymbol{\hat{x}}$, the above equations give:

\begin{align}
    \label{eq:Velocity_Verification_x}
      U_x = \frac{F_x}{6\pi a \eta_0}; \qquad  U_y = U_z = 0  \\
\label{eq:Orientation_Verification_x}
    \omega_x = \omega_y =  \omega_z = 0
\end{align}
We compare the results of the simulation against Eq.~\eqref{eq:Velocity_Verification_x} in Fig \ref{fig:Sphere_Verification}(b) where a good match is seen. {The relative error in the simulation with respect to the simulation results is $< 1\%$}
\end{enumerate}

% \begin{figure}[t]
% \centering
% \subfloat[]{\includegraphics[width=0.5\linewidth]{Validation_A.eps}}
% \subfloat[]{\includegraphics[width=0.5\linewidth]{Validation_B.eps}}

% \caption{Verification of the code used for simulation by simulating a sphere sedimenting in a fluid with prescribed viscosity gradient in (a) $y$ direction and (b) $x$ direction. For all the cases, the external force is a unit vector acting in $x$ direction, while the external torques is $0$. The results of the theory are from \citep{Datt_Elfring_Viscosity_Gradient}, expanded in Sec.~\ref{sec:code_verify}.}
% \label{fig:Sphere_Verification}
% \end{figure}
\section{Semi analytical theory}
\label{sec:theory}

\subsection{Introduction and motivation}
\label{sec:analytical_development}
The simulations described in the previous section solve the rigid body motion of the particle, but are computationally intensive.  At each timestep, one has to evaluate six volume integrals in Eq.~\eqref{eq:PolyForceTorque} to obtain the viscosity-stratified force and torque.  Furthermore, a new time sweep has to be performed if one examines a different viscosity gradient direction and magnitude. 

An alternative approach to obtain the same dynamics is to develop a semi-analytical theory based on the symmetry of the problem.   Such a theory will give the general form of the particle’s motion in terms of three undetermined constants, which in turn can be found by performing simulations at three specific configurations.  The result of this analysis is that one can cheaply obtain the particle’s motion for an arbitrary set of particle orientations, forcing, and viscosity gradients.  

What we are doing is essentially finding the general form of the mobility tensor when a viscosity gradient is present.  Thus, the analysis below will not only give general information about the force-rotation coupling of these orientable particles, but can also give results for the case when a torque is applied – for example, the torque-translation coupling.  A description is below.

\subsection{General form of mobility tensor}

The governing momentum and continuity Eqns.~\eqref{eq:stress_balance} - \eqref{eqn:force_torque_defn} are linear in the external force and torque $(\boldsymbol{F}, \boldsymbol{T})$.  Thus, the translational and rotational velocities  $(\boldsymbol{U},\boldsymbol{\omega})$ are also linear in these quantities and obey the following relationship:

% \begin{equation}
% \left(\begin{array}{c}
% \boldsymbol{U} \\
% \omega \\
% \end{array}\right)=\left(\begin{array}{cc}
% \boldsymbol{A} & \boldsymbol{B}   \\
% \boldsymbol{C} & \boldsymbol{D}  \\
% \end{array}\right)\left(\begin{array}{c}
% \boldsymbol{F} \\
% \boldsymbol{T} \\
% \end{array}\right)
% \end{equation}

\begin{equation} \label{eqn:grand_mobility}
\left(\begin{array}{c}
\boldsymbol{U} \\
\omega \\
\end{array}\right)=\left(\begin{array}{cc}
\boldsymbol{A} & \boldsymbol{B}   \\
\boldsymbol{B}^{T} & \boldsymbol{D}  \\
\end{array}\right) \cdot \left(\begin{array}{c}
\boldsymbol{F} \\
\boldsymbol{T} \\
\end{array}\right)
\end{equation}
Here, $(\boldsymbol{A},\boldsymbol{B},\boldsymbol{D})$ are mobility tensors that are non-dimensionalized by $\left(\frac{U}{F},\frac{U}{FR},\frac{U}{FR^2}\right) = \left( \frac{1}{6\pi \eta_0 R}, \frac{1}{6\pi \eta_0 R^2}, \frac{1}{6\pi \eta_0 R^3}\right)$, respectively .  In a constant viscosity fluid, these tensors are only a function of the particle shape and orientation, characterised by the aspect ratio parameter $A_R$ and the orientation vector $\boldsymbol{p}$. If a viscosity gradient is present, the tensors will also be a function of the non-dimensional viscosity gradient $\boldsymbol{\nabla}\eta = \beta \boldsymbol{\hat{d}}$.  Note:  the the off-diagonal terms of the matrix in Eq. \eqref{eqn:grand_mobility} are transposes of each other as can be proved by the reciprocal theorem (not shown here).{Additionally, the mobility tensors $(\boldsymbol{A},\boldsymbol{D})$ are symmetric for arbitrary shaped particle in viscosity stratified fluid, as has been shown with due rigor with aid of reciprocal theorem earlier in literature \citep{Oppenheimer_Stone_2016}}

In the limit of $\beta \ll 1$, we expand the mobility tensors in a Taylor series as follows:

\begin{equation}
    \left\{{A}_{ij},{B}_{ij},{D}_{ij}\right\} =\left\{{A}^{(0)}_{ij},{B}^{(0)}_{ij},{D}^{(0)}_{ij}\right\}+\beta\left\{{A}^{(1)}_{ij},{B}^{(1)}_{ij},{D}^{(1)}_{ij}\right\}
\end{equation}

At leading order ($O(1)$), the tensors are the same as those for the particle in a constant viscosity fluid.  These quantities are well-characterized and formulas are given in Appendix B for a general ellipsoid.  Specifically, for the case where the particle has an orientation vector $\boldsymbol{p}$, they take the form:

% \begin{align}
%    \boldsymbol{A}^{0} &=\frac{1}{6\pi R\eta_0} \left[c_1\left(\boldsymbol{I}-\boldsymbol{p}\boldsymbol{p}\right)+c_2\boldsymbol{p}\boldsymbol{p}\right]  \\
%    \boldsymbol{D}^{0} &=\frac{1}{8\pi R^3\eta_0} \left[c_3\left(\boldsymbol{I}-\boldsymbol{p}\boldsymbol{p}\right)+c_4\boldsymbol{p}\boldsymbol{p}\right]   \\
%    \boldsymbol{B}^{0} &=0,
% \end{align}

\begin{subequations} \label{eq:mobility_tensor_leading_order}
\begin{align}
\label{eq:mobility_tensor_Newtonian_1}
   {A}^{(0)}_{ij} &= c_1\left(\delta_{ij}-{p_i}{p_j}\right)+c_2{p_i}{p_j}  \\  \label{eq:mobility_tensor_Newtonian_2}
   {D}^{(0)}_{ij} &= c_3\left(\delta_{ij}-{p_i}{p_j}\right)+c_4{p_i}{p_j}   \\
   {B}^{(0)}_{ij} &=0
\end{align}    
\end{subequations}
% \begin{align}
% \label{eq:mobility_tensor_Newtonian_1}
%    {A}^{(0)}_{ij} &= c_1\left(\delta_{ij}-{p_i}{p_j}\right)+c_2{p_i}{p_j}  \\  \label{eq:mobility_tensor_Newtonian_2}
%    {D}^{(0)}_{ij} &= c_3\left(\delta_{ij}-{p_i}{p_j}\right)+c_4{p_i}{p_j}   \\
%    {B}^{(0)}_{ij} &=0,
% \end{align}
where $c_1,c_2,c_3$ and $c_4$ are functions of the {aspect} ratio parameter and are given in Appendix B.

At $\mathcal{O}(\beta)$, the motion will be linear in $\boldsymbol{\nabla} \eta$.  Thus, in non-dimensional form, the mobility tensors take the following structure:

\begin{subequations}
\label{eq:Mobility_Forms}
\begin{align}
    A_{ij}^{(1)} &=\alpha_{ijk}\hat{d}_k\\
    D_{ij}^{(1)} &=\beta_{ijk}\hat{d}_k \\ 
    \label{eq:Mobility_Forms_B}
     B_{ji}^{(1)} &= M_{ikj}\hat{d}_k 
\end{align}
\end{subequations}
where $\hat{d}_k$ is the direction of the viscosity gradient.  Therefore the problem of finding the mobility matrices $(A_{ij}^{(1)},B_{ij}^{(1)}, {D}_{ij}^{(1)})$ reduces to the problem of finding $\alpha_{ijk}$,$\beta_{ijk}$ and $M_{ijk}$. For a spheroid, these third order tensors depend on the orientation product $p_ip_j$, since fore-aft symmetry dictates that changing $p_i$ to $-p_i$ will not alter the results. Noting that $(\alpha_{ijk}, \beta_{ijk})$ are third order true tensors, and such tensors cannot be formed from $p_i p_j$, we obtain the result:
\begin{equation}
   \alpha_{ijk} = \beta_{ijk} =0
\end{equation}

The above relationship means that at $\mathcal{O}(\beta)$, the force-velocity coupling and torque-angular velocity coupling are unchanged. However, as we will see next, the force-rotation coupling and torque-velocity coupling will change.
$B_{ji}$ is a pseudo tensor since it connects a pseudo vector (angular velocity) with a true vector (force). Therefore, $M_{ikj}$ is a third order pseudo tensor, which depends on the orientation product $p_i p_j$. The general form of $M_{ijk}$ is given below as:
\begin{equation}
\label{eq:M_general}
    M_{ijk} =\lambda_1\epsilon_{ijk}+\lambda_2p_i\epsilon_{jkq}p_q+\lambda_3p_j\epsilon_{ikq}p_q+\lambda_4p_k\epsilon_{ijq}p_q
\end{equation}
where $\lambda_1, \lambda_2, \lambda_3, \lambda_4$ are dimensionless coefficients that depend only on the aspect ratio parameter $A_R$.  One can show that without loss of generality $\lambda_2=0$ (see Appendix D) and therefore the problem reduces to finding the coefficients $\lambda_1,\lambda_3,\lambda_4$. In other words, Eq.~\eqref{eq:M_general} reduces to 
\begin{equation}
\label{eq:M_general2}
     M_{ijk} =\lambda_1\epsilon_{ijk}+\lambda_3p_j\epsilon_{ikq}p_q+\lambda_4p_k\epsilon_{ijq}p_q
\end{equation}
 {We note that at the time of this article being prepared for publication, another manuscript solving a similar problem, arrived at a similar formulation for the mobility tensor (see Eq. 27 in \cite{GSE_2023})}.

In summary, the mobility relationships up to $\mathcal{O}(\beta)$ reduce to:
\begin{subequations} \label{eqn:mobility_tot}
\begin{equation} \label{eqn:trans_mobility}
    U_i = A_{ij}^{(0)}F_j+\beta M_{jki}\hat{d}_kT_j    
\end{equation}
\begin{equation} \label{eqn:rot_mobility}
    \omega_i =\beta M_{ikj}\hat{d}_kF_j+D_{ij}^{(0)}T_j
\end{equation}    
\end{subequations}
where $A_{ij}^{(0)}$,$D_{ij}^{(0)}$ are the known mobility tensors for a spheroid without a viscosity gradient, given by Eq.~\eqref{eq:mobility_tensor_leading_order}, while $M_{ijk}$ is the cross-coupling term given by Eq.~\eqref{eq:M_general2}.  The unknown coefficients for the tensor $M_{ijk}$ are $(\lambda_1,\lambda_3,\lambda_4)$, which are functions of the aspect ratio parameter $A_R$ for the spheroid.{We also note, from Eq.~\eqref{eq:M_general2} in conjunction with Eq.~\eqref{eq:Mobility_Forms_B}, that $\lambda_1,\lambda_3,\lambda_4$ are related to the eigenvalues of $\boldsymbol{B}$, and in some cases of the orientation of the viscosity gradient vector, are exactly equal to the eigenvalues of $\boldsymbol{B}$.  For a more detailed discussion on this interpretation, please refer to \cite{Wittens_2020}.} The next section discusses how we determine these coefficients.

 \subsection{Determining coefficients $\lambda_1,\lambda_3,\lambda_4$ for the mobility matrix $M_{ijk}$(force-rotation and torque-translation coupling)}
 \label{sec:Design_Simulations}

\begin{figure}
\centering
\includegraphics[width=0.5\linewidth]{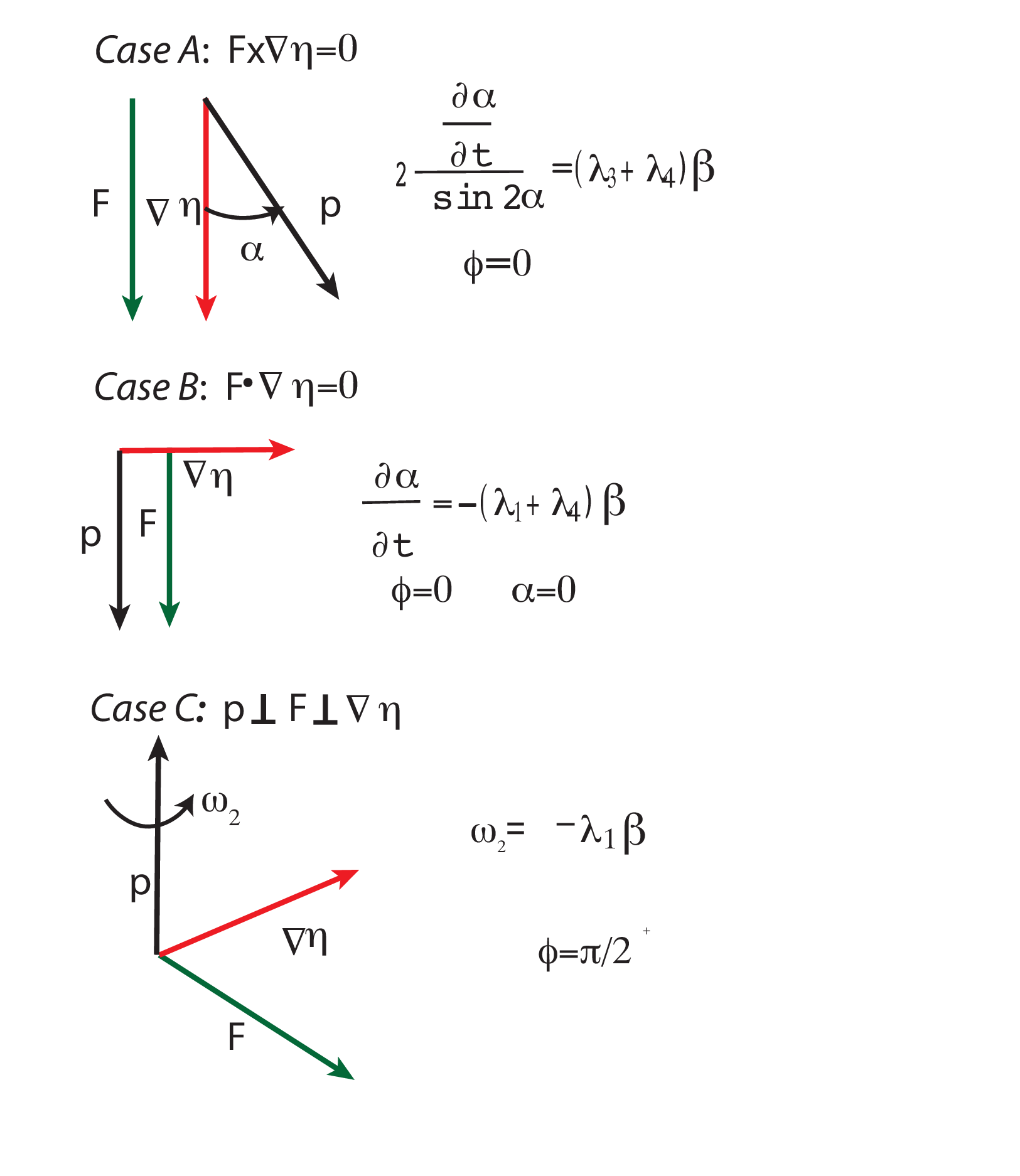}
\caption{Simulations carried out to estimate the parameters $(\lambda_1,\lambda_3,\lambda_4)$ in the third order pseudo tensor $M_{ijk}$ given by Eq.~\eqref{eq:M_general2}.  The orientation angles $(\alpha, \phi)$ are defined in Figs. \ref{fig:Schematic_Parallel} and \ref{fig:Schematic_Perpendicular} respectively.  }
\label{fig:Case_A_B_C}
\end{figure}

% \begin{figure}
% \centering
% \includegraphics[width=0.6\linewidth]{Theory_Verify.eps}
% \caption{\textcolor{red}{Verification of theory by plotting  of Eq.~\eqref{eq:evolution_caseA} for different values of $\alpha$, for a prolate spheroid with external force $\boldsymbol{F}$ and viscosity gradient $\boldsymbol{\nabla} \eta$ in the positive $z$-direction {with $\beta =0.1$}. This situation corresponds to Case A shown in Fig.~\ref{fig:Case_A_B_C}a.}}
% \label{fig:Theory_Verification}
% \end{figure}

% Figure \ref{fig:Lambda_Estimation} then concludes the numerical estimation of $\lambda$'s. 

\begin{figure}
\centering
\subfloat[Prolate spheroid]{\includegraphics[width=\linewidth]{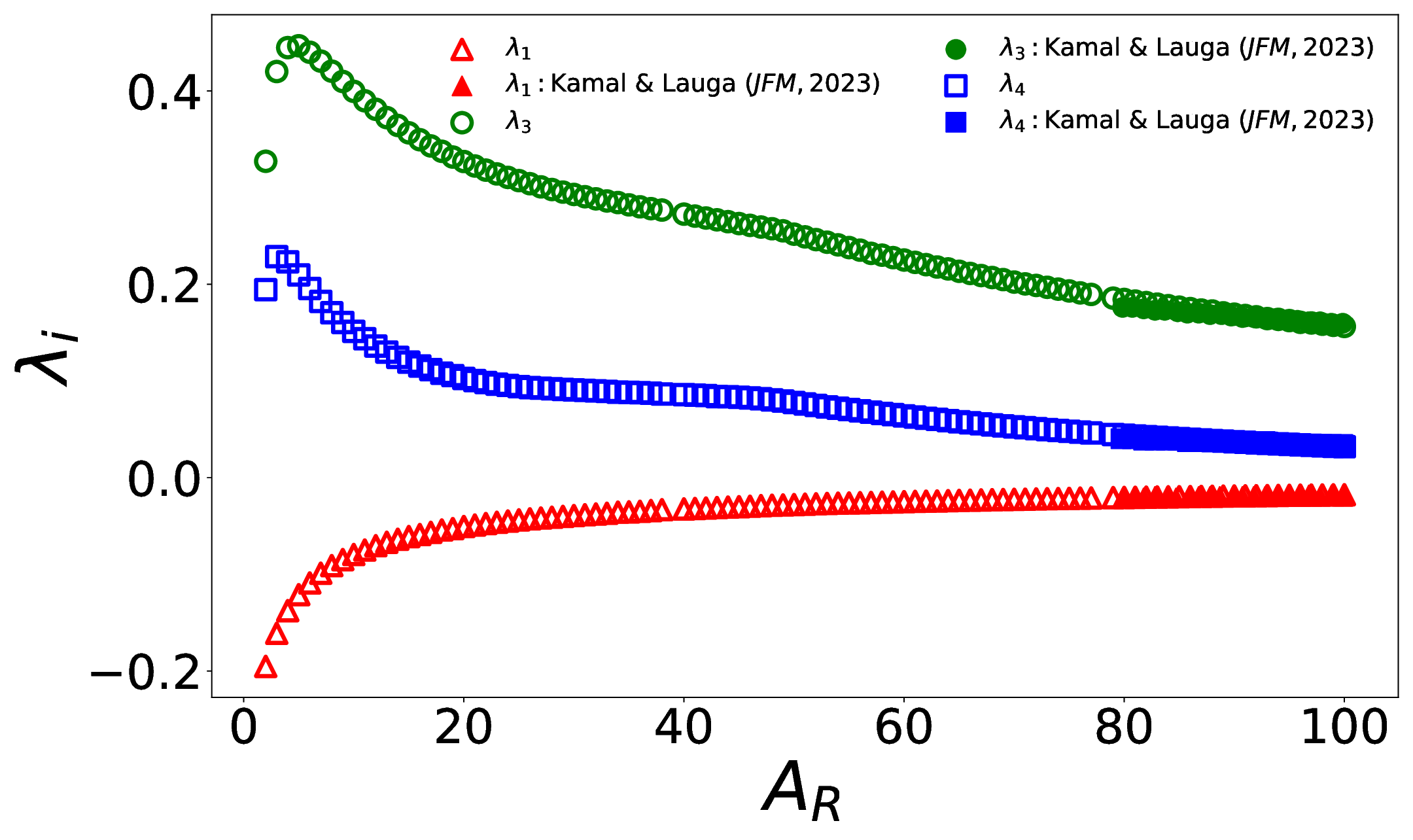}}
\hspace{5mm}
\subfloat[Oblate spheroid]{\includegraphics[width=\linewidth]{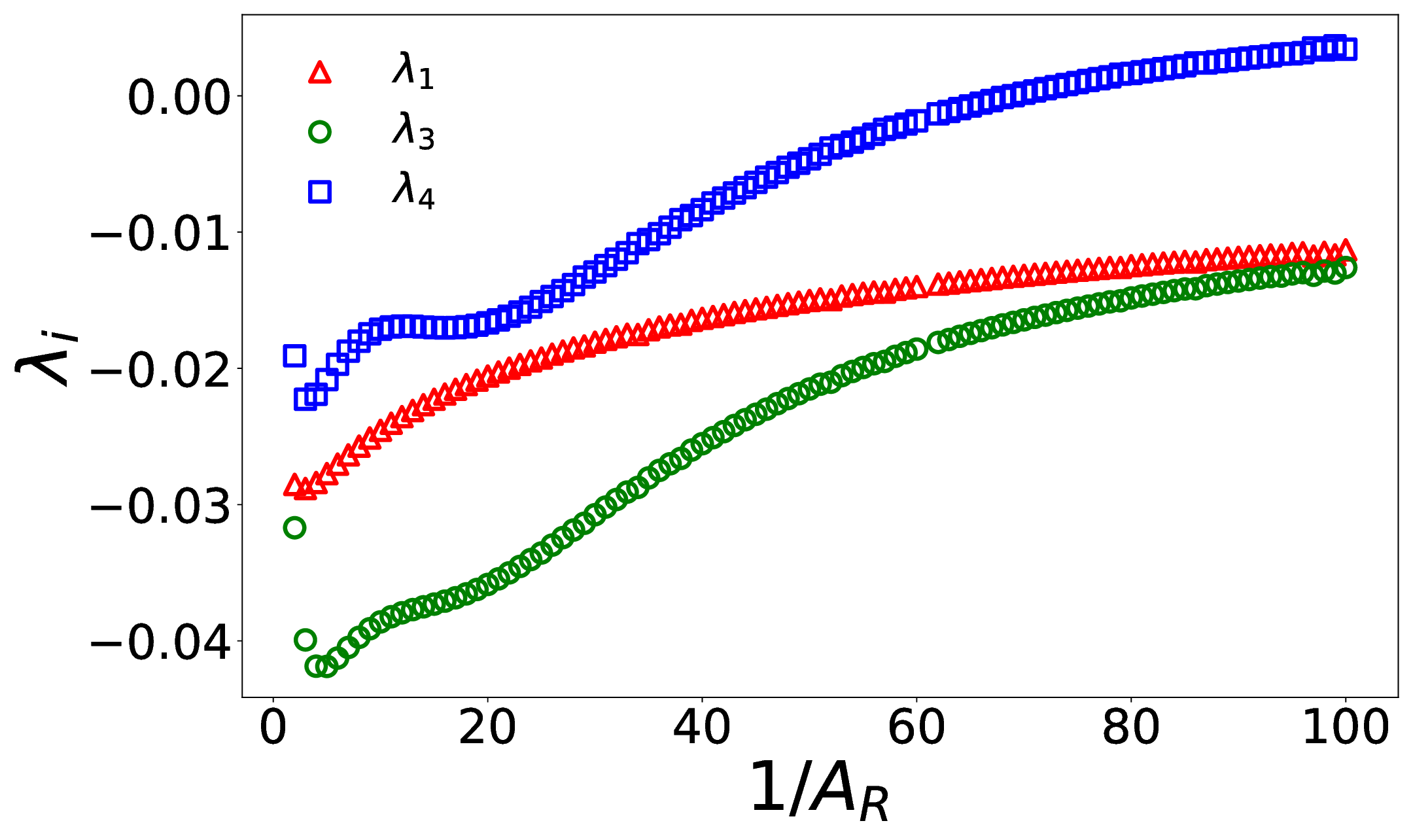}}
\caption{{Computed values of $(\lambda_1, \lambda_3, \lambda_4)$ for prolate and oblate spheroids for different values of aspect ratio parameters $A_R$.The empty symbols denote the results derived in this paper, while the filled symbols, for prolate spheroids, denote the results from \cite{KLauga_2023}, who used the slender body theory to derive their results} }
\label{fig:Lambda_Estimation}
\end{figure}

 Fig \ref{fig:Case_A_B_C} outlines the simulations we perform to obtain the coefficients $(\lambda_1,\lambda_3,\lambda_4)$ for $M_{ijk}$ in Eq.~\eqref{eq:M_general2}.  We examine a torque-free particle $(T_i  = 0)$ and quantify its angular velocity $\omega_i$ for the three specific geometries listed below.  We note that the angular velocity can cause two effects – it can change the spheroid’s orientation or it can keep the orientation the same but spin it along its axis.  The rate of change of the orientation is given by:
 \begin{equation}
 \label{eq:projector_rotation}
    \frac{d p_i}{d t} =\epsilon_{ijk}\omega_j p_k 
\end{equation} 
%  \begin{subequations}
%  \label{eq:projector_rotation}
% \begin{align}
%     \dot{p}_i &= \frac{d p_i}{d t} =\epsilon_{ijk}\omega_j p_k 
%     \end{align}
% \end{subequations} 
while the rate of spinning is:
\begin{equation}
\label{eq:projector_spinning}
    \Omega =\omega_ip_i
\end{equation}

These quantities are computed for the cases below:\\

\begin{enumerate}
    \item \underline{Case A}: $\boldsymbol{\nabla} \eta \times \boldsymbol{F} =0$: Here, we examine the situation in Fig ~\ref{fig:Case_A_B_C}a where the external force and viscosity gradient are in the same direction – i.e., $\boldsymbol{F} = \boldsymbol{\hat{d}} = \boldsymbol{\hat{z}}$.  The particle has its orientation in the $x-z$ plane with an angle $\alpha$ with respect to the force direction – i.e., $\boldsymbol{p} =[\sin \alpha,0,\cos \alpha]$. Using Eqs.  \eqref{eq:M_general2}, \eqref{eqn:rot_mobility}, and \eqref{eq:projector_rotation}, one finds the angular velocity to be:

    \begin{equation}
    \label{eq:evolution_caseA}
    {\omega_2=\frac{d \alpha}{d t} = \frac{1}{2} \beta\left(\lambda_3+\lambda_4\right) \sin({2\alpha})}
    \end{equation}

    Thus, performing one simulation at a specific polar angle and viscosity gradient magnitude (say $\alpha=\pi/4, \beta=0.1)$ allows us to obtain $(\lambda_3+\lambda_4)$.  {To test this theory, we carried out simulations of prolate spheroids undergoing sedimentation at aspect ratio $A_R =3$ and $A_R = 5$.  The simulations were performed for many different values of angles $\alpha$ and non-dimensional viscosity gradient $\beta =0.1$, and they found that for a given aspect ratio, the quantity $2\frac{\frac{d \alpha}{d t}}{\beta \sin({2\alpha})}$ was constant for all values of $\alpha$ (equal to $0.0649$ for $A_R=3$ and $0.0656$ for $A_R =5$).  This behavior is consistent with the expression listed above. }\\

    \item \underline{Case B}: $\boldsymbol{\nabla} \eta \cdot \boldsymbol{F} =0, \boldsymbol{p} \times \boldsymbol{F} = 0$:  We examine the situation in Fig ~\ref{fig:Case_A_B_C}b where the external force and viscosity gradient are perpendicular to each other – i.e., $\boldsymbol{F}=\boldsymbol{\hat{z}}$, $\boldsymbol{\hat{d}}= \boldsymbol{\hat{x}}$.  The particle has its orientation along the force direction -- i.e., $\boldsymbol{p}=[0,0,1]$ -- which corresponds to the polar and azimuthal angles of $\alpha = \phi = 0$ in Fig \ref{fig:Schematic_Perpendicular}.  Using Eqs.  \eqref{eq:M_general2}, \eqref{eqn:rot_mobility}, and \eqref{eq:projector_rotation}, one finds the angular velocity to be:

%     \begin{subequations}
% \begin{align}
%     \label{eq:evolution_caseB_3}
%     \frac{\partial \alpha}{\partial t}
%     &=-\cos{\phi}\left[\lambda_1-\lambda_3\sin^2\alpha+\lambda_4\cos^2{\alpha}\right]\beta \\
%     \label{eq:evolution_equation_caseB_4}
%     \frac{\partial \phi}{\partial t} &=\sin{\phi}\cot{\alpha}\left[\lambda_1+\lambda_4\right]\beta
% \end{align}
% \end{subequations}

\begin{equation}
    {\omega_2 = \frac{d \alpha}{d t}|_{\alpha = \phi = 0^{\circ}} = -\beta (\lambda_1 + \lambda_4) }
\end{equation}

 Performing one simulation at a specific value of $\beta$ (e.g.,  $\beta=0.1$) allows us to obtain {$(\lambda_1+\lambda_4)$}. \\

 \item \underline{Case C}: $\boldsymbol{\nabla} \eta \perp \boldsymbol{F} \perp \boldsymbol{p}$: 
 We examine case in Fig ~\ref{fig:Case_A_B_C}c where the orientation, viscosity gradient, and force are all perpendicular to each other – i.e., $\boldsymbol{F}=\boldsymbol{\hat{z}}$, $\boldsymbol{\hat{d }} = \boldsymbol{\hat{x}}$, $\boldsymbol{p}= \boldsymbol{\hat{y}}$.  Here, the particle will spin but not change orientation.  Using Eqs. \eqref{eq:M_general2}, \eqref{eqn:rot_mobility}, and \eqref{eq:projector_spinning}, we find the spinning rate to be:  

 \begin{equation}
\label{eq:evolution_equation_caseC_final}
    \Omega = \omega_2 =-\beta \lambda_1
\end{equation}

Performing one simulation at a specific value of $\beta$ (e.g.,  $\beta=0.1$) allows us to obtain $\lambda_1$.\\
\end{enumerate}

The three simulations listed above yield a linear system of equations for the coefficients $(\lambda_1,\lambda_3,\lambda_4)$ that can be solved.  Fig \ref{fig:Lambda_Estimation} shows the values of the coefficients for different values of the aspect ratio parameter $A_R$, for both prolate and oblate spheroids. For comparison, we have also plotted the $\lambda'$ s from the \cite{KLauga_2023}, who invoked the slender body theory to solve the problem of sedimenting slender bodies in inertialess flows of Newtonian fluids with spatially varying viscosity. As can be seen from our plots, a good match is observed between our results for prolate spheroids at high $A_R$ and that given by  \cite{KLauga_2023}.

Once these coefficients are tabulated, one has a general form for the rigid body motion (Eq. \eqref{eqn:mobility_tot}) for spheroids that can be solved for arbitrary viscosity gradient, orientation, aspect ratio, and external force/torque.

%To summarize, we developed the theory up to the point where the mobility of the sedimenting spheroid was determined completely by Eq.\eqref{eq:Mobility_Force_Torque}, which introduces the tensor $M_{ijk}$, whose expression is given in Eq.~\eqref{eq:M_general2}, barring three undetermined coefficients namely $\lambda_1,\lambda_3,\lambda_4$. To determine these $3$ coefficients , we designed three ($3$) cases of numerical simulations in Sec.\ref{sec:Design_Simulations}, which have been summarised in Fig.\ref{fig:Case_A_B_C}.
%Once these coefficients are tabulated, one has a general form for the rigid body motion (Eq.~\eqref{eq:Mobility_Force_Torque}) that can be solved for arbitrary viscosity gradient, spheroid geometry, and external force/torque.

\section{Results and illustrative examples} \label{sec:results}

In Sec.~\ref{sec:theory}, we developed a theory to describe the rigid body motion of a spheroid in a spatially varying viscosity field. { In the limit of weak or linearly varying viscosity field,} the general form of the translational and rotational velocities is given in Eq.~\eqref{eqn:mobility_tot}, where $A_{ij}^{(0)}$ and $D_{ij}^{(0)}$ are the standard mobility tensors for force/translation and torque/rotation in Stokes flow, and $M_{ijk}$ is a newly introduced coupling tensor between force/rotation and torque/translation that arises due to viscosity gradients.  The tensor $M_{ijk}$ is given in Eq. ~\eqref{eq:M_general2} in terms of three coefficients $(\lambda_1, \lambda_3, \lambda_4)$ that are only functions of the spheroid aspect ratio parameter $A_R$.  These coefficients are estimated numerically using the reciprocal theorem (see Sec.~\ref{sec:simulation} and Fig. \ref{fig:Lambda_Estimation}). 

In this section, we investigate the spheroid’s dynamics for some special cases and discuss the physics that arise.  Details are below. {Please note that all the results and examples are valid only up to the first order in $\beta$ deviation from the constant viscosity, unless otherwise mentioned.}

% We have developed a comprehensive theory to delineate the sedimentation characteristics of arbitrary spheroids in quiescent flow of Newtonian fluids with viscosity varying spatially in a linear fashion, in any direction. The theory development has been described in detail in Sec.~\ref{sec:theory}, and Eq.~\eqref{eq:projector_rotation}, in conjunction with Eq.~\eqref{eq:Mobility_Force_Torque} and Eq.~\eqref{eq:M_general2}, describes the evolution of the projector of the spheroids in arbitrarily directed spatially varying linear viscosity field. The three undetermined (by theory) coefficients in the equation are then estimated numerically using a reciprocal theorem based approach (see Sec.~\ref{sec:simulation} and Fig.\ref{fig:Lambda_Estimation}). 

% In this section, we investigate some special cases and observe the results of our theory through some illustrative examples.

\subsection{Viscosity gradient is along or opposite the force direction}
\subsubsection{Governing equations}
Let us examine the situation in Fig. \ref{fig:Schematic_Parallel} where the external force is in the positive $z$-direction, and the viscosity gradient is either parallel to the force (positive $z$-direction) or anti-parallel to the force (negative $z$-direction).  In this case, the particle orientation only has one degree of freedom, namely the polar angle $\alpha$ measured from the $z$-axis. Without loss of generality, we will state that $\boldsymbol{p}$ lies in the $x-z$ plane, and thus $\boldsymbol{p} = [\sin\alpha, 0, \cos\alpha]$. From our theory (Eqs. \eqref{eq:M_general2}, \eqref{eqn:rot_mobility}, and \eqref{eq:projector_rotation}), the orientation angle obeys the following equation:

{\begin{subequations} 
\begin{align}
\label{eqn:angle_parallel}
\frac{d \alpha}{dt} &= \pm \frac{1}{2} \beta (\lambda_3 + \lambda_4) \sin (2\alpha) \\
\Rightarrow |\tan (\alpha)|&=|\tan (\alpha_0)|e^{\beta(\lambda_3+\lambda_4)(t-t_0)}
\end{align}
\end{subequations}
where $\pm$ illustrates the cases where the viscosity gradient is parallel ($+$) or anti-parallel ($-$) to the force {and $\alpha_0$ is the initial angle at $t=t_0$}.  The translational motion of the particle obeys:
\begin{equation} \label{eqn:trans_parallel}
\frac{dx}{dt} = \frac{1}{2}(c_2- c_1) \sin(2\alpha); \qquad \frac{dz}{dt} = c_1 \sin^2 \alpha + c_2 \cos^2 \alpha
\end{equation}
where $c_1$ and $c_2$ are the mobility coefficients for spheroid translation in Stokes flow (given in Appendix B in dimensional form).  Major conclusions are given below.

\subsubsection{Particle takes a stable orientation depending on its shape and viscosity gradient direction}

\begin{figure}
\centering
\subfloat[Prolate, parallel]{\includegraphics[width=0.45\linewidth]{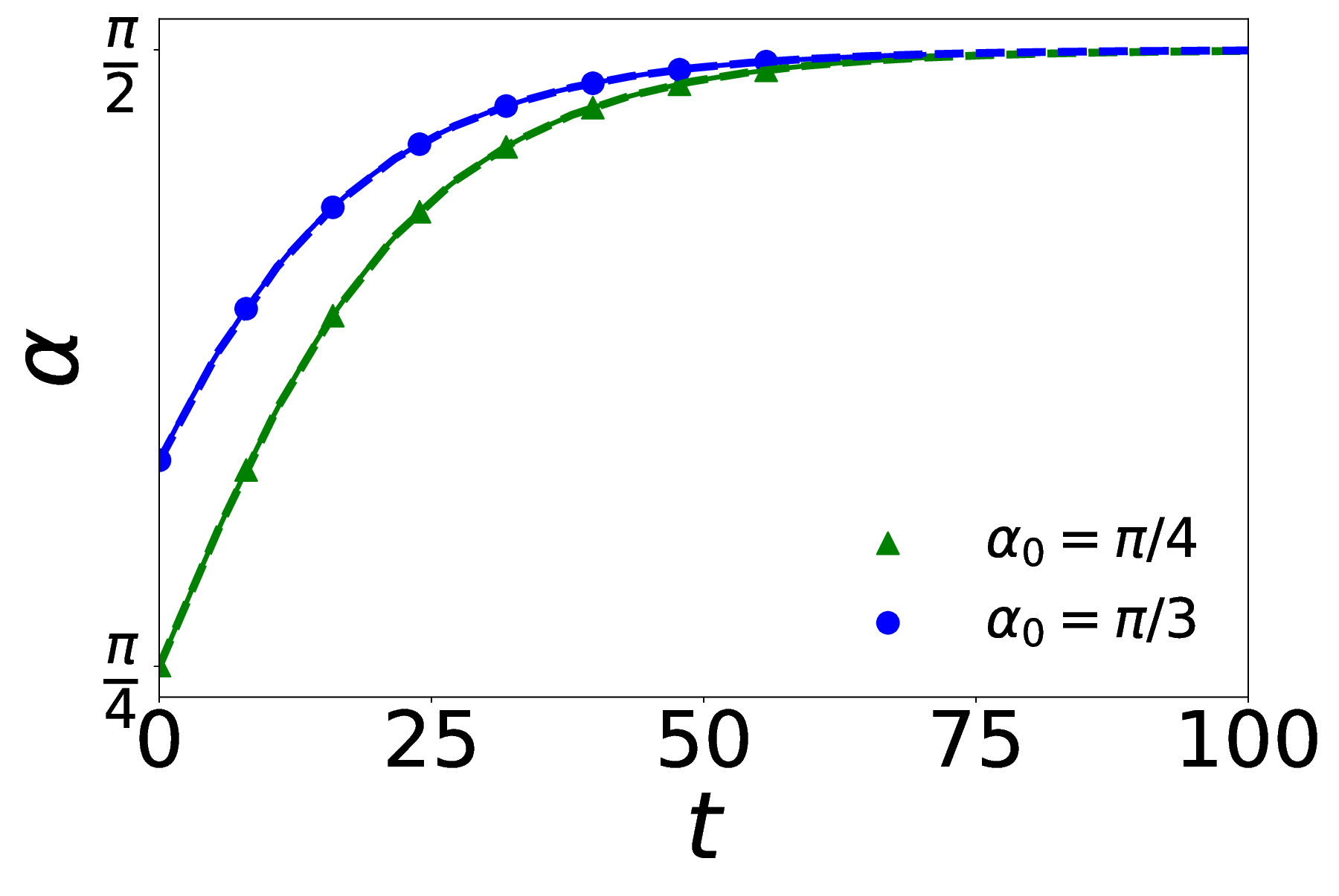}}
\subfloat[Oblate, parallel]{\includegraphics[width=0.45\linewidth]{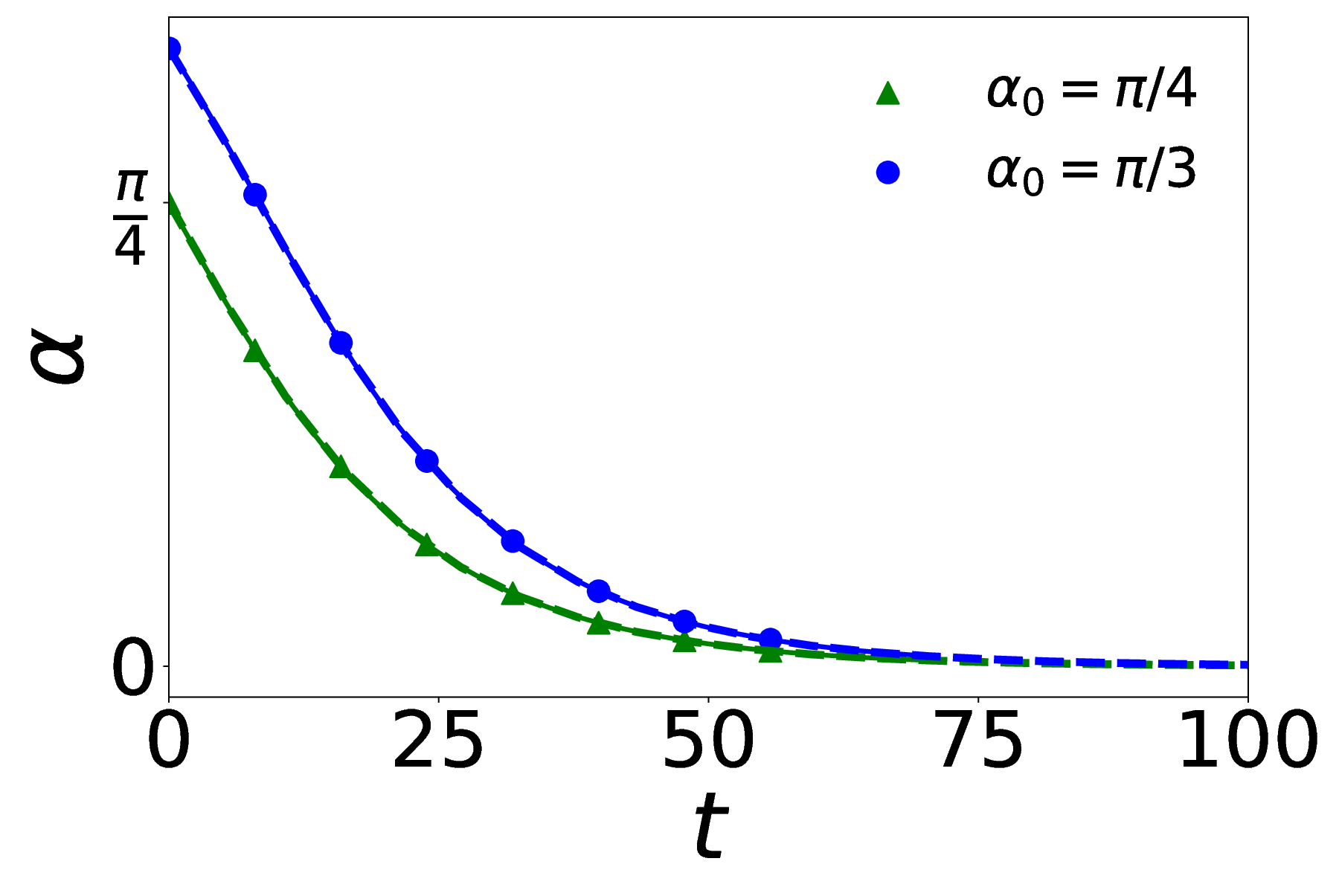}}
\hfill
\subfloat[Prolate, anti-parallel]{\includegraphics[width=0.45\linewidth]{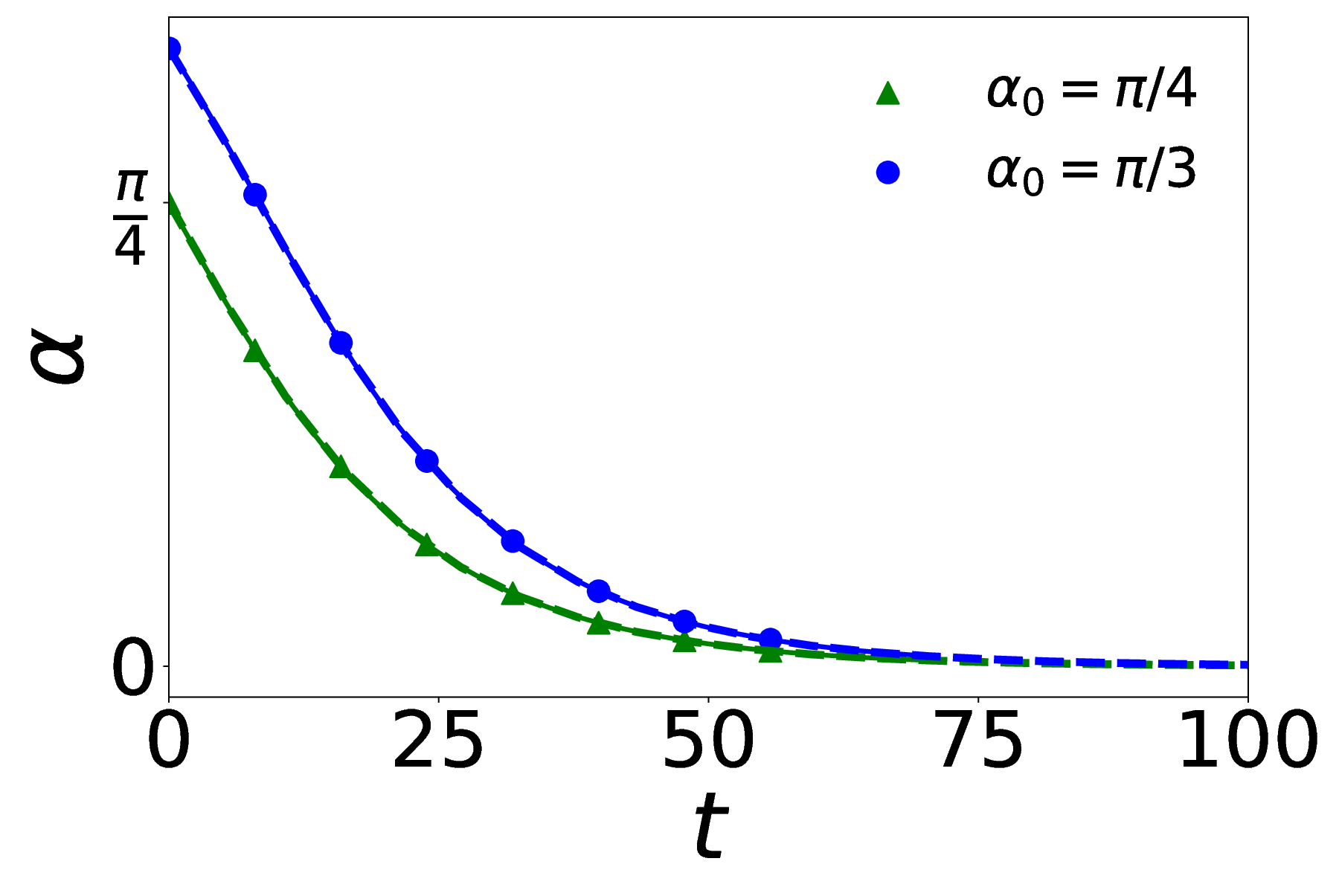}}
\subfloat[Oblate, anti-parallel]{\includegraphics[width=0.45\linewidth]{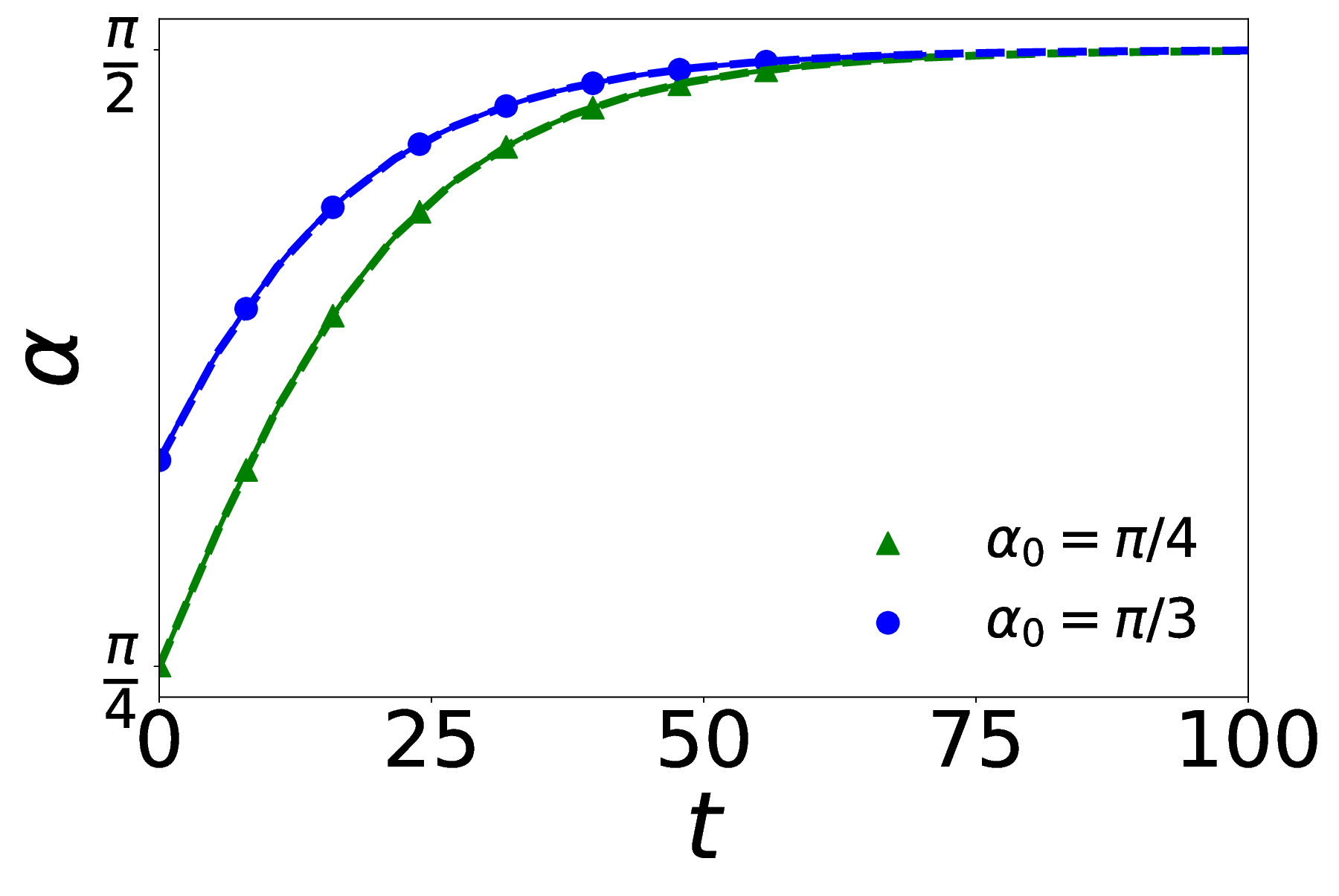}}
\caption{Orientation angle $\alpha$ vs. time for prolate and oblate spheroids when the external force $\boldsymbol{F}$ and viscosity gradient $\boldsymbol{\nabla} \eta$ are parallel or anti-parallel to each other.  The left figures (a,c) correspond to prolate spheroids with $A_R =5$, while those the right figures (b,d) correspond to oblate spheroids with $A_R =1/5$.  The top row (a,b) is the case when the $\boldsymbol{F}$ and $\boldsymbol{\nabla }\eta$ are in the same direction, while the bottom row (b,d) is the case when they are in opposite directions.  The solid curves are from full numerical simulations based on the reciprocal theorem, while the dashed curves are from the {symmetry based} theory (solving Eq. \eqref{eqn:angle_parallel}).  The dimensionless viscosity gradient is $\beta = 0.1$.}
\label{fig:Orient_X}
\end{figure}

\begin{figure}
\centering
\subfloat[Prolate spheroid]{\includegraphics[width=0.45\linewidth]{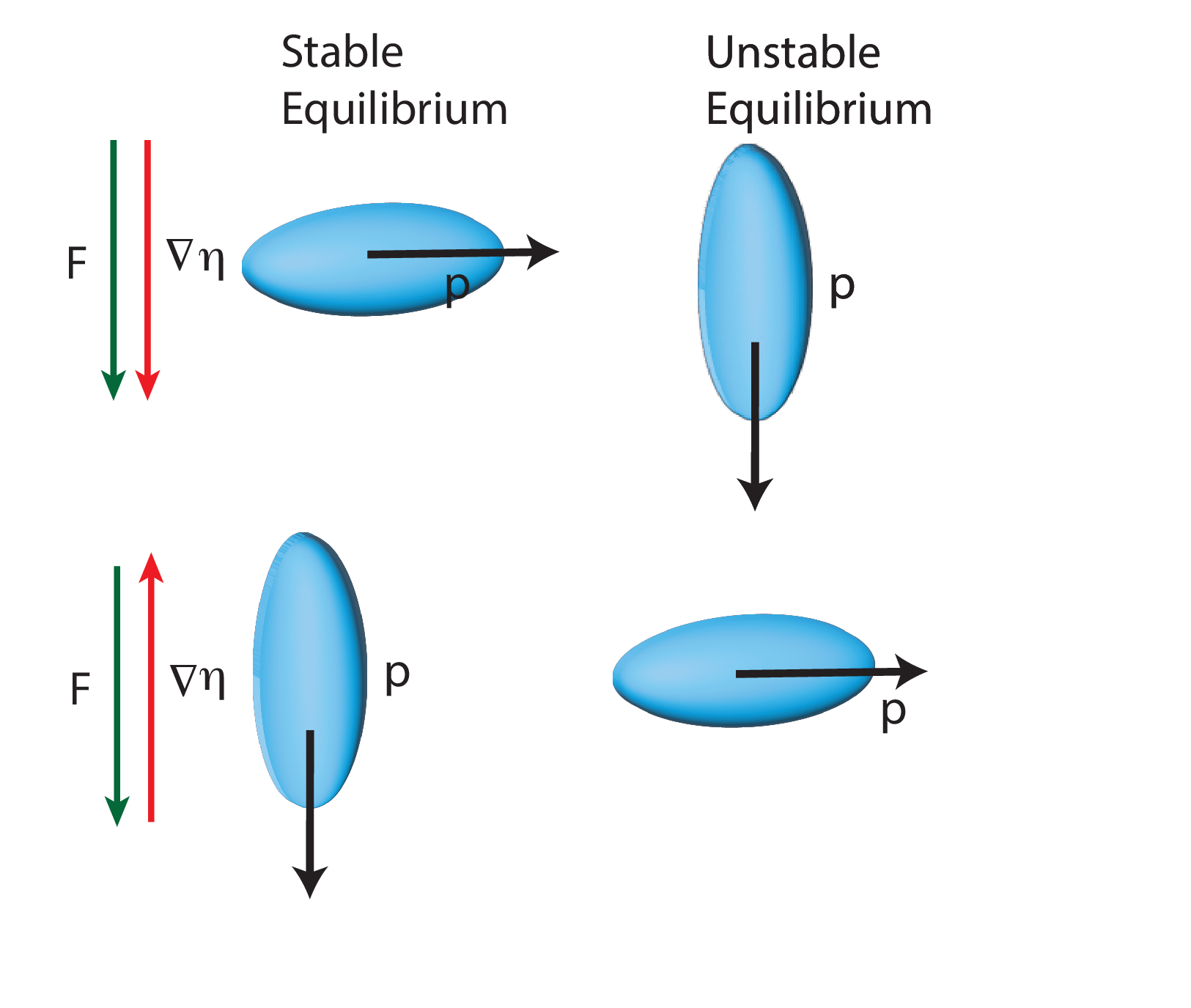}}
\subfloat[Oblate spheroid]{\includegraphics[width=0.45\linewidth]{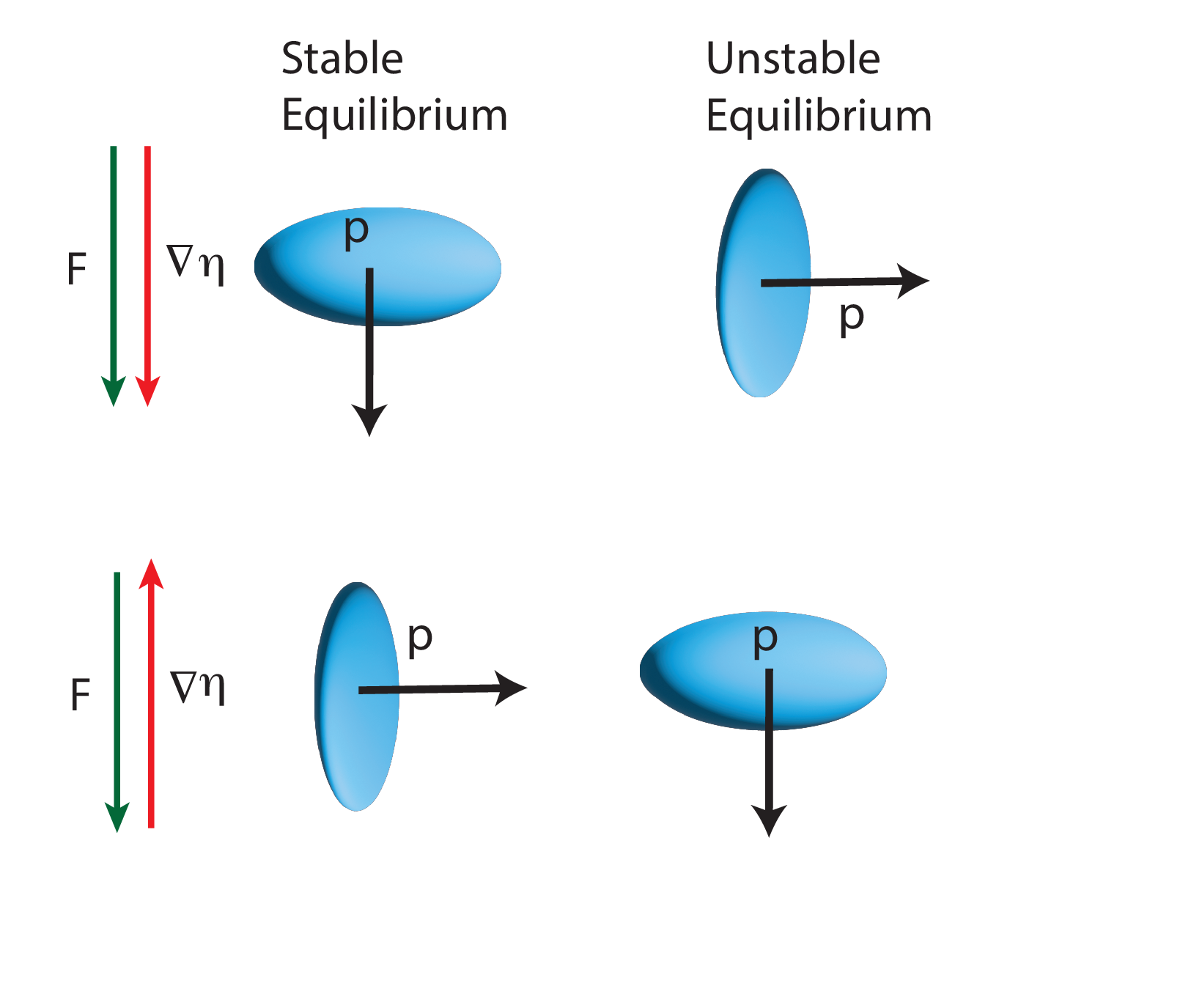}}
\caption{Steady configurations attained by (a) prolate and (b) oblate spheroids when the external force $\boldsymbol{F}$ and viscosity gradient $\boldsymbol{\nabla}\eta$ are co-linear. The top row is for the case when the external force and the viscosity gradient are in the same direction, while the bottom row is when they are in the opposite direction.}
\label{fig:Stability_X}
\end{figure}

\begin{figure}
\centering
\subfloat[Parallel]{\includegraphics[scale=0.21,angle =-90]{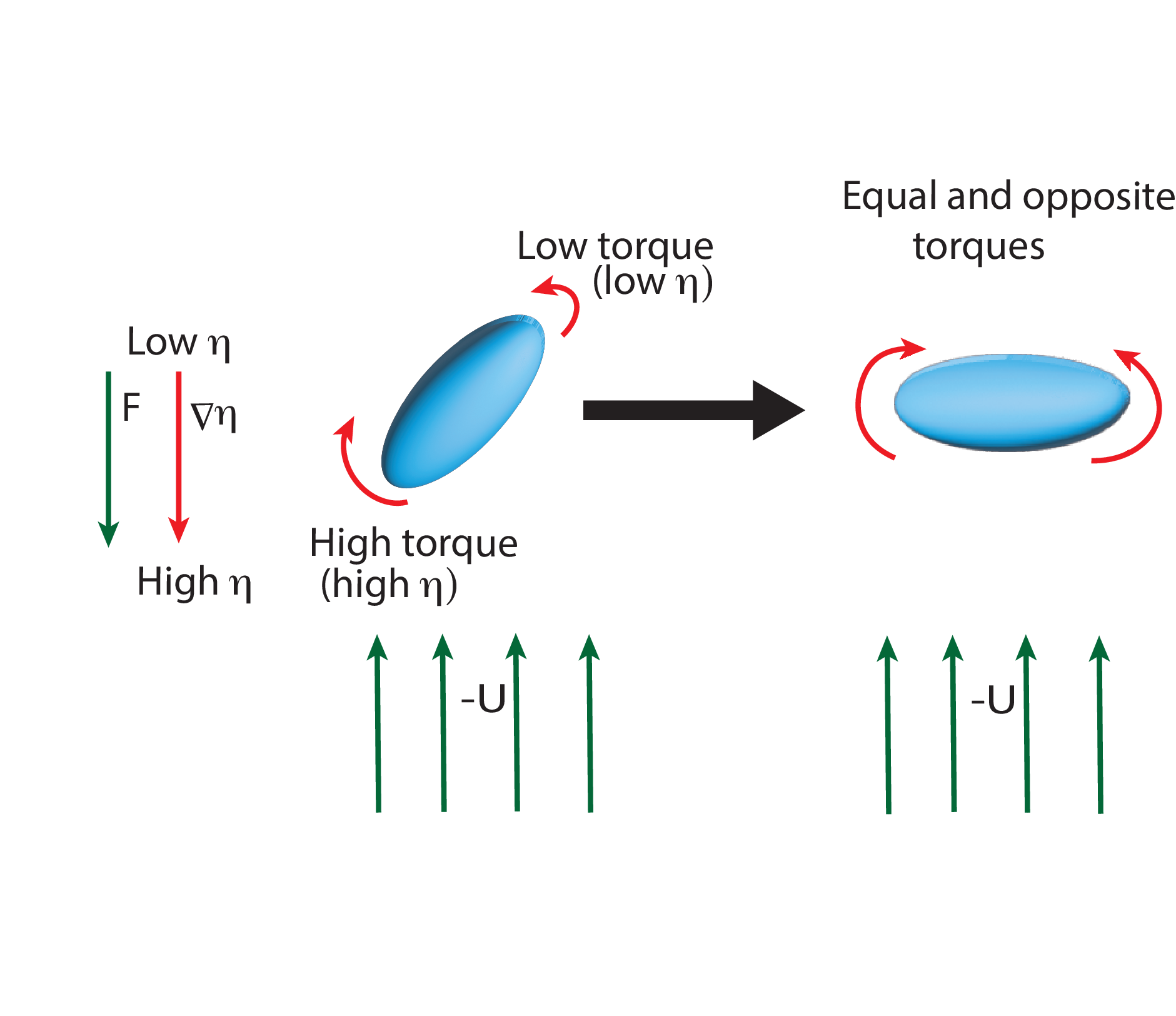}}
\hfill
\subfloat[Anti-parallel]{\includegraphics[scale =0.21, angle =-90]{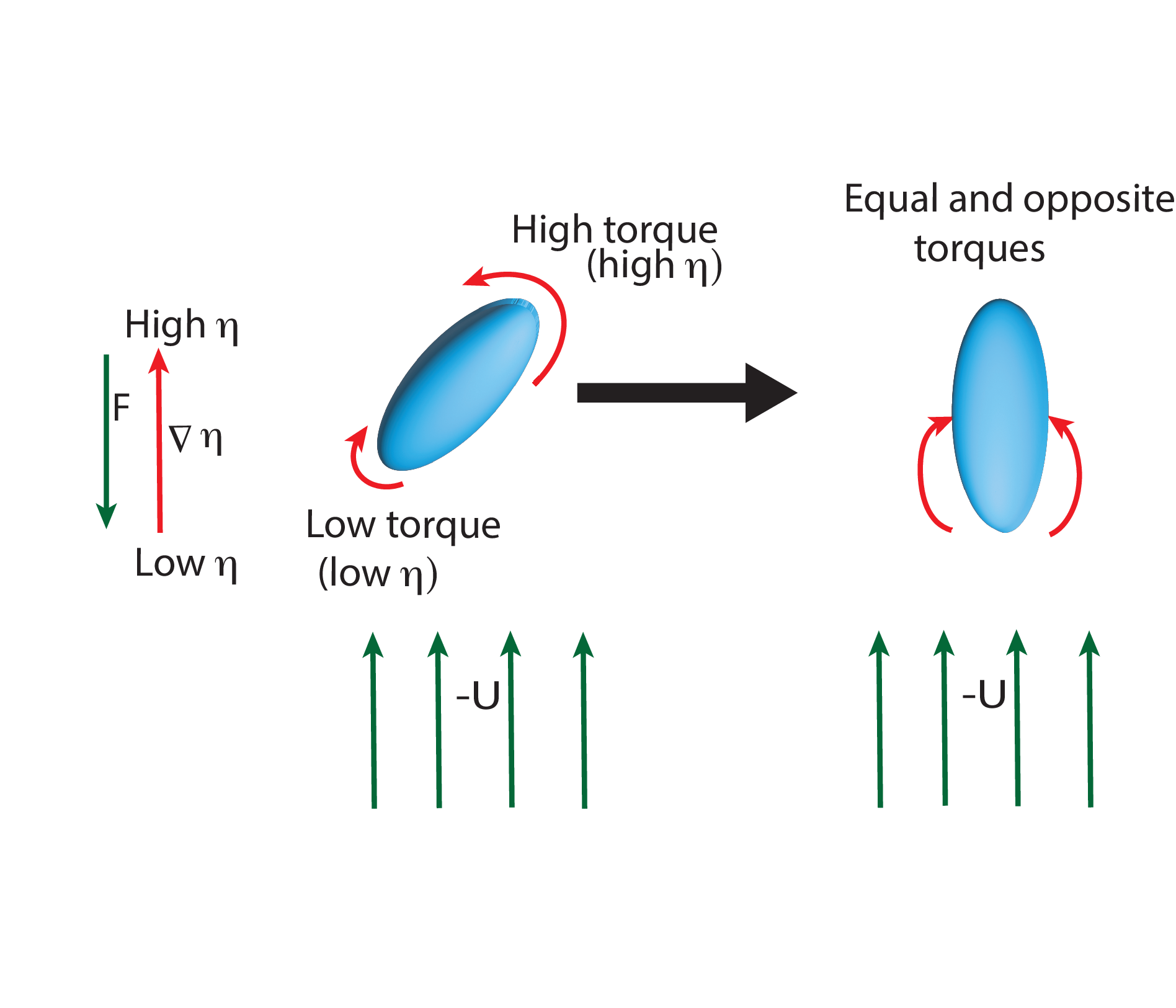}}

\caption{Illustration of unequal torques created on a prolate spheroid when the force and viscosity gradient are co-linear. The left figure (a) is when the viscosity gradient and force are in the same direction, while the right figure (b) is when they are in opposite directions.{This schematic is shown in particle's frame of reference.} }
\label{fig:Schematic_Explanation_1}
\end{figure}

Fig.~\ref{fig:Orient_X} plots the evolution of $\alpha$ with respect to time for prolate and oblate spheroids, for the cases when the force $\boldsymbol{F}$ and viscosity gradient $\boldsymbol{\nabla} \eta$ are parallel and anti-parallel to each other. For each set of conditions, two curves are given – one arising from the {symmetry based} theory (dashed curve, Eq. \eqref{eqn:angle_parallel}), and another from full numerical simulations where the reciprocal theorem is used at every time step (solid curve).  The overlap is indistinguishable, thereby validating our theory. The second observation we make is that that irrespective of the initial orientation and viscosity gradient direction (parallel or anti parallel), both prolate and oblate spheroids evolve to a steady configuration of $\alpha$.  This observation is very different than what is observed in Stokes flow where the orientation stays at its initial angle at all times \citep{Leal2007}.

Fig.~\ref{fig:Stability_X} summarizes the steady orientations observed for different particle shapes and viscosity gradient directions. When the external force and the viscosity gradient are parallel to each other, the prolate spheroid adopts a stable configuration where the projector is perpendicular to external force, while the oblate spheroid orients itself such the projector is along the same direction as the external force. In both of these cases, the spheroid (whether prolate or oblate) has its shortest axis oriented along the direction of the viscosity gradient. On the other hand, when the spheroid is falling in the direction of decreasing viscosity (i.e., $\boldsymbol{F}$ and $\boldsymbol{\nabla}\eta$ are anti-parallel), the prolate spheroid attains a stable configuration where the projector is oriented along the force direction, whilst the oblate spheroid orients the projector perpendicular to the force direction. In both these cases, the longest axis of the particle (whether prolate and oblate) will be along the force direction.  

To provide a physical understanding of this behavior, we refer to Fig. \ref{fig:Schematic_Explanation_1}. Here, as observed from the reference frame of the particle, the flow around the prolate spheroid bifurcates into two parts about the stagnation point and engenders both a clockwise and counter-clockwise hydrodynamic torque. Fig. \ref{fig:Schematic_Explanation_1}(a) illustrates the magnitude of the torques for the case when the viscosity gradient is in the same direction as the force, while Fig. \ref{fig:Schematic_Explanation_1}(b) illustrates the case when the viscosity gradient is in the opposite direction.  The pictures illustrate that the the unequal torques push the particle toward the stable orientations discussed above.  

Lastly, we note that Fig. \ref{fig:Stability_X}  summarizes the unstable, steady orientations that can occur for different combinations of viscosity gradient and particle shape.  These orientations only exist if the initial condition is at a specific angle, and can only be observed in exceptionally rare cases.

\subsubsection{Particle translation is different than in Stokes flow}

\begin{figure}
\centering
\subfloat[Prolate, $A_R = 5$]{\includegraphics[width=0.5\linewidth]{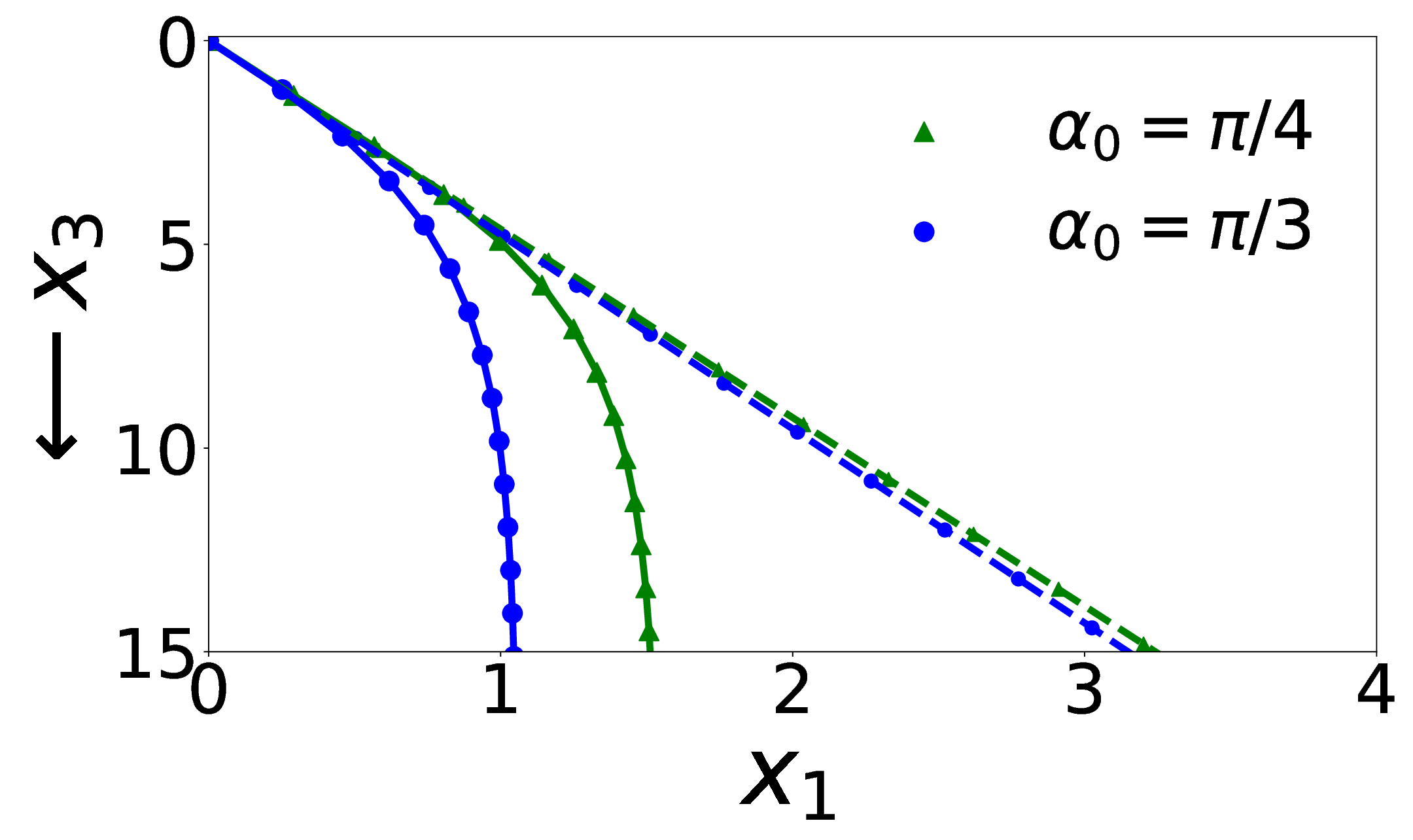}}
\subfloat[Oblate, $A_R = 1/5$]{\includegraphics[width=0.5\linewidth]{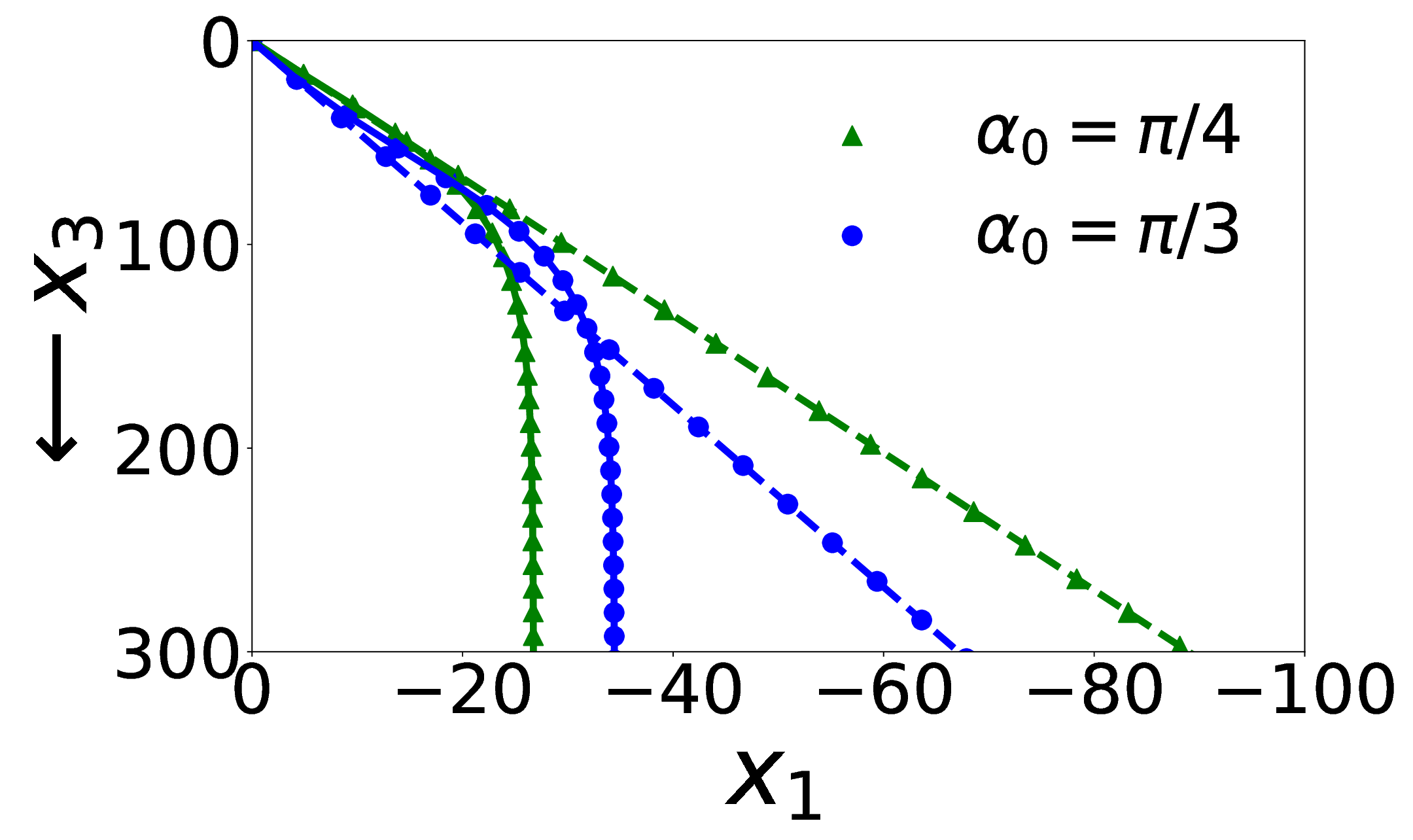}}

\caption{Particle trajectories for (a) prolate and (b) oblate spheroids when the external force and viscosity gradient are in the same direction ($\boldsymbol{F} = \boldsymbol{\hat{z}}, \boldsymbol{\nabla}\eta = \beta \boldsymbol{\hat{z}}$). The dashed curves correspond to when no viscosity gradient is present ($\beta = 0$), while the solid curve is when a viscosity gradient is present ($\beta = 0.1$). Different colors {and symbols} correspond to different initial starting angles $\alpha_0$.  The prolate spheroid has $A_R = 5$ while the oblate spheroid has $A_R = 1/5$.}
\label{fig:Translation}
\end{figure}

Beyond orientational kinematics, we are also interested in the translation of the spheroid.  In a constant viscosity fluid with zero inertia, it is well-known that the particle stays at its initial orientation \citep{Leal2007}.  If the initial angle is $\alpha = 0$, $\frac{\pi}{2}$, or $\pi$, the particle will sediment vertically, while if $\alpha$ is not these values, the particle will drift in a straight, diagonal path.  The direction in which the particle sediments is dictated by the resistances parallel and perpendicular to its orientation vector $\boldsymbol{p}$.

When a viscosity gradient is present, the translational velocity $\boldsymbol{U}$ obeys the same differential equation as the Stokes flow case (Eq. \eqref{eqn:trans_parallel}), since we found that the viscosity gradient does not alter the force/translation coupling (see Eq. \eqref{eqn:mobility_tot}).  Thus, on the surface, it appears that the particle trajectory \textit{may seem} unchanged due to the presence of a spatially varying viscosity field.  However, upon closer inspection, we see that the differential equation (Eq. \eqref{eqn:mobility_tot}) depends on the particle’s orientation angle $\alpha$, which itself is altered due to the viscosity gradient as discussed in the previous section.  Thus, the viscosity gradient plays an indirect role in altering the translational dynamics.

Fig.~\ref{fig:Translation} plots the trajectories of oblate and prolate spheroids for different values of the initial orientation angle $\alpha_{0}$. For $\alpha_0 \neq 0$, $\frac{\pi}{2}$, and $\pi$, we observe motion in the sedimentation direction (3-direction) as well as a cross stream drift (1-direction).  For the case when no viscosity gradient is present, the particle moves in a straight, diagonal path.  When a viscosity gradient is present, the trajectory is no longer a straight line.  The cross-stream drift eventually stops when the spheroid reaches a stable orientation, beyond which the spheroid sediments vertically in the $3$-direction. Since the spheroid ceases to drift once the stable orientation is reached, a spheroid whose initial orientation is further away from its stable orientation will drift further than a spheroid whose initial orientation is closer to its stable orientation. Therefore, for a prolate spheroid, a particle with initial orientation $\alpha_0 =\pi/4$ will drift further than one with $\alpha_0 = \pi/3$, since the stable orientation is $\alpha =\pi/2$ (see Fig.~\ref{fig:Translation}(a)). Conversely, for an oblate spheroid, the particle with an initial orientation $\alpha_{0} =\pi/3$ will drift further than one with $\alpha_{0} =\pi/4$, since the stable orientation is at $\alpha =0$.

\subsection{Viscosity gradient is perpendicular to the external force}
\subsubsection{Governing equations}
We will now examine the situation in Fig. \ref{fig:Schematic_Perpendicular} where the external force is in the positive $z$-direction ($\boldsymbol{F} = \boldsymbol{\hat{z}}$), and the viscosity gradient is perpendicular to the force ($\boldsymbol{\nabla}\eta = \beta \boldsymbol{\hat{x}}$).  The spheroid’s orientation can point in any direction, and we state it takes the form $\boldsymbol{p} = [\sin\alpha \cos\phi, \sin\alpha \sin\phi, \cos\alpha]$, where $\alpha$ and $\phi$ are the polar and azimuthal angles, respectively.  From our theory (Eqs. \eqref{eq:M_general2}, \eqref{eqn:rot_mobility}, and \eqref{eq:projector_rotation}), the orientation angles evolve as follows:

\begin{subequations} \label{eqn:angles_perp}
\begin{equation} \label{eqn:alpha_perp}
\frac{d \alpha}{dt} = -\beta \left(\lambda_1 -\lambda_3 \sin^2\alpha + \lambda_4 \cos^2\alpha \right) \cos\phi 
\end{equation}
\begin{equation} \label{eqn:phi_perp}
\frac{d \phi}{d t} = \beta (\lambda_1 + \lambda_4) \cot \alpha \sin\phi
\end{equation}
\end{subequations}
where $\lambda_1, \lambda_3,$ and $\lambda_4$ are the force-rotation mobility coefficients determined in Sec. \ref{sec:theory}.  The translation of the particle obeys the following:

\begin{equation} \label{eqn:trans_perp}
\frac{dx}{dt} = \frac{1}{2} (c_2- c_1) \sin(2\alpha) \cos\phi; \qquad \frac{dy}{dt} = \frac{1}{2} (c_2- c_1) \sin(2\alpha) \sin\phi; \qquad \frac{dz}{dt} = c_1 \sin^2 \alpha + c_2 \cos^2 \alpha
\end{equation}
where $c_1$ and $c_2$ are the mobility coefficients for particle translation in Stokes flow (given in Appendix B in dimensional form).  Major conclusions are given below.

\subsubsection{Particle can take a steady orientation different than the force and viscosity gradient directions}

\begin{figure}
\centering
\subfloat[Prolate spheroid]{\includegraphics[width =0.5\linewidth]{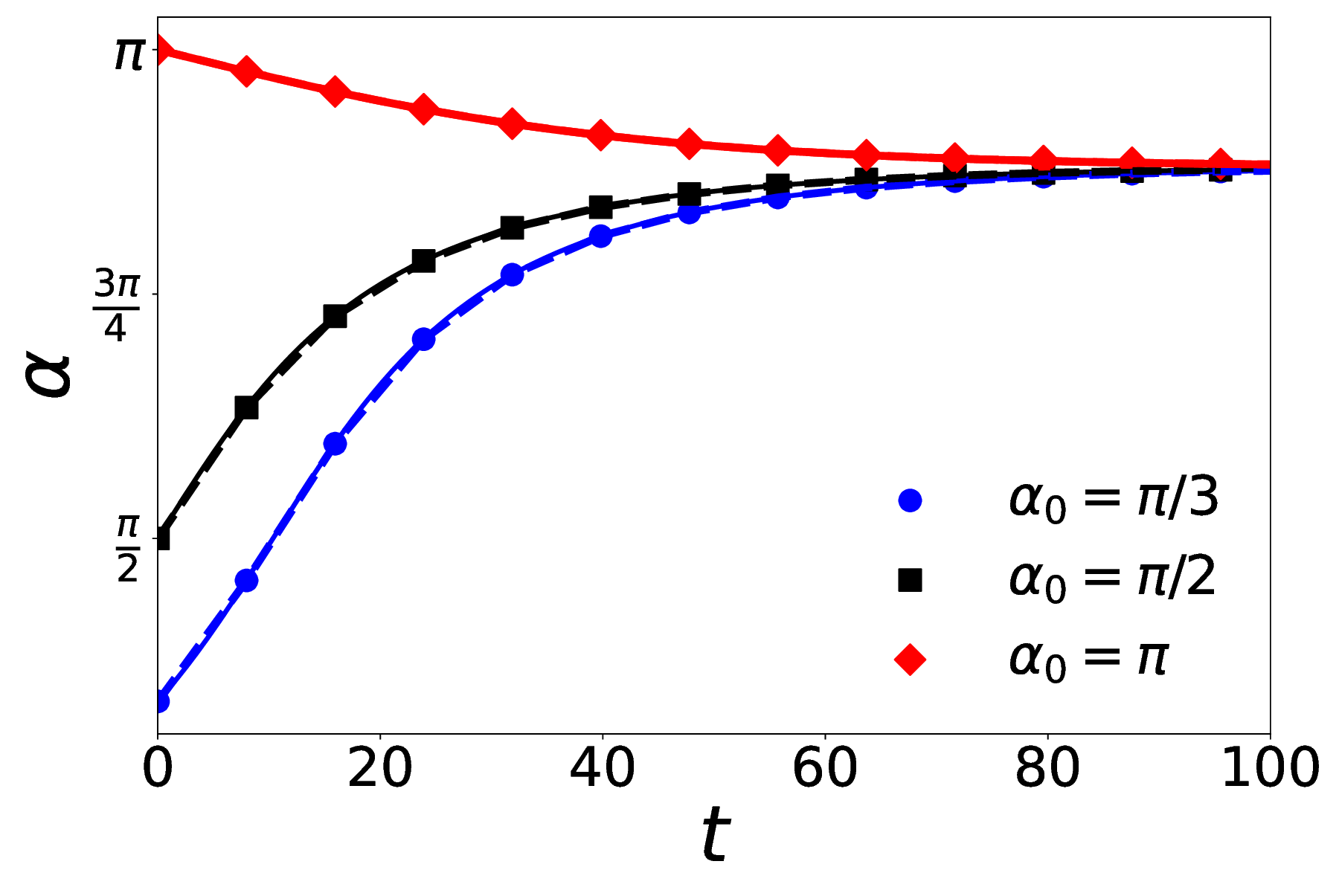}}
\subfloat[Oblate spheroid]{\includegraphics[width =0.5\linewidth]{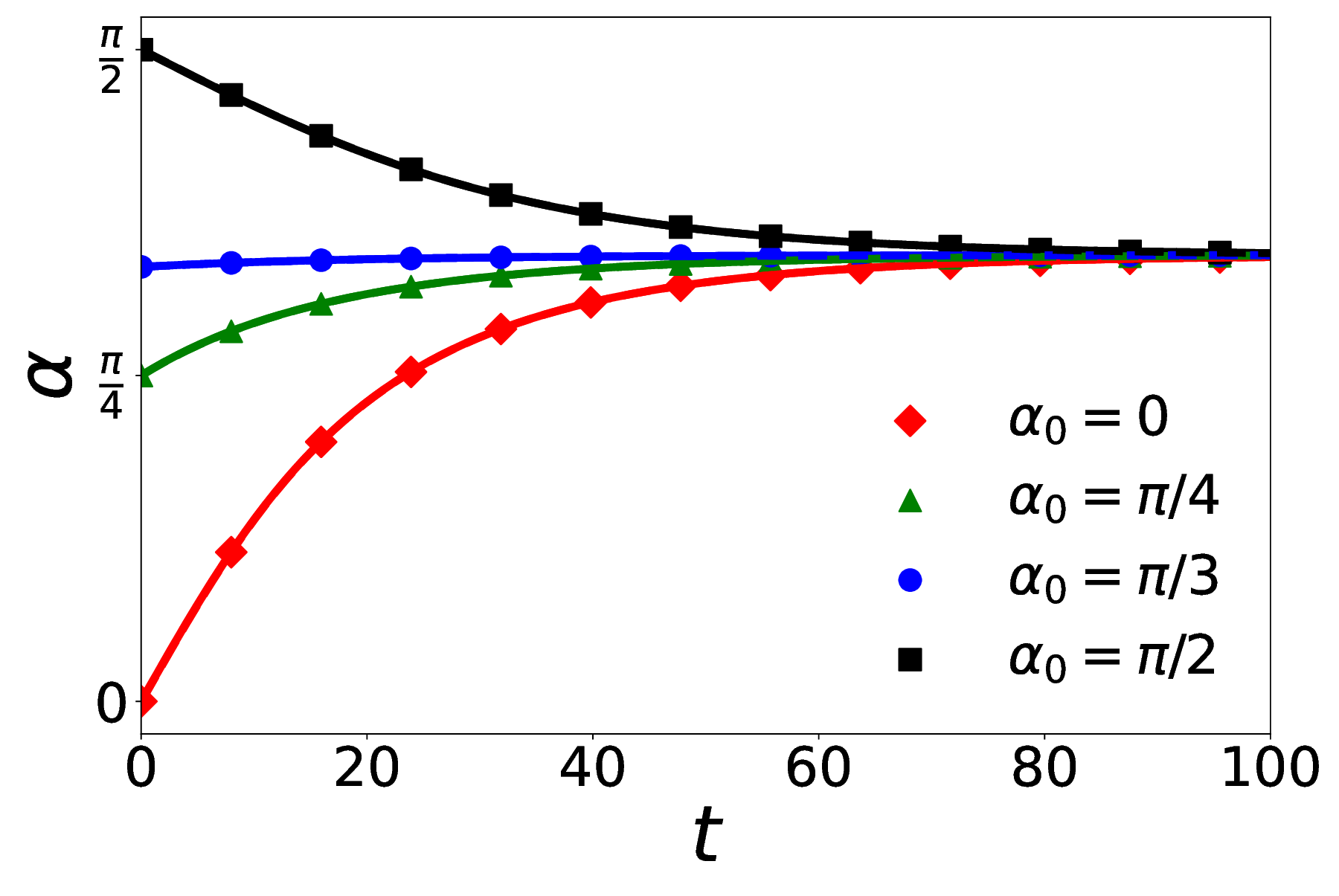}}
\caption{Orientation angle $\alpha$ vs. time for prolate ($A_R =5$) and oblate ($A_R =1/5$) spheroids when the external force and viscosity gradient are perpendicular ($\boldsymbol{F} = \boldsymbol{\hat{z}}, \boldsymbol{\nabla}\eta = \beta \boldsymbol{\hat{x}}$).  The dimensionless viscosity gradient is $\beta =0.1$, and the particle initially starts in the plane of $\boldsymbol{F}$ and $\boldsymbol{\nabla }\eta$ (i.e., $\phi_0 = 0$).  Solid curves are from full numerical simulations based on the reciprocal theorem, while the dashed curves are from the {symmetry based} theory (solving Eq. \eqref{eqn:angles_perp}). }
\label{fig:Orient_Y}
\end{figure}

\begin{figure}
\centering
\subfloat[Prolate spheroid]{\includegraphics[scale=0.23]{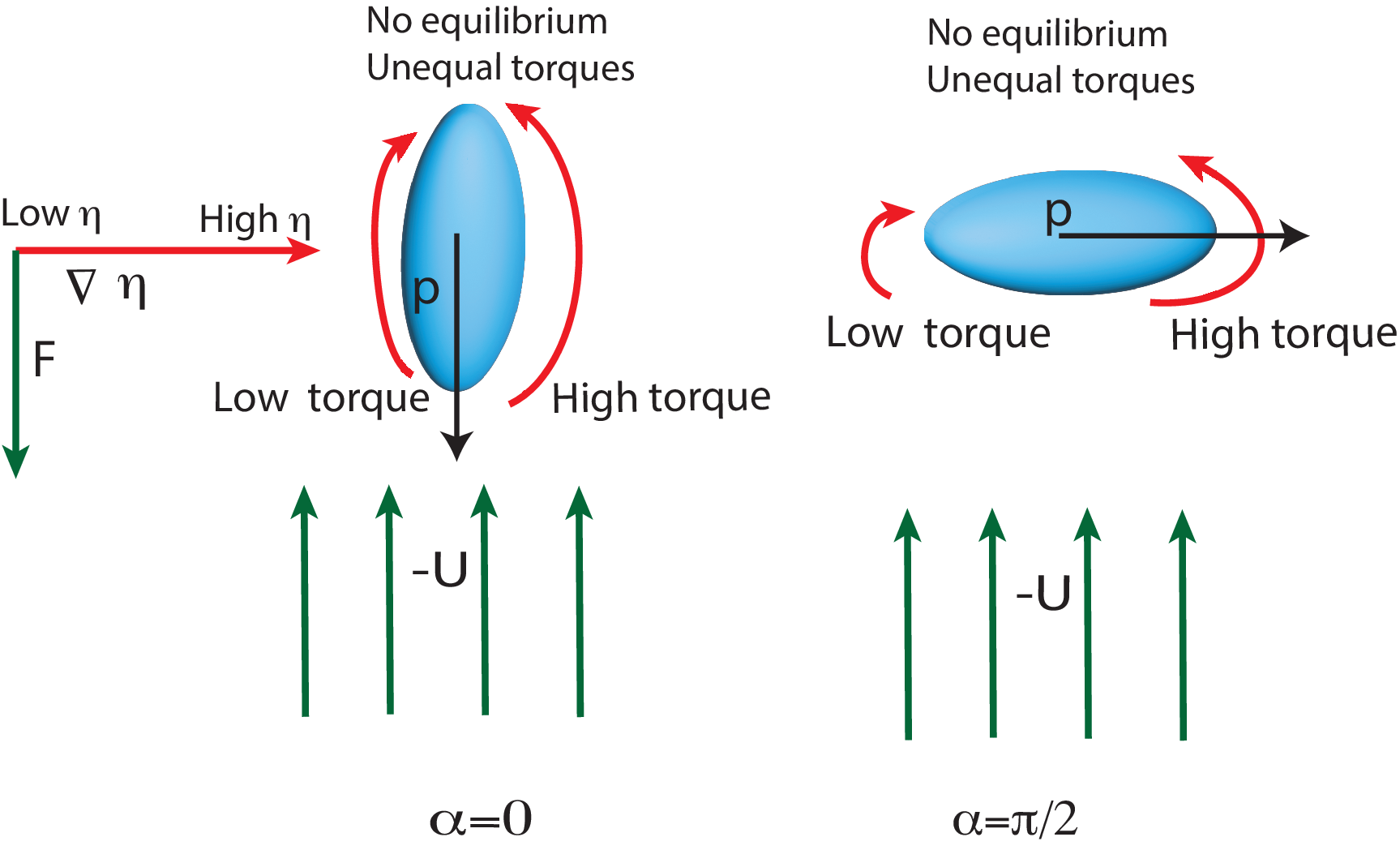}}
\hfill
\subfloat[Oblate spheroid]{\includegraphics[scale =0.23]{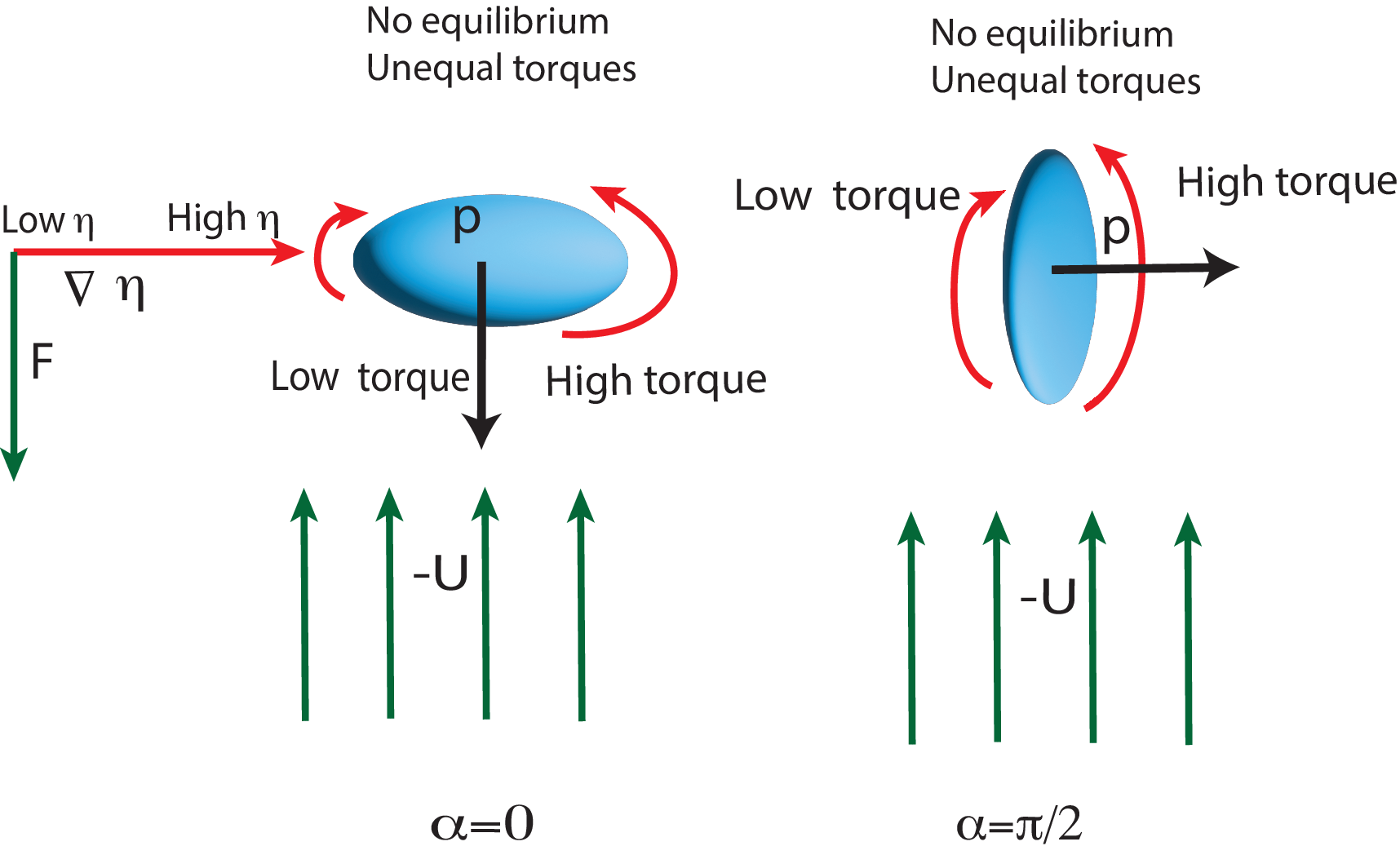}}

\caption{Schematic explaining the absence of steady orientations at $\alpha =0$ and $\alpha =\pi/2$ for (a) prolate  and (b) oblate spheroids when the external force and viscosity gradient are perpendicular. This schematic is shown in the particle's frame of reference.  }
\label{fig:Schematic_Explanation_2}
\end{figure}

\begin{figure}
\centering
\subfloat[Prolate, $\beta = 0.1$]{\includegraphics[width=0.45\linewidth]{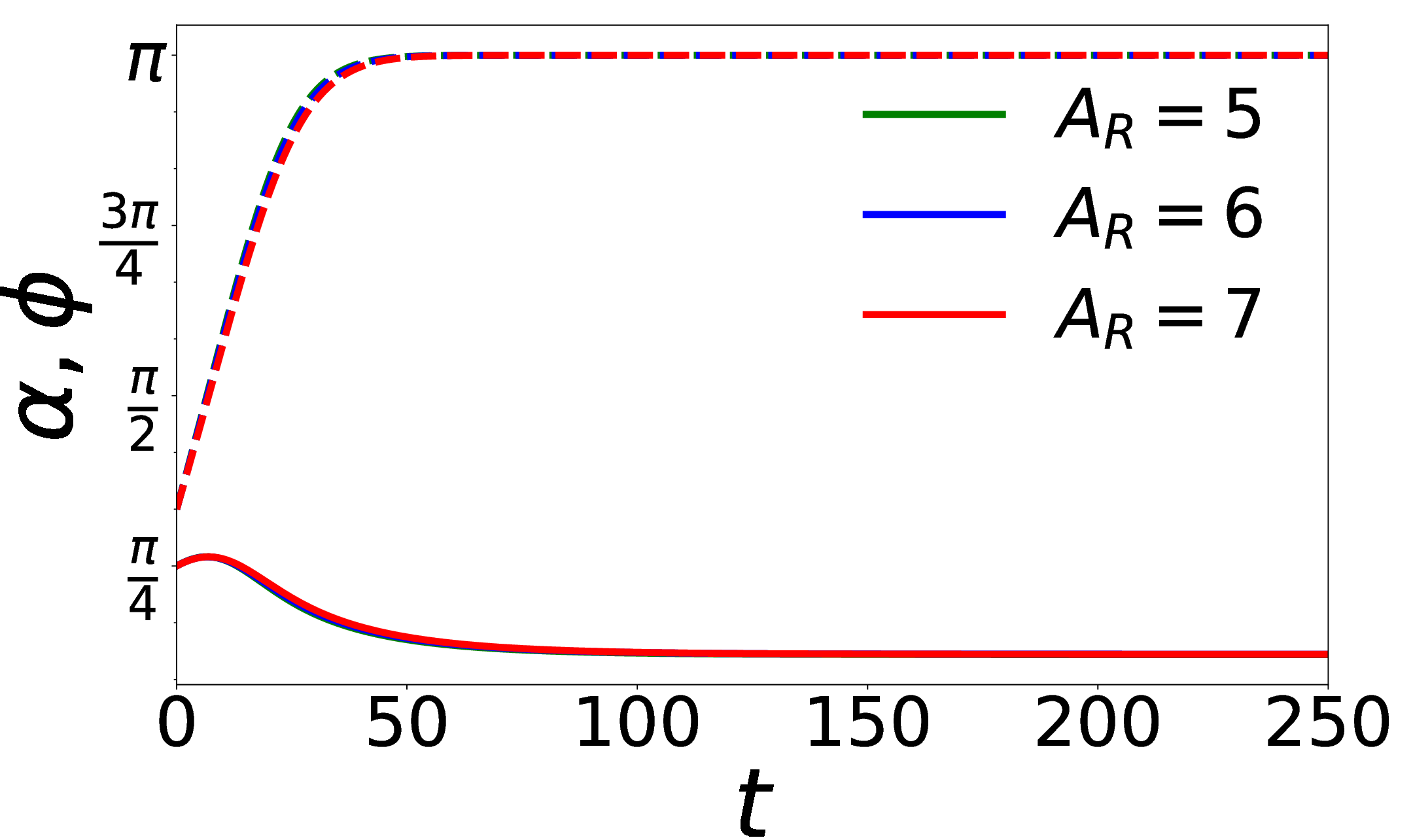}}
\subfloat[Oblate, $\beta = 0.1$]{\includegraphics[width=0.45\linewidth]{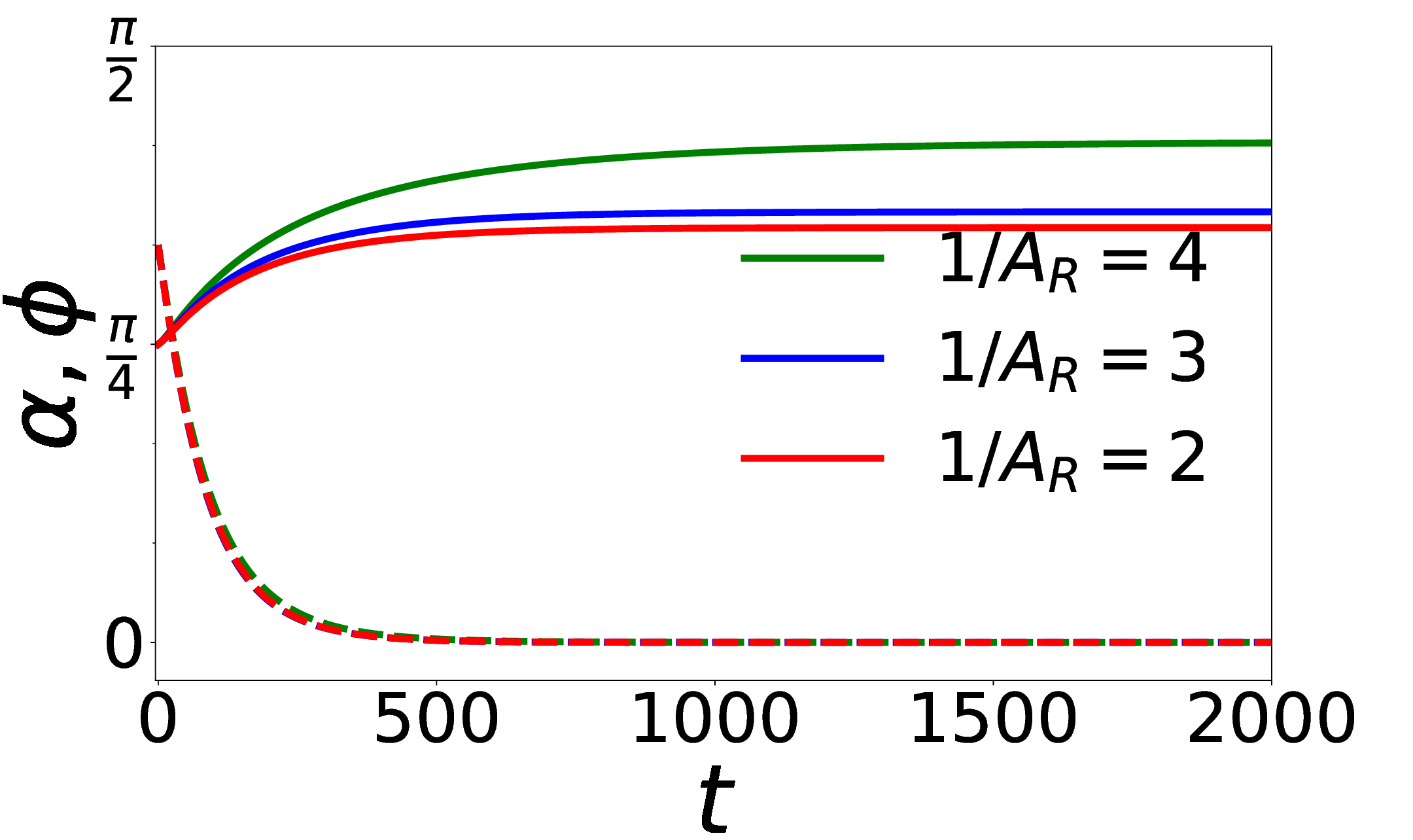}}
\hfill
% \subfloat[prolate, $\beta = -0.1$], {\includegraphics[width=0.45\linewidth]{Prolate_Y_AR_Alpha_Phi_Positive_Initial.eps}}
% \subfloat[oblate, $\beta = -0.1$]{\includegraphics[width=0.45\linewidth]{Oblate_Y_AR_Alpha_Phi_Positive_Initial.eps}}
\caption{Orientation angles $\alpha(t)$ and $\phi(t)$ for prolate and oblate spheroids when the external force and viscosity gradient are perpendicular ($\boldsymbol{F} = \boldsymbol{\hat{z}}, \boldsymbol{\nabla}\eta = \beta \boldsymbol{\hat{x}}$).  The dashed curves show the evolution of $\phi$, while the solid curves show the evolution of $\alpha$. For all cases, the initial orientation is given by the ordered pair $(\phi_0,\alpha_0 )=(\pi/3,\pi/4)$ and the dimensionless viscosity gradient is $\beta =0.1$.  The results show that $\phi \rightarrow 0$ or $\pi$, and hence the particle becomes co-planar with $\boldsymbol{F}$ and $\boldsymbol{\nabla}\eta $.}.
\label{fig:Orient_Y_NoPlane}
\end{figure}

We will first discuss the case when the particle starts in the same plane as $\boldsymbol{F}$ and $\boldsymbol{\nabla} \eta$ – in other words $\phi_0 = 0$.  From Eq. \eqref{eqn:phi_perp}, we see that $d \phi/dt = 0$ for this angle, so the particle stays at $\phi = 0$ and only the polar angle $\alpha$ will change.  Fig. \ref{fig:Orient_Y} plots $\alpha$ versus time for both prolate and oblate spheroids, for the specific case of $A_R = 5$ and $A_R = -1/5$, respectively.  First of all, we note that the results from the {symmetry based} theory (solid curve, Eq. \eqref{eqn:angles_perp}) are virtually indistinguishable from the full numerical simulation (dashed curve), indicating the validity of our theory.  Secondly, for all starting conditions, we observe the particle converges to one steady orientation.  However, this steady orientation is not $\alpha =0$, $\alpha = \pi$, or $\alpha =\pi/2$, which was the case when the force and viscosity gradient vectors were co-linear.

We elucidate this point more clearly in Fig. \ref{fig:Schematic_Explanation_2}. Here, we observe that neither $\alpha =0,\pi/2,$ or $\pi$ are steady configurations because the counter-clockwise torque is different than the clockwise torque at these specific angles.  Some general trends are described below for prolate and oblate particles:

\begin{itemize}
    \item \underline{Prolate spheroids}:  For prolate spheroids, we observe from Fig.~\ref{fig:Schematic_Explanation_2} that the difference between the counter-clockwise and clockwise torques is smaller for $\alpha =0$ and $\pi$ (where the long axis is along the force direction) compared to $\alpha =\pi/2$ (where the long axis is along the viscosity gradient direction). Therefore, the steady orientation is closer to $\alpha =0$ and $\pi$ than to  $\alpha =\pi/2$, and continues to approach $\alpha =0$ or $\pi$ as the aspect ratio increases. \textit{In the limiting case of needle like particles where $A_R \to \infty$ the steady orientation reaches $\alpha =n\pi$}. Between the two configurations of $\alpha =0+\Delta$ and $\alpha =\pi-\Delta$ (where $\Delta$ is a positive constant depending on aspect ratio), $\alpha =\pi-\Delta$ is the stable configuration, while $\alpha =0+\Delta$ is unstable (see Fig.\ref{fig:Orient_Y}(a)).

    \item \underline{Oblate spheroids}:  For oblate spheroids, the difference in hydrodynamic torques is larger at $\alpha =0$ compared to $\alpha =\pi/2$, because in the former case the longer axis is oriented along the viscosity gradient direction. Therefore, for oblate spheroids, the equilibrium orientation configuration is closer to $\alpha = \pi/2$ than to $\alpha = 0$. \textit{In the limiting case of a thin disc where $A_R \rightarrow 0$, the stable orientation is at $\alpha =\pi/2$}. Between the two configurations of $\alpha =\pi/2 \pm \xi$ (where $\xi$ is a positive constant depending on the aspect ratio), $\alpha = \pi/2 -\xi$ is the stable orientation, while $\alpha = \pi/2 +\xi$ is an unstable orientation (see Fig.~\ref{fig:Orient_Y}(b))
\end{itemize}

The results discussed above illustrate the dynamics when the initial particle orientation is co-planar with $\boldsymbol{F}$ and $\boldsymbol{\nabla} \eta$ – i.e., $\phi_0 = 0$ or $\phi = \pi$.  Fig. \ref{fig:Orient_Y_NoPlane} plots the orientation angles $\phi$ and $\alpha$ over time when the starting angle is no longer co-planar with $\boldsymbol{F}$ and $\boldsymbol{\nabla} \eta$ – i.e., $\phi_0 \neq 0$ or $\pi$.  We see that at long times, the angle $\phi \rightarrow 0$ or $\pi$ -- i.e., the orientation ends up in the same plane as $\boldsymbol{F}$ and $\boldsymbol{\nabla} \eta$.  The angle $\alpha$ also converges to the same result as before.  Thus, we conclude that the steady orientation angles discussed previously are stable to out of plane perturbations.  
\subsubsection{Stable orientation for different aspect ratios}

Fig. \ref{fig:Orient_Y_Stable_Prolate_AR} plots the steady orientation angles for prolate and oblate spheroids for different aspect ratio parameters.  The steady orientations occur when $\frac{d \alpha}{dt} = 0$ and $\frac{d \phi}{d t} = 0$ in Eq. \eqref{eqn:angles_perp}, which corresponds to the criterion:
{\begin{subequations}
\begin{align}
\phi = n\pi  \qquad \lambda_1 - \lambda_3 \sin^2\alpha + \lambda_4 \cos^2 \alpha & = 0 \Rightarrow \sin^2{\alpha} =\frac{\lambda_4+\lambda_1}{\lambda_3+\lambda_4} \\
& 0<\frac{\lambda_4+\lambda_1}{\lambda_3+\lambda_4} <1
\end{align}
\end{subequations}}

In the above equation, ($\lambda_1, \lambda_3,\lambda_4$) are the mobility coefficients for force-rotation coupling that were calculated in Sec. \ref{sec:theory}, which are only functions of the aspect ratio parameter $A_R$.  Fig. \ref{fig:Orient_Y_Stable_Prolate_AR}  show that for a wide range of $A_R$, the above criterion is satisfied and a steady angle $\alpha_{se}$ exists.  The stable orientation $\alpha_{se}$ for a prolate spheroid is closer to $\pi$ compared to an oblate spheroid, while the oblate spheroid has a stable angle closer to $\pi/2$.
%This can also be verified by performing a linear stability analysis on Eq. \eqref{eqn:angles_perp} for $\phi = 0$ or $\pi$. 

%\subsubsection{Not all spheroids have a steady orientation}%

\begin{figure}
\centering
\subfloat[Prolate]{\includegraphics[width=0.45\linewidth]{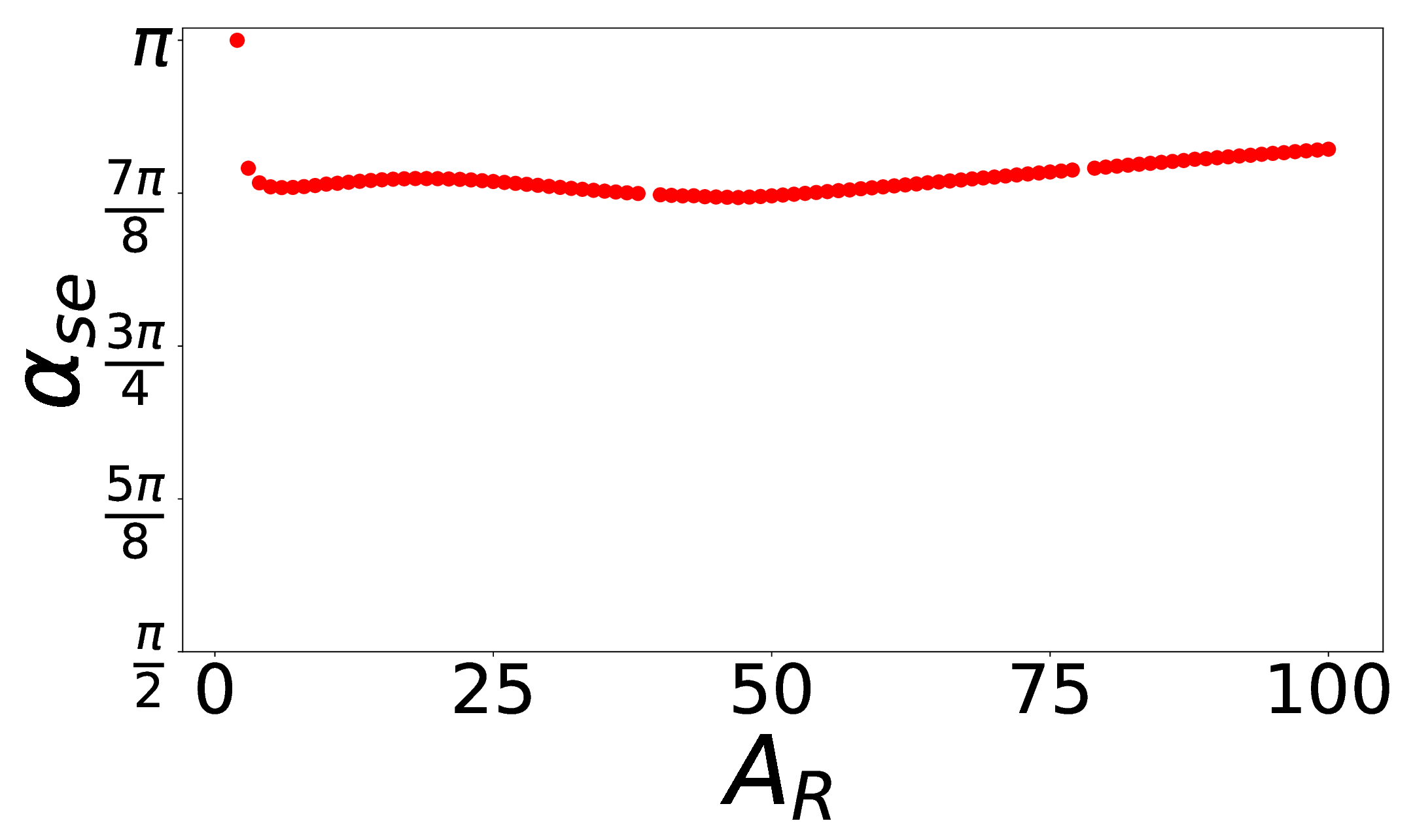}}
\subfloat[Oblate]{\includegraphics[width=0.45\linewidth]{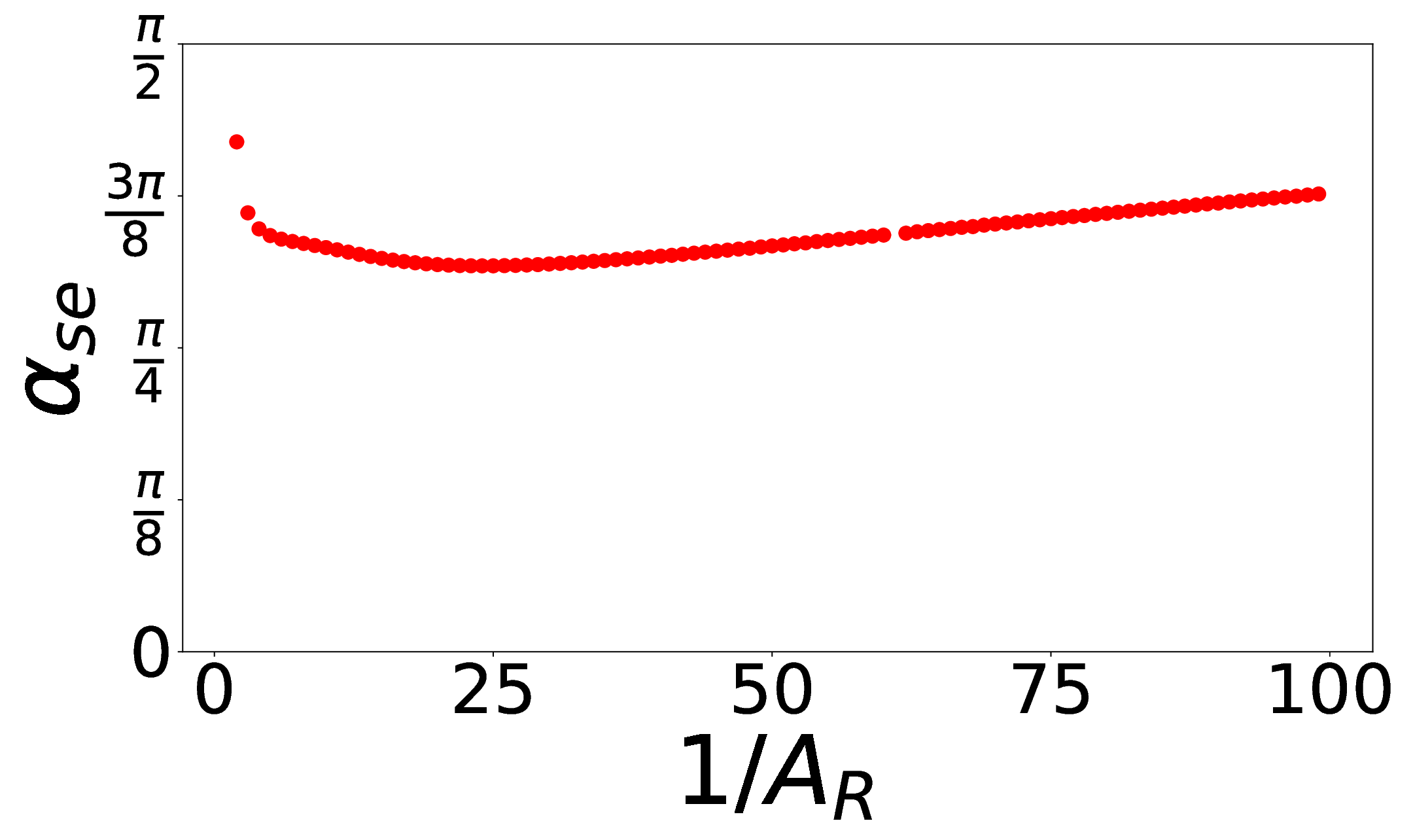}}
\caption{Stable orientations $\alpha_{se}$ for prolate and oblate spheroids of different aspect ratio parameters $A_R$ when the external force and viscosity gradient are perpendicular to each other ($\boldsymbol{F} = \boldsymbol{\hat{z}}$ $\boldsymbol{\nabla}\eta = \beta \boldsymbol{\hat{x}}$, $\beta = 0.1$).}
\label{fig:Orient_Y_Stable_Prolate_AR}
\end{figure}

%\begin{figure}
%\centering
%\subfloat[Prolate, $A_R =13$]{\includegraphics[width=0.45\linewidth]{Prolate_Y_Positive_Analytical2_unstableR.eps}}
%\subfloat[Oblate, $A_R =1/13$]{\includegraphics[width=0.45\linewidth]{Oblate_Y_Positive_Analytical2_unstableR.eps}}
%\caption{Tumbling of (a) prolate spheroids and (b) oblate spheroids when the external force and viscosity gradient are perpendicular to each other ($\boldsymbol{F} = \boldsymbol{\hat{z}}$ $\boldsymbol{\nabla}\eta = \beta \boldsymbol{\hat{x}}$, $\beta = 0.1$). For the aspect ratio parameter shown in this figure ($A_R =13$ for prolate and $A_R =1/13$ for oblate), there is no stable orientation and the spheroids continue to tumble.}
\label{fig:Orient_Y_Unstable}
%\end{figure}

%\begin{figure}
%\centering
%\includegraphics[width=0.6\linewidth]{Unstable_TimePeriod_Ar.eps}
%\caption{{Time period of tumbling of prolate spheroids for the case when there is no stable orientation when the viscosity gradient and external force are perpendicular to each other ($\boldsymbol{F} = \boldsymbol{\hat{z}}$ $\boldsymbol{\nabla}\eta = \beta \boldsymbol{\hat{x}}$, $\beta = 0.1$, $\alpha_0 =\pi/4$)}}
%\label{fig:Time_Period_Unstable_Prolate}
%\end{figure}
\begin{figure}
\centering
\subfloat[Prolate $A_R=5$]{\includegraphics[width=0.45\linewidth]{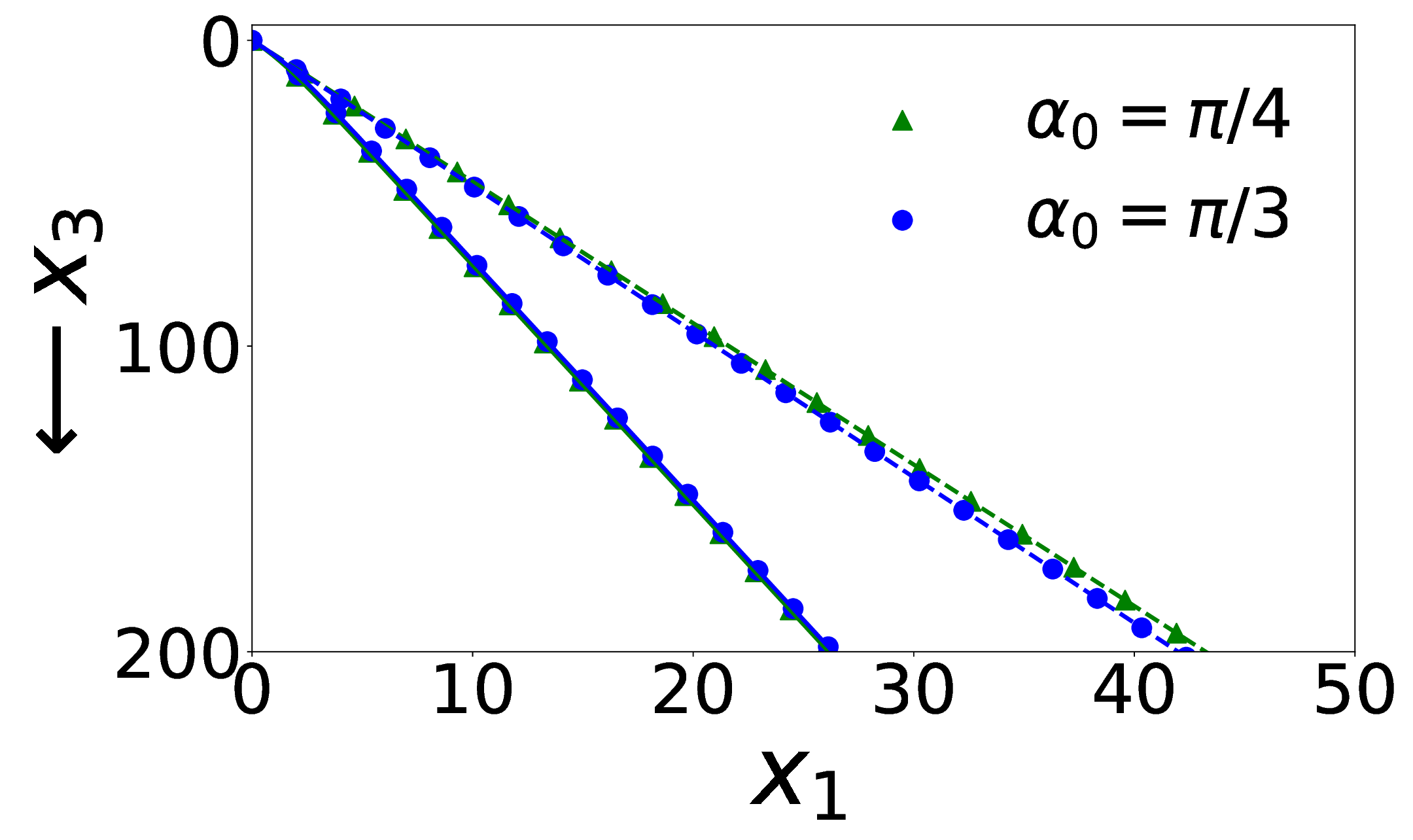}}
\subfloat[Oblate $1/A_R=5$]{\includegraphics[width=0.45\linewidth]{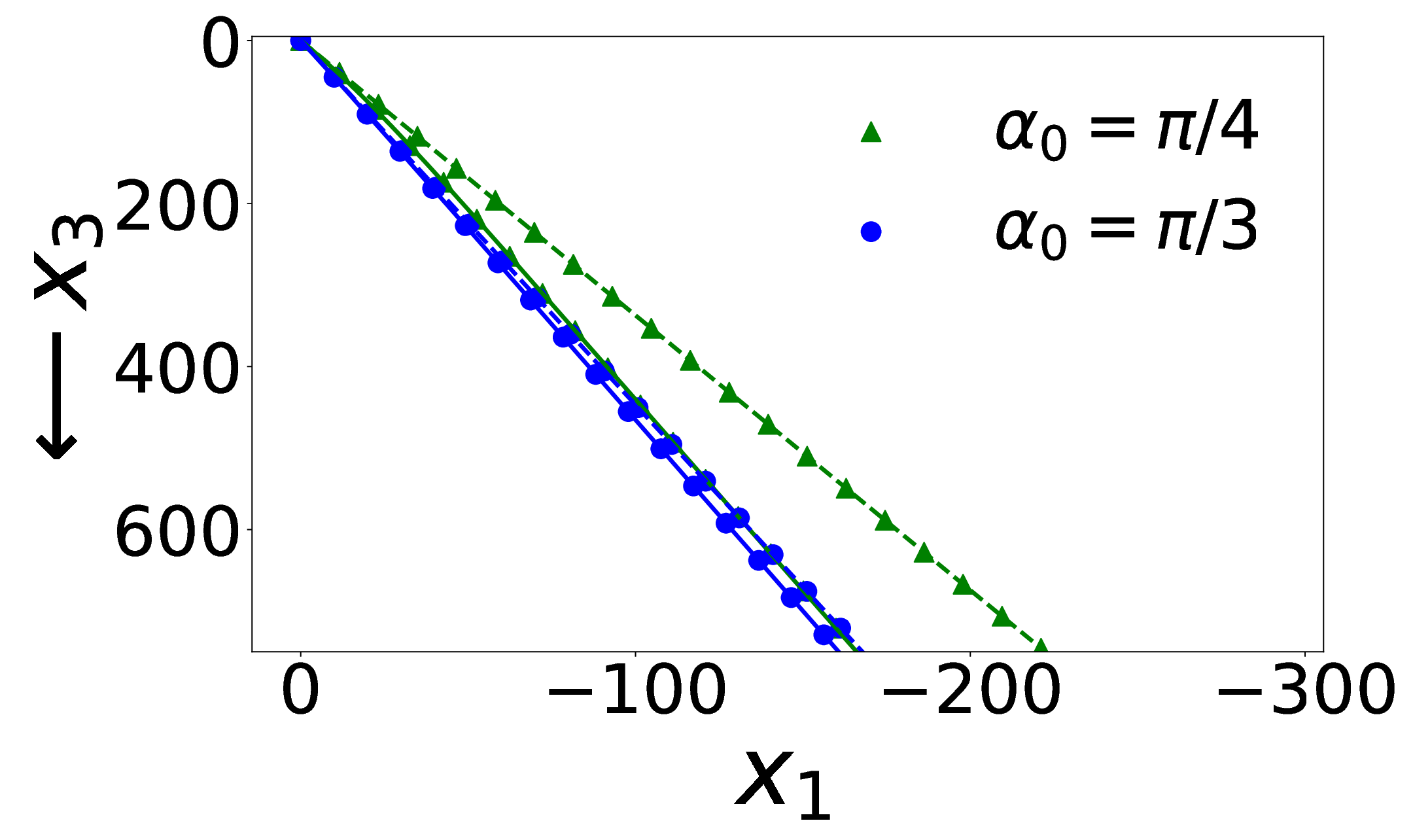}}
\caption{Particle trajectories for  spheroids with (a) $A_R =5$ and (b) $1/A_R =5$ when the external force and viscosity gradient are perpendicular ($\boldsymbol{F} = \boldsymbol{\hat{z}}, \boldsymbol{\nabla}\eta = \beta \boldsymbol{\hat{x}}$). The dashed curves correspond to when no viscosity gradient is present ($\beta = 0$), while the solid curve is when a viscosity gradient is present ($\beta = 0.1$). Different color curves correspond to different initial starting angles $\alpha_0$.}
\label{fig:Translation_Y}
\end{figure}

\subsubsection{Translation dynamics}
Fig.~\ref{fig:Translation_Y} shows the spheroid’s translation trajectories for the case when the force and viscosity gradient are perpendicular to each other for both prolate and oblate spheroids.  For both these spheroids, when the particle has a stable orientation ($A_R =5$), the particle at long times will move in a straight, diagonal line – i.e., sediment downwards and also have a component along the viscosity gradient direction.  This diagonal motion qualitatively looks similar to the motion when the spheroid is in a constant viscosity fluid \citep{Leal2007}.  However, in a constant viscosity fluid, the angle of motion is determined by the particle’s initial angle, whereas in this case, all particles will eventually move with the same trajectory, regardless of starting angle. 
%Conversely, in Fig.~\ref{fig:Translation_Y} (b) when the particle is at an aspect ratio that does not have a steady orientation, the particle will tumble throughout its sedimentation.  In this case, the particle’s motion will sediment in the gravity direction, but its trajectory will oscillate in the viscosity gradient direction (1-direction), with the oscillation period scaling with the tumbling time. 

\subsection{General case:  general direction for viscosity gradient}
\subsubsection{Governing equations}
We now consider the most general case where $\boldsymbol{F}$ and $\boldsymbol{\nabla}\mathbf{\eta}$ are neither parallel or orthogonal to each other, but are inclined at an angle $\theta$ to each other. The external force points in the positive $z$-direction $\boldsymbol{F} = \hat{\boldsymbol{z}}$, while the viscosity gradient is as follows:
\begin{equation}
\boldsymbol{\nabla}\eta = \beta \boldsymbol{\hat{d}} = \beta \cos \theta \boldsymbol{\hat{z}} + \beta \sin \theta \boldsymbol{\hat{x}}
\end{equation}

Similar to before, the orientation vector is $\boldsymbol{p} = [\sin \alpha \cos\phi, \sin\alpha \cos\phi, \cos\alpha]$, where $\alpha$ and $\phi$ are the polar and azimuthal angles.  To determine how these angles evolve over time, we note that the dynamics are a linear superposition of the cases described previously.  In other words,

\begin{subequations}
\begin{align} \label{eqn:angle_general}
\frac{d \alpha}{d t} &= \frac{d \alpha}{d t}|_{\parallel} \cos \theta +  \frac{d \alpha}{d t}|_{\perp} \sin \theta \\
\frac{d \phi}{d t} &= \frac{d \phi}{d t}|_{\parallel} \cos \theta +  \frac{d \phi}{d t}|_{\perp} \sin \theta
\end{align}    
\end{subequations}
where $\frac{d \alpha}{d t}_{\parallel}$ and $\frac{d \alpha}{d t}_{\perp}$ are the variations in the polar angle from viscosity gradients parallel and perpendicular to the external force, given by Eq. \eqref{eqn:angle_parallel} (using the positive sign) and Eq. \eqref{eqn:alpha_perp}, respectively.  The corresponding terms $\frac{d \phi}{d t}_{\parallel}$ and $\frac{d \phi}{d t}_{\perp}$ are the same quantities for the azimuthal angle, which is zero for $\frac{d \phi}{d t}_{\parallel}$ and Eq. \eqref{eqn:phi_perp} for $\frac{d \phi}{d t}_{\perp}$.  The equation for particle translation is the same as Eq. \eqref{eqn:trans_perp}.  

\subsubsection{Steady orientation angles}

\begin{figure}
\centering
\subfloat[Prolate spheroid]{\includegraphics[width =0.48\linewidth]{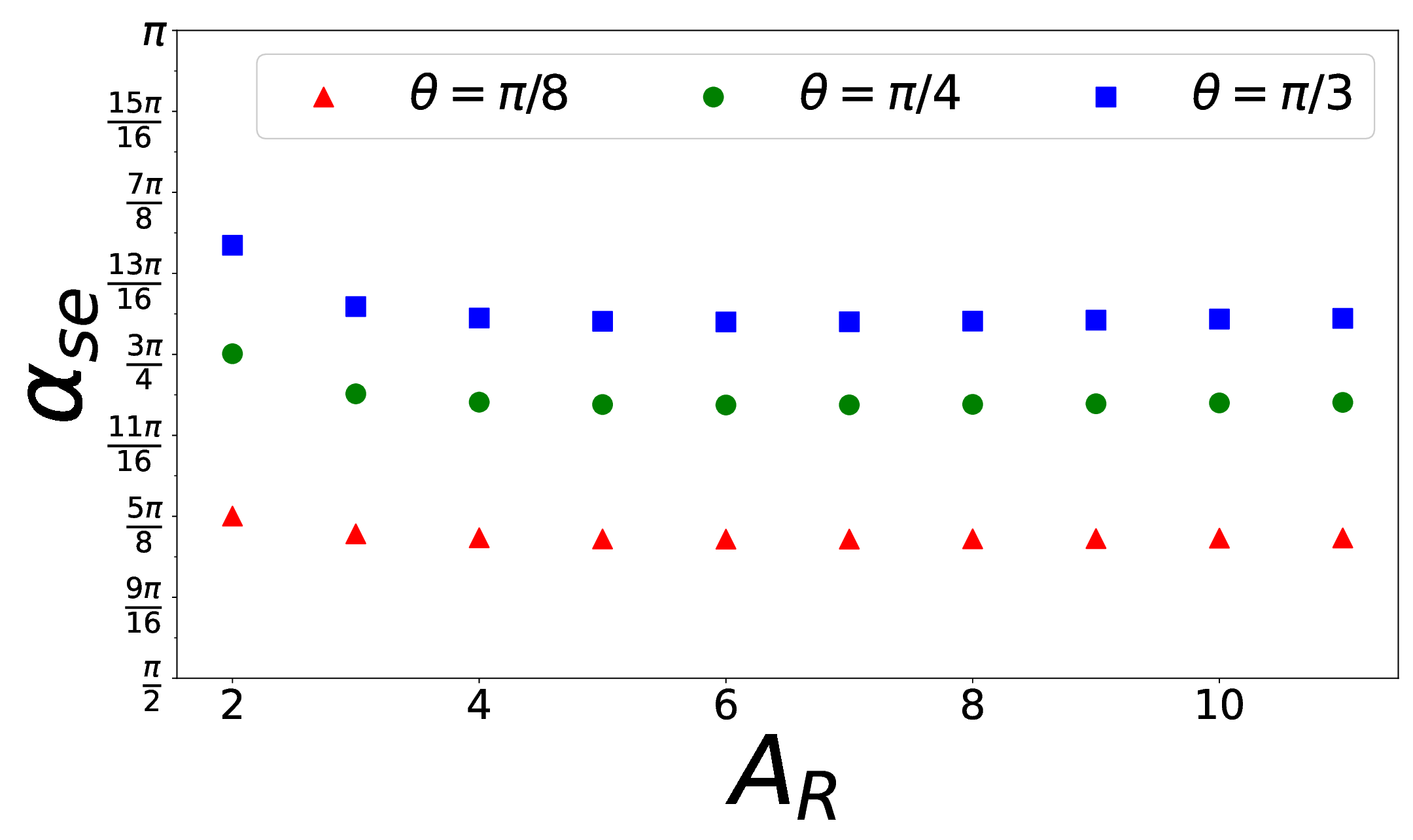}}
\hfill
\subfloat[Oblate spheroid]{\includegraphics[width =0.48\linewidth]{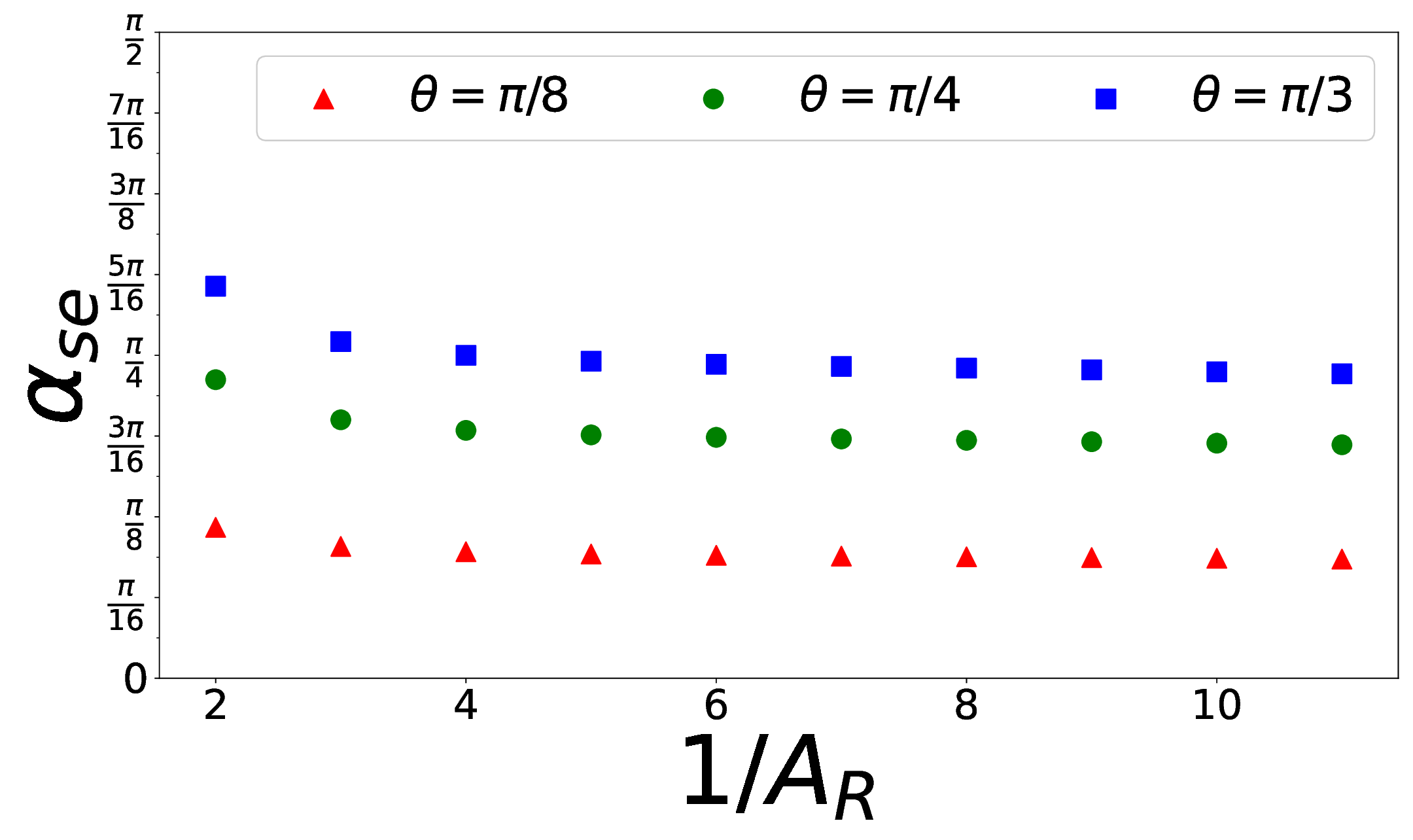}}

\caption{Stable orientation angles $\alpha_{se}$ for prolate and oblate spheroids when the viscosity gradient $\boldsymbol{\nabla}\eta$ and the external force $\boldsymbol{F}$ are inclined at an angle $\theta$ to each other.}
\label{fig:Random_Equilibrium}
\end{figure}

If a steady orientation angle exists, it will be in the plane spanned by $\boldsymbol{F}$ and $\boldsymbol{\nabla} \eta$ as discussed previously – i.e., $\phi = 0$.  We set $\phi = 0$ and determine the conditions under which $\frac{d \alpha}{dt} = 0$ in Eq. \eqref{eqn:angle_general}.  The criterion for a steady orientation angle is:

\begin{equation}
\label{eq:evolution_net_random}
\frac{1}{2} \left(\lambda_3+\lambda_4\right) \sin(2\alpha)\cos{\theta}-\left(\lambda_1-\lambda_3\sin^2\alpha+\lambda_4\cos^2{\alpha}\right)\sin{\theta} = 0.
\end{equation}

For illustration, Fig.~\ref{fig:Random_Equilibrium} plots the steady orientation angles $\alpha_{se}$ for different values of the angle $\theta$ between the external force $\boldsymbol{F}$ and $\boldsymbol{\nabla}\mathbf{\eta}$.  The results are plotted for prolate and oblate spheroids with aspect ratio parameter $A_R = 5$ and $A_R = 1/5$, respectively.  We observe that $\alpha_{se}$ varies between $\pi/2$ and $\pi$ for prolate spheroids and between $0$ and $\pi/2$ for oblate spheroids. As discussed in the previous sections, these limits are the stable orientations for very high aspect ratio spheroids when the viscosity gradients are parallel and perpendicular to external force. For example, as $\theta \to 0$, we see $\alpha_{se} \rightarrow \pi/2$  for prolate spheroids and $0$ for oblate spheroids, which are the stable orientation for these particles when the viscosity gradient is parallel to the external force.

%{For illustrative purpose, we also plot the unstable orientational trajectory in the $\alpha,\phi$ spherical coordinate system in Fig. \ref{fig:Unstable_Trajectory_3D} for different combinations of aspect ratio $A_R$ and $\theta$. We observe that for the cases shown here, the spheroids start from the same initial position $\alpha_0,\phi_0$, follow different trajectories and ultimately descend to $\phi =0$ or $(1-3)$ plane, where they keep rotating and do not reach a stable orientation. }

%{Lastly, Fig. ~\ref{fig:Random_Phase} provides a phase diagram that describes when a steady orientation exists for different particle shapes and viscosity gradient directions}.   When the viscosity gradient is parallel ($\theta = 0$) or anti-parallel ($\theta = \pi$) to the force, there always exists a stable, steady orientation, whereas when the viscosity gradient is perpendicular to the force ($\theta = \pi/2$), there is a range of aspect ratio parameters $A_R$ where steady behavior does not exist.  At other angles, we observe intermediate behavior between the two limits as illustrated in the figure.

\subsection{Discussion of applicability of model and incorporating disturbance viscosity} \label{sec:applicability}

In this paper, we assumed the viscosity field around the particle is a linear function of space and is independent of the flow and the particle geometry. In reality, however, the viscosity field has a more complicated spatial dependence, as it is linked to a scalar field like temperature  or concentration that depends on the aforementioned quantities ({see \citep{Oppenheimer_Stone_2016} for the related problem of a hot sphere in a temperature coupled viscosity field}). In this section, we make suggestions on how to incorporate these effects into the analysis and what changes can be expected to the main results.

For illustrative purposes, let us consider a particle in a fluid subject to a temperature gradient $\nabla T$ far away from the particle.  The fluid’s viscosity depends linearly on temperature -- i.e., $\eta - \eta_0 = \frac{d \eta}{dT} (T - T_0)$, and thus the viscosity field also varies spatially around the particle.  If the thermal Peclet number is small and the temperature profile is steady, the temperature field will satisfy Laplace’s equation inside and outside the particle:
 (see \citep{Dassios_Ellipsoidal} for details):
\begin{equation}
\nabla^2 T^{out} =0;    \qquad \qquad   \nabla^2 T^{in} =0;
\end{equation}

This equation is subject to the following boundary conditions:  (a) $T^{out} \rightarrow T_0 + \nabla T \cdot \boldsymbol{x}$ far away from the particle ($|\boldsymbol{x}| \rightarrow \infty)$, and (b) on the particle surface, the temperatures and fluxes are continuous – i.e., $T^{in} = T^{out}$ and $\left( \boldsymbol{n} \cdot \nabla T^{out} \right) = k_r \left( \boldsymbol{n} \cdot \nabla T^{in} \right)$, where $k_r$ is the conductivity ratio between the particle and fluid phase.  Once one solves the temperature profile, one can obtain the viscosity field $\eta(\boldsymbol{x})$ and then solve for the particle motion in this field.  The rigid body motion will still follow the same procedure discussed earlier in the paper -- i.e., one performs a perturbation expansion for the viscosity and finds the correction to the rigid body motion via the reciprocal theorem using an extra stress tensor $\tau_{ij}^{ex} = (\eta(\boldsymbol{x}) - \eta_0) \gamma_{ij}^{(0)}$.  The mobility tensors described in Sec. \ref{sec:theory} will take the same form, except the numerical values for the force/rotation mobility coefficients $(\lambda_1, \lambda_3, \lambda_4)$ will be different.  For the special cases when one neglects the presence of the particle in the transport equation, or if the conductivity ratio is $k_r = 1$, the viscosity field will be linear everywhere, and we will recover the results described earlier in the manuscript.  {Otherwise, and in general, the viscosity field will have a more complicated spatial and even temporal dependence, emanating from the motion of the particle. Such a viscosity field, which is coupled with the particle motion, posits at least a theoretical possibility of a more complicated orientational dynamics than that has been presented in this paper. However, based on the work carried out previously by \cite{Vaseem_Elfring_Viscosity} for spheres, we speculate that such a motion-coupled viscosity field will not introduce any novel effects but only bolster the trends in orientational dynamics discussed in our paper. Of course, this hypothesis will need to be tested, via a full fledged analysis. The outlines of such an analysis have been presented in this subsection already. A key mathematical ingredient of such an analysis, if taken up later in a different paper, will include the solution to Laplace's equation around an ellipsoidal particle. In Appendix E, we outline how one solves Laplace’s equation around an ellipsoidal particle.}
 
\section{Conclusion} \label{sec:conclusion}

In this paper, we study a spheroid sedimenting in Newtonian fluid with a viscosity field that varies linearly in space.  We employ the principles of linearity, reversibility, symmetry to delineate the mobility relationships for this problem. In the limit of small viscosity gradients, we find that the force/velocity and torque/rotation couplings remain unchanged from the Stokes flow limit.  However, the viscosity gradient gives rise to an additional force/rotation and torque/velocity coupling, which is characterized by a third order tensor $M_{ijk}$. The reduced analytical form of this tensor is given by Eq.~\eqref{eq:M_general2}, up to three undetermined coefficients. The values of these coefficients are determined numerically, under the aegis of a reciprocal theorem-based simulation, for a wide range of particle aspect ratios.

Illustrative examples and specific results of our theory are discussed next. Unlike in Stokes flow where the particle orientation stays at its initial orientation during sedimentation, we find that viscosity gradients alter the orientation over time. When the viscosity gradient is along the external force direction, both prolate and oblate spheroids reach a stable orientation where the longest axis is perpendicular to the viscosity gradient. When the viscosity gradient is opposite the external force, the spheroids reach a stable orientation where the longest axis is along to the viscosity gradient.  We also show that for most initial orientation angles, the spheroid aquires a drift in a direction transverse to its (main) sedimentation direction until its orientation stabilizes, at which point it moves downward. 

When the viscosity gradient and the external force are perpendicular, the plane defined by the viscosity gradient and force is a plane of stability, and the spheroid, irrespective of its initial orientation, will eventually become co-planar with the force and the viscosity gradient. For the limiting case of a needle-like particle, the prolate spheroid will orient its projector in the direction of the force, while conversely, for the limiting case of a flat disk, the oblate spheroid will orient its projector in the direction of the viscosity gradient.  Finally, we note in the general case when the viscosity gradient and external force are neither parallel or perpendicular to each other, the dynamics of the particle is a linear combination of the cases discussed above.

Throughout the analysis, we have neglected the coupling between the viscosity field and the flow or particle motion. Guidelines for incorporating this coupling are presented, using ellipsoidal harmonics to solve the Laplace equation in low $Pe$ limit.  However, based on previous literature \citep{Vaseem_Elfring_Viscosity} we believe that such an analysis may not yield any novel results not yet accounted for.

{We have taken a perturbative approach to solving this problem, where only the terms first order in viscosity gradient are accounted for. Due to the limitation of the perturbative approach , the analysis here is valid for the following two cases: a) when the viscosity gradient is weak , so that the terms of 
 $\mathcal{O}(\beta^2)$ can be neglected. b) when the viscosity gradient is linear, but not necessarily weak, so that the terms of $\mathcal{O}(\beta^2)$ are identically zero. For this second case, where the viscosity gradient is linear, but not weak, we expect that the steady state behavior -- namely the stable orientation of the spheroids -- will remain unchanged. Stronger, but still linear,  viscosity gradients will change the rate at which the stable orientation is attained, but not the value of the steady orientation \textit{per se.}}  Additionally, a spheroid is a typical axisymmetric particle with no isotropy but fore-aft symmetry.  {The mechanisms of hydrodynamic torque acting on the spheroid and the resultant stable orientation as discussed in this paper may therefore prove useful in understanding the dynamics of other orientable particles, which are common in nature and in the industry.}  The current problem may also be a stepping stone towards the analysis of more complex systems, for instance flows with linear and quadratic components, or with density (in addition to viscosity) stratification, among others. {Lastly, we have shown the existence of torque-translation coupling, in addition to the force-rotation coupling, due to the linear viscosity gradient. This coupling opens the exciting possibility of controlling the motion (translation) of  the spheroid by subjecting it to external torque, say via a rotating magnetic field (\citep{Morozov1, Morozov2})}.

\section*{Acknowledgment}
The authors would like to acknowledge support from the American Chemical Society Petroleum Research Fund (DNI-ACS PRF 61266-DNI9), as well as support from the Michael and Carolyn Ott Endowment at Purdue University.

The authors would also like to thank Prof. Gwynn Elfring and his team (specifically Dr. Vaseem Shaik and Mr. Jiahao Gong) at University of British Columbia for constructive feedback and discussion on this manuscript.
\section*{Declaration of Interests}
The authors report no conflict of interest.

\section{Appendix}
\subsection{Appendix A -- Disturbance velocity for an ellipsoid in Stokes flow}
Consider a reference frame at an ellpsoid’s center of mass with axes aligned along the particle’s principle axes.  From \citep[pg.~55]{KimKarilla2005}, the Stokes velocity field around the ellipsoid from external force and torque is the following:

 \begin{subequations} \label{eqn:disturb_vel}
\begin{align}
    v_{i} = \frac{1}{16\pi\eta_0} \sum_{j=1}^3 F_j \left[ \delta_{ij} G_0 - x_j \frac{\partial G_0}{\partial x_i} + \frac{a_j^2}{2} \frac{\partial^2 G_1}{\partial x_i \partial x_j}\right] \\
    v_{i} = \frac{3}{64\pi\eta_0} \sum_{j=1}^3 \left(\boldsymbol{T\times \nabla}\right)_j \left[ \delta_{ij} G_1 - x_j \frac{\partial G_1}{\partial x_i} + \frac{a_j^2}{4} \frac{\partial^2 G_2}{\partial x_i \partial x_j}\right]
\end{align}
\end{subequations}
In these formulas, no summation is assumed for repeated indices unless explicitly stated.  To obtain formulas for $v_{ki}^{trans}$ and $v_{ki}^{rot}$ in the reciprocal theorem, we substitute into Eq. \eqref{eqn:disturb_vel} the force and torque that comes from unit translation and rotation, respectively.

In the above expressions, the expression for $G_n$ is:

\begin{equation}
    G_n(x,y,z) = \int_{\lambda}^{\infty} \left( \frac{x^2}{a^2 + t} + \frac{y^2}{b^2 + t}  + \frac{z^2}{c^2 + t} - 1 \right)^n \frac{dt}{\Delta(t)}
\end{equation}
with $\Delta(t) = \sqrt{(a^2 + t)(b^2 + t)(c^2 + t)}$ and $\lambda(x,y,z)$ being the positive root of 

\begin{equation}
    \frac{x^2}{a^2 + t} + \frac{y^2}{b^2 + t}  + \frac{z^2}{c^2 + t} = 1
\end{equation}

\subsection{Appendix B -- Resistance formulae for an ellipsoid in Stokes flow}

Let us consider a reference frame with the origin at the center of mass of an ellipsoid and the Cartesian axes aligned along the principle axes.

We denote the ellipsoid’s semi-axes as ($a_1,a_2,a_3$) = ($a,b,c$).  In dimensional form, the resistance tensors $R^{FU}$ and $R^{T\omega}$ are diagonal, while the cross-coupling term $R^{F\omega}=R^{TU}=0$.  The diagonal elements are:
\begin{equation}
R_{11}^{F U}=\frac{12 \eta_0 V}{\chi_0+\alpha_1 a_1^2} ; \quad R_{11}^{T \omega}=\frac{4 \eta_0 V\left(a_2^2+a_3^2\right)}{\alpha_2 a_2^2+\alpha_3 a_3^2}
\end{equation}
where $V=\frac{4 \pi}{3} a_1 a_2 a_3$ is the particle volume, and $\left(\chi_0, \alpha_1, \alpha_2 \alpha_3\right)$ are elliptic integrals defined below:
$$
\begin{gathered}
\chi_0=\frac{3}{4 \pi} V \int_0^{\infty} \frac{d t}{\Delta(t)} \\
\alpha_i=\frac{3}{4 \pi} V \int_0^{\infty} \frac{d t}{\left(\mathrm{a}_{\mathrm{i}}^2+\mathrm{t}\right) \Delta(t)} \\
\Delta(t)=\sqrt{\left(a_1^2+t\right)\left(a_2^2+t\right)\left(a_3^2+t\right)}
\end{gathered}
$$
The other elements of the diagonal tensors are obtained by index cycling.

The mobility matrix is the inverse of the resistance matrix, and hence given by the inverse of the diagonal elements above. For the special case when the particle is a spheroid with $a_1 \neq a_2=a_3$, the coefficients $c_1-c_4$ for the mobility matrix in Eq.~\eqref{eq:mobility_tensor_Newtonian_1} and Eq.~\eqref{eq:mobility_tensor_Newtonian_2} are:
$$
c_1=\frac{1}{R_{22}^{F U}} ; \quad c_2=\frac{1}{R_{11}^{F U}} ; \quad c_3=\frac{1}{R_{22}^{T \omega}} ; \quad c_4=\frac{1}{R_{11}^{T \omega}}
$$
These coefficients have analytical formulae (see pgs 64 and 68 in Kim and Karilla).  Using the notation in this paper, we obtain for prolate and oblate spheroids:\\

\begin{itemize}
    \item \underline{Prolate spheroids}
    \begin{subequations}
        \begin{align}
            c_1 &= \frac{1}{6\pi \eta_0 a} \frac{1}{Y_A}; & 
            &Y_A = \frac{16}{3} e^3 \left[2e + (3e^2 -1)L\right]^{-1}  \\     
        c_2 &= \frac{1}{6\pi \eta_0 a} \frac{1}{X_A} & &X_A = \frac{8}{3} e^3 \left[-2e + (1+e^2)L\right]^{-1}\\
       c_3 &= \frac{1}{8\pi \eta_0 a^3} \frac{1}{Y_C} & &Y_C = \frac{4}{3} e^3 (2-e^2) \left[-2e + (1+e^2)L\right]^{-1} \\
        c_4 &= \frac{1}{8\pi \eta_0 a^3} \frac{1}{X_C} & &X_C = \frac{4}{3} e^3 (1-e^2) \left[2e - (1-e^2)L\right]^{-1}
        \end{align}
    \end{subequations}
    where $e = \sqrt{1 - \frac{b^2}{a^2}}$ is the spheroid's eccentricity and $L = \ln \left( \frac{1+e}{1-e} \right)$.   To get the non-dimensional form used in the manuscript, we multiply $c_1$ and $c_2$ by $6\pi\eta_0 R = 6\pi \eta_0 (ab^2)^{1/3}$, and multiply $c_3$ and $c_4$ by $6\pi\eta_0 R^3 = 6\pi \eta_0 ab^2$.\\
    
    \item \underline{Oblate spheroids}
        \begin{subequations}
        \begin{align}
            c_1 &= \frac{1}{6\pi \eta_0 b} \frac{1}{Y_A}; &             &Y_A = \frac{8}{3} e^3 \left[(2e^2 + 1) C - e\sqrt{1-e^2}\right]^{-1} \\        
        c_2 &= \frac{1}{6\pi \eta_0 b} \frac{1}{X_A} & &X_A = \frac{4}{3} e^3 \left[(2e^2 - 1) C + e\sqrt{1-e^2}\right]^{-1} \\
        c_3 &= \frac{1}{8\pi \eta_0 b^3} \frac{1}{Y_C} & &Y_C = \frac{2}{3} e^3 (2-e^2) \left[e\sqrt{1-e^2} - (1 - 2e^2) C \right]^{-1} \\
        c_4 &= \frac{1}{8\pi \eta_0 b^3} \frac{1}{X_C} & &X_C = \frac{2}{3} e^3 \left[C - e\sqrt{1-e^2}\right]^{-1}
        \end{align}
    \end{subequations}
    where $e = \sqrt{1 - \frac{a^2}{b^2}}$ is the spheroid's eccentricity and $C = \cot^{-1} \left( \frac{\sqrt{1-e^2}}{e} \right)$.  To get the non-dimensional form used in the manuscript, we multiply $c_1$ and $c_2$ by $6\pi\eta_0 R = 6\pi \eta_0 (ab^2)^{1/3}$, and multiply $c_3$ and $c_4$ by $6\pi\eta_0 R^3 = 6\pi \eta_0 ab^2$.
\end{itemize}

\subsection{Appendix C -- Reciprocal theorem and $O(\beta)$ solution}

To delineate the $\mathcal{O}(\beta)$ correction to the particle kinematics -- i.e., obtain the solution for ($U_i^{(1)}, \omega_i^{(1)}$) -- there are two approaches possible. The brute force approach is to solve the velocity and stress field around the particle, and then integrate the stress on the particle's surface to find the viscosity-stratified force and torque. However, this approach is tedious and analytically intractable for complicated geometries. Instead, we circumvent the calculation of the velocity and the stress field around the particle and directly obtain the viscosity-stratified force and torque using the reciprocal theorem \citep{lealadvanced}.

First, we note that the fluid stress field at $\mathcal{O}(\beta)$  has two parts:
\begin{equation}
 \sigma_{ij}^{(1)} = \dot{\gamma}_{ij}^{(1)} -p^{(1)}\delta_{ij}+\tau_{ij}^{ex}.
 \end{equation}
 One is the Newtonian part given by $\dot{\gamma}_{ij}^{(1)} -p^{(1)}\delta_{ij}$.   The other part is viscosity-stratified, denoted as $\tau_{ij}^{\text{ex}}$ and given by:
\begin{align}
\tau_{ij}^{ex} = (\hat{d}_kx_k)\dot{\gamma}_{ij}^{(0)}
\end{align}
We note the important observation that the viscosity-stratified stress at $\mathcal{O}(\beta)$ depends on the strain rate at leading order.

In the spirit of the reciprocal theorem, we define an auxiliary problem wherein the same particle, at the same location and same orientation, is sedimenting in a Newtonian fluid with a constant (spatially invariant) viscosity. The quantities pertaining to the auxiliary problem are denoted by the $\emph{aux}$ superscript. Therefore, the external force and the torque acting on the particle in the auxiliary problem is given by $F_i^{aux},T_i^{aux}$ and its rigid body motion is given by $U_i^{aux},\omega^{aux}_i$. The flow field around the particle is $v_i^{aux}$, while the stress field is $\sigma_{ij}^{aux}$, expressed as:
\begin{equation}
    \sigma_{ij}^{aux} = \dot{\gamma}_{ij}^{aux} - p^{aux}\delta_{ij}
\end{equation}

Since the stress field of the auxiliary problem and that of the $\mathcal{O}(\beta)$ problem are divergence free:
\begin{equation}
    \frac{\partial \sigma_{ij}^{aux}}{\partial x_j} =\frac{\partial \sigma_{ij}^{(1)}}{\partial x_j} =0,
\end{equation}
or,
\begin{equation}
    v_i^{(1)}\frac{\partial \sigma_{ij}^{aux}}{\partial x_j} =v_i^{aux}\frac{\partial \sigma_{ij}^{(1)}}{\partial x_j} =0,
\end{equation}
Using the product rule, the above equation reduces to:
\begin{equation}
    \frac{\partial v_i^{(1)}\sigma_{ij}^{aux}}{\partial x_j} - \frac{\partial v_i^{aux}\sigma_{ij}^{(1)}}{\partial x_j} =\sigma_{ij}^{aux}\frac{\partial v_i^{(1)}}{\partial x_j}-\sigma_{ij}^{(1)}\frac{\partial v_i^{aux}}{\partial x_j}
\end{equation}
We now substitute the expressions $\sigma_{ij}^{aux} = \dot{\gamma}_{ij}^{aux} - p^{aux}\delta_{ij}$ and $\sigma_{ij}^{(1)} = \dot{\gamma}_{ij}^{(1)}-p^{(1)}\delta_{ij}+\tau_{ij}^{ex,(0)}$ to the right hand side.  Using the identities $\frac{\partial v_i}{\partial x_i} = \frac{\partial v_i^{aux}}{\partial x_i} = 0$ and $\dot{\gamma}_{ij}^{(1)}\frac{\partial v_i^{aux}}{\partial x_j} = \dot{\gamma}_{ij}^{aux}\frac{\partial v_i^{(1)}}{\partial x_j}$, we obtain:
\begin{equation}
    \frac{\partial v_i^{(1)}\sigma_{ij}^{aux}}{\partial x_j} - \frac{\partial v_i^{aux}\sigma_{ij}^{(1)}}{\partial x_j} = -\tau_{ij}^{ex,(0)}\frac{\partial v_i^{aux}}{\partial x_j},
\end{equation}
Next, we integrate the above equation over the volume outside the particle and use the divergence theorem.  This procedure yields:
\begin{equation}
  \int_{\mathcal{S}} n_j v_i^{(1)}\sigma_{ij}^{aux}dS = \int_{\mathcal{S}}n_jv_i^{aux}\sigma_{ij}^{(1)}dS + \int_{V}\tau_{ij}^{ex,(0)}\frac{\partial v_i^{aux}}{\partial x_j}dV,
\end{equation}
where $\mathcal{S}$ is the surface of the particle and $n_j$ is the normal to the particle surface pointing inside the fluid. On particle surface,  $v_i^{(1)}$ and $v_i^{aux}$ are rigid body motion -- i.e., $v_i^{(1)} = U_i^{(1)} + \epsilon_{ijk} \omega_j^{(1)}x_k$ and $v_i^{aux} = U_i^{aux} + \epsilon_{ijk} \omega_j^{aux}x_k$. Substituting these expressions into the surface integrals yield:
\begin{equation} \label{eqn:last_step_reciprocal}
      -F_i^{aux} U_i^{(1)}-T_i^{aux}\omega_i^{(1)} = \int_{V}\tau_{ij}^{ex,(0)}\frac{\partial v_i^{aux}}{\partial x_j}dV
\end{equation}
Note when deriving the above expression, we made use of the fact that the force and torque acting on the particle at $O(\beta)$ is zero ($F_i^{(1)} = T_i^{(1)} = 0$).  Lastly, let us write the auxillary force and torque as a linear combination of the rigid body velocities using the resistance tensors for the particle:

\begin{equation} \label{eqn:aux_resistance}
\begin{aligned}
    F_i^{aux} &= R_{ij}^{FU} U_j^{aux} + R_{ij}^{F\omega} \omega_j^{aux} \\
    T_i^{aux} &= R_{ij}^{TU} U_j^{aux} + R_{ij}^{T\omega} \omega_j^{aux}
\end{aligned}    
\end{equation}
where in the above equation, the resistance tensors satisfy the following symmetry relationships:  $\boldsymbol{R}^{FU} = (\boldsymbol{R}^{FU})^T$, $\boldsymbol{R}^{T\omega} = (\boldsymbol{R}^{T\omega})^T$, and $\boldsymbol{R}^{F\omega} = (\boldsymbol{R}^{TU})^T$.  We will also write the auxillary velocity field in the volume integral for Eq. \eqref{eqn:last_step_reciprocal} as a linear combination of the rigid body motions:

\begin{equation} \label{eqn:aux_vel_linear_comb}
    v_i^{aux} = v_{ik}^{trans} U_k^{aux} + v_{ik}^{rot} \omega_k^{aux}
\end{equation}
where $v_{ik}^{rot}$ and $v_{ik}^{rot}$ are the velocity fields in the $i$ direction induced by unit translation or rotation in the $k$ direction.  Substituting Eqs. \eqref{eqn:aux_resistance} and \eqref{eqn:aux_vel_linear_comb} into \eqref{eqn:last_step_reciprocal} and eliminating $U_i^{aux}$ and $\omega_i^{aux}$ yields the final result (Eq. \eqref{eq:Resistance_Order_Beta}) stated in the manuscript.

\subsection{Appendix D -- Simplification of mobility tensor $M_{ijk}$}
Here, we show that Eq.~\eqref{eq:M_general} is equivalent to Eq.~\eqref{eq:M_general2}. To that end, we re-write Eq.~\eqref{eq:M_general} as:
\begin{equation}
\label{eq:M_general_appendix}
    M_{ijk} =\lambda_1\underbrace{\epsilon_{ijk}}_{\text{Term 1}}+\lambda_2\underbrace{p_i\epsilon_{jkq}p_q}_{\text{Term2}}+\lambda_3\underbrace{p_j\epsilon_{ikq}p_q}_{\text{Term3}}+\lambda_4\underbrace{p_k\epsilon_{ijq}p_q}_{\text{Term4}}
\end{equation}
 Without any loss of generality, we assume a particular orientation of the projection vector $p_i$ namely $p_i =\delta_{i1}$ (and so on). Therefore, Eq.~\eqref{eq:M_general_appendix} may be written as:
 \begin{equation}
\label{eq:M_general_appendix}
    M_{ijk} =\lambda_1\underbrace{\epsilon_{ijk}}_{\text{Term 1}}+\lambda_2\underbrace{\delta_{i1}\epsilon_{jk1}}_{\text{Term2}}+\lambda_3\underbrace{\delta_{j1}\epsilon_{ik1}}_{\text{Term3}}+\lambda_4\underbrace{\delta_{k1}\epsilon_{ij1}}_{\text{Term4}}
\end{equation}

Say,

\begin{subequations}
\begin{align}
     M_{ijk}^{(1)} &= \text{Term 1} =\epsilon_{ijk} \\
   M_{ijk}^{(2)}   &= \text{Term 2} =\delta_{i1}\epsilon_{jk1} \\
   M_{ijk}^{(3)}   &= \text{Term 3} =\delta_{j1}\epsilon_{ik1} \\
  M_{ijk}^{(4)}   &= \text{Term 4} =\delta_{k1}\epsilon_{ij1} 
\end{align}
\end{subequations}

To expand the different tensors, term by term, we find that the only nonzero terms in  $ M_{ijk}^{(1)} $ are:
\begin{subequations}
\label{eq:M1_Expand}
\begin{align}
   M_{123}^{(1)}  = M_{312}^{(1)} = M_{231}^{(1)} =1 \\
    M_{132}^{(1)}  = M_{321}^{(1)} = M_{213}^{(1)} =-1  
\end{align}
\end{subequations}

Similarly, the nonzero terms in $ M_{ijk}^{(2)} $ are given as:
\begin{subequations}
\label{eq:M2_Expand}
    \begin{align}
         M_{123}^{(2)} =1 \\
         M_{132}^{(2)} =-1
    \end{align}
\end{subequations}
and the nonzero terms in $ M_{ijk}^{(3)} $ are given as:
\begin{subequations}
\label{eq:M3_Expand}
    \begin{align}
        M_{213}^{(3)} = -1 \\
        M_{312}^{(3)} =1
    \end{align}
\end{subequations}
whilst, the nonzero terms in $ M_{ijk}^{(4)} $ are given as:
\begin{subequations}
\label{eq:M4_Expand}
    \begin{align}
        M_{231}^{(4)} = 1 \\
        M_{321}^{(4)} = -1
    \end{align}
\end{subequations}

From the visual inspection of Eqs.~\eqref{eq:M1_Expand},~\eqref{eq:M2_Expand},~\eqref{eq:M3_Expand},~\eqref{eq:M4_Expand}, we obtain the following relationship:

\begin{equation}
    M_{ijk}^{(1)} = M_{ijk}^{(2)}+M_{ijk}^{(3)}+M_{ijk}^{(4)},
\end{equation}
which means out of the four terms $M_{ijk}^{(1)}$,$M_{ijk}^{(2)}$,$M_{ijk}^{(3)}$ and $M_{ijk}^{(4)}$, only $3$ are linearly independent, and therefore, without loss of generality, we can remove $M_{ijk}^{(2)}$ ($=\text{Term2}$) from Eq.~\eqref{eq:M_general_appendix} (or Eq.~\eqref{eq:M_general}), which leads to, with slight change in notation, Eq.~\eqref{eq:M_general2}.

\subsection{Appendix E -- Solving Laplace equation around an ellipsoid with a far field temperature gradient}

Here we outline how to solve Laplace’s equation around an ellipsoidal particle.  We will use ellipsoidal harmonics, a technique is widely used in electrostatics, and the results in papers \citep{Ellipsoidal_Harmonics1} directly apply here.  Let us consider a frame of reference where the Cartesian coordinate system $(x,y,z)$ aligns with the semi-major axes $(a,b,c)$ of the ellipsoid, with $a \geq b \geq c$.  If the far-field temperature is:

\begin{equation}
T^{\infty}(\boldsymbol{x}) = T_0 + \frac{\partial T}{\partial x} x + \frac{\partial T}{\partial y} y + \frac{\partial T}{\partial z} z
\end{equation}
the solution outside the ellipsoid takes the following form:
\begin{equation} \label{eqn:soln_ellipsoidal_harmonics}
T(\boldsymbol{x}) = T^{\infty}(\boldsymbol{x}) + \sum_{p=1}^3 B_{1p} \mathcal{F}_1^p(\boldsymbol{x})
\end{equation}
In the above equation, $\mathcal{F}_1^p(\boldsymbol{x}) = F_1^p(\xi) E_1^p(\mu) E_1^p(\nu) $ are decaying ellipsoidal harmonics using the ellipsoidal coordinate system $(\xi, \mu,\nu)$, where $\xi = a$ denotes the surface of the ellipsoid.  The functions $E_1^p$ and $F_1^p$ are Lame functions of the first and second kind, defined in publication \citep{Ellipsoidal_Harmonics1}.  In Eq. \eqref{eqn:soln_ellipsoidal_harmonics}, the coefficients $B_{1p}$ are the following:

\begin{subequations}
\begin{align}
B_{11} &= \frac{abc}{3} \frac{1}{kh} \frac{ (1 – k_r)}{1 + L_1^1(a) (k_r -1)} \frac{\partial T}{\partial x}  \\
B_{12} &= \frac{abc}{3} \frac{1}{h\sqrt{k^2 – h^2}} \frac{ (1 – k_r)}{1 + L_1^2(a) (k_r -1)} \frac{\partial T}{\partial y} \\
B_{13} &= \frac{abc}{3} \frac{1}{k \sqrt{k^2 – h^2}} \frac{ (1 – k_r)}{1 + L_1^3(a) (k_r -1)} \frac{\partial T}{\partial z}
\end{align}    
\end{subequations}
where $k = \sqrt{a^2 – c^2}$ and $h = \sqrt{a^2 – b^2}$.  The geometric factors $L_1^{1,2,3}(a)$ take the following form \citep{Ellipsoidal_Harmonics1}:

\begin{equation}
L_1^{1,2,3}(a) = abc \int_a^{\infty} \frac{ d \xi’}{ \left[ E_1^{1,2,3}(\xi’) \right]^2 \sqrt{\left( \xi'^{2} – h^2 \right) \left( \xi'^{2} – k^2\right)} }
\end{equation}

\bibliographystyle{jfm.bst}
\bibliography{references}

\end{document}